\documentclass[num-refs,compress]{wiley-article}
\usepackage{siunitx}
\usepackage{jf-names}
\usepackage{caption}
\usepackage{subcaption}
\usepackage{xspace}
\usepackage{gensymb}
\usepackage{color}
\usepackage{tikz}
\usepackage{relsize}
\usetikzlibrary{arrows, decorations.pathmorphing, backgrounds, positioning,fit,petri, shapes, calc}
\usepackage{pgf}

\papertype{Original Article}
\paperfield{Journal Section}

\newcommand{\argmin}[1]{\underset{#1}{\text{argmin}}\hspace{2mm}}

\newcommand{\mattnorm}[1]{\left \| {#1} \right \|}
\newcommand{\mattnormsquare}[1]{\left \| {#1} \right \|_2^2}

\makeatletter
\newcommand*{\defeq}{\mathrel{\rlap{%
                     \raisebox{0.3ex}{$\m@th\cdot$}}%
                     \raisebox{-0.3ex}{$\m@th\cdot$}}%
                     =}
    
\makeatother

\newcommand{\matt}[1] {{\color{red} #1}}
\newcommand{\ben}[1] {{\color{blue} #1}}

\title{Training a Neural Network for Gibbs and Noise Removal in Diffusion MRI}
\author[1]{Matthew~J.~Muckley}
\author[1,2]{Benjamin~Ades-Aron}
\author[1]{Antonios~Papaioannou}
\author[1]{Gregory~Lemberskiy}
\author[1]{Eddy~Solomon}
\author[1]{Yvonne~W.~Lui}
\author[1]{Daniel~K.~Sodickson}
\author[1]{Els~Fieremans}
\author[1]{Dmitry~S.~Novikov}
\author[1]{Florian~Knoll}

\affil[1]{Center for Advanced Imaging Innovation and Research (CAI$^2$R), Department of Radiology, New York University School of Medicine, New York, NY, 10016 United States}
\ben{\affil[2]{NYU Tandon School of Engineering, Brooklyn, NY, 11201 United States}}

\corraddress{Matthew J. Muckley, Department of Radiology, NYU School of Medicine, New York, NY, 10016, United States}
\corremail{Matthew.Muckley@nyulangone.org}

\fundinginfo{NIH: NIBIB, Awards R01 EB024532 and P41 EB017183; NINDS, Award R01 NS088040}

\runningauthor{Muckley et al.}

\begin{document}
\maketitle

\begin{abstract}
\textbf{Purpose:} To develop and evaluate a neural network-based method for Gibbs artifact and noise removal.

\noindent \textbf{Methods:} A convolutional neural network (CNN) was designed for artifact removal in diffusion-weighted imaging data. Two implementations were considered: one for magnitude images and one for complex images. Both models were based on the same encoder-decoder structure and were trained by simulating MRI acquisitions on synthetic non-MRI images.

\noindent \textbf{Results:} Both machine learning methods were able to mitigate artifacts in diffusion-weighted images and diffusion parameter maps. The CNN for complex images was also able to reduce artifacts in partial Fourier acquisitions. 

\noindent \textbf{Conclusion:} The proposed CNNs extend the ability of artifact correction in diffusion MRI. The machine learning method described here can be applied on each imaging slice independently, allowing it to be used flexibly in clinical applications.
\end{abstract}

\section{Introduction}
\label{sec:introduction}
The Gibbs phenomenon \cite{Wilbraham1848,gibbs1898fourier} and noise are artifacts that affect all magnetic resonance imaging scans. The traditional solution for both low signal-to-noise ratio (SNR) and Gibbs artifacts has been to apply a smoothing filter to the image. Although smoothing can reduce the variance of noise, it cannot directly correct for the Rician-biased signal present in low-SNR regimes, nor can it accurately reproduce images that are Gibbs-artifact-free. Methods using Gegenbauer polynomials have been shown to be able to detect edges and accurately estimate images up to the edge for the purpose of Gibbs removal \cite{archibald2002method}. Subvoxel shifting is another method that can substantially reduce Gibbs ringing in acquisitions without partial Fourier encoding \cite{kellner2016gibbs}. Model-based reconstruction methods \cite{block2008suppression,veraart2016gibbs,basu2006rician,varadarajan2015majorize} can improve SNR and/or reduce Gibbs ringing. In parametric protocols with repetitive acquistions such as diffusion MRI, correlations across the diffusion directions can be used for denoising. Recently, a random matrix theory-based MP-PCA method has been developed that takes advantage of these correlations for noise estimation \cite{Veraart_2015} and removal \cite{veraart2016denoising}. Also for diffusion, the DESIGNER pipeline \cite{ades2018evaluation} based on MP-PCA and subvoxel shifting has been demonstrated and validated.

A drawback of many of these methods is that they rely on models that are not fully general. MP-PCA requires the distribution of the noise to be unaltered by processing, as well as many diffusion directions in order to take advantage of the corresponding correlations \cite{Veraart_2015,veraart2016denoising}, but many clinical protocols only use a small number of diffusion directions. Meanwhile, subvoxel shifting does not perform well on partial Fourier acquisitions \cite{kellner2016gibbs}.
Partial Fourier imaging is a complex case due to the interaction between blurring effects from the undersampling and ringing effects from the Fourier series approximation of the image. The combination of these effects is difficult to model explicitly due to the presence of ringing artifacts of varying frequency and phase \cite{kellner2016gibbs}. 

Machine learning offers a framework for building models that can be applied to simultaneously remove a variety of artifacts in an acquisition-flexible manner. One potential class of models suitable for these tasks is convolutional neural networks (CNNs) \cite{lecun1998gradient,krizhevsky2012imagenet}. CNNs are models composed of simple building blocks --- often convolution kernels only 3 pixels wide followed by an activation function. When many convolution kernels are used together, they can approximate highly nonlinear functions of many variables. The use of simple components makes training such models ideally suited for parallelization and efficient implementation on modern computer architectures. Several recent papers have demonstrated the applicability of CNNs for medical imaging in areas including image reconstruction \cite{hammernik2018learning, schlemper2018deep, ye2018deep, zhu2018image, Knoll2019}, image quality transfer \cite{tanno2017bayesian}, super-resolution \cite{chaudhari2018super}, segmentation \cite{deniz2018segmentation}, and artifact correction \cite{schnurr2019simulation}.

Diffusion MRI is an intriguing test application for Gibbs and noise removal models. Both effects can have drastic effects on diffusion parameter maps. The calculation of higher order diffusion parameters can require the application of \textit{b}-values up to $b = 2500\,$s/mm$^2$ \cite{fieremans2011white}, which, coupled with strong diffusion gradients and long echo times, leads to low signal-to-noise ratio (SNR) \cite{basser2000statistical}. The well-known $\leq 9$\% intensity variation due to Gibbs ringing \cite{Wilbraham1848,gibbs1898fourier} may translate into $\sim 100$\% errors in the estimated signal moments or cumulants, in particular leading to the so-called ``black voxels'' (masked outliers) in kurtosis maps \cite{veraart2016gibbs}. Thus, systemic defects --- such as Gibbs or effects from machine learning-based processing --- are substantially amplified in the corresponding parameter maps.

Here, our aim is to develop and evaluate a machine learning approach for Gibbs and noise removal. We show that the framework of CNNs allows the simultaneous removal of these artifact sources in an acquisition-flexible manner --- specifically, we extend the ability for Gibbs artifact removal to the partial Fourier setting and noise removal to the setting with few diffusion directions. We accomplish this by training the model entirely on simulations of MRI acquisitions with images from the ImageNet data set \cite{deng2009imagenet}. Prior to the evaluation stage of our experiments, networks trained this way have never seen an MR image, mitigating overfitting risk. Our results suggest that, with CNN processing, high-quality parameter maps can be calculated from abbreviated acquisitions with a little more than half of the encoding time required for traditional full acquisitions, opening up new possibilities for MRI. The method can be applied independently to each imaging slice, allowing application to clinical protocols with few repetitive scans. Finally, the simulation framework can be adapted to other acquistion protocols, allowing generalization to other applications.

\section{Methods}
\label{sec:methods}
In MRI, images are reconstructed by sampling a Fourier series at a discrete set of locations where the Fourier series coefficients are the acquired k-space points. This suffers from the Fourier series approximation error that manifests as the Gibbs phenomenon \cite{Wilbraham1848,gibbs1898fourier}. In practice, the MRI scanner does not produce the true, noise-free Fourier series coefficients, but a noise-corrupted version of them. Here, we denote the true image at a given resolution as $\mathbf{m}_d$ and its noisy, Gibbs-corrupted version as $\mathbf{x}_d$ where $d$ is an integer that indicates the $d$th image. Our goal is to learn a functional, $f_{\theta}(\cdot)$, that estimates $\mathbf{m}_d$ from $\mathbf{x}_d$ where $\theta$ are the parameters of the functional. We accomplish this by solving the following optimization problem:
\begin{equation}
\label{eq:optim}
    \hat{\theta} = \argmin{\theta} \sum_{d=1}^D \mattnormsquare{\mathbf{m}_d - f_{\theta} \left ( \mathbf{x}_d \right)},
\end{equation}
where $D$ is the size of the training data set.

Solving (\ref{eq:optim}) requires a data set of image pairs $(\mathbf{m}_d, \mathbf{x}_d), d \in [1, ..., D]$ or a means of generating them. We choose to simulate both $\mathbf{m}_d$ and $\mathbf{x}_d$ from a latent, high-resolution image that comes from the ImageNet data set \cite{deng2009imagenet}.
\begin{figure}[htb]
    \tikzstyle{decision} = [diamond, draw, fill=blue!20, 
    text width=4.5em, text badly centered, node distance=3cm, inner sep=0pt]
    \tikzstyle{block} = [rectangle, draw, 
        text width=11em, text centered, rounded corners, minimum height=2em]
    \tikzstyle{line} = [draw, -latex']
    \tikzstyle{cloud} = [draw, ellipse,fill=red!20, node distance=3cm,
        minimum height=2em]
	\centering
	\begin{tikzpicture}[node distance = 0.7cm, auto]
		\node [block] (init) {ImageNet Input ($256^2$) \\ Random Flipping \\ Random Transpose \\ Random Phase \\ Ellipsoid Cropping};
		\node [block, below of=init, yshift=-1.8cm] (fft) {Forward FFT \\ Gibbs Crop (to $100^2$) \\ Add Noise \\ Partial Fourier Mask \\ Inverse FFT};
        \node [block, below of=fft, xshift=1.65cm, yshift=-.9cm, text width=22.5em] (imnorm2) {Image Normalization};
        \node [block, below of=imnorm2, yshift=-.2cm, xshift=1.65cm] (abs) {Absolute Value \\ Spline Resizing (to $100^2$)};
        \node [block, below of=abs, yshift=-.2cm] (finalm) {$\mathbf{m}_d$};
		\node [block, left of=finalm, node distance=3.3cm] (finalx) {$\mathbf{x}_d$};
		\path [line] ([xshift=-.5cm]init.south) -- ([xshift=-.5cm]fft.north);
        \path [line] (fft) -- ([xshift=-1.65cm]imnorm2.north);
        \path [line] ([xshift=1.65cm]imnorm2.south) -- (abs.north);
		\path [line] ([xshift=-1.65cm]imnorm2.south) -- (finalx);
		\path [line] ([xshift=.5cm]init.south)  -- ([xshift=.5cm,yshift=-0.1cm]init.south) -| ([xshift=1.65cm]imnorm2.north);
		\path [line] (abs) -- (finalm);
	\end{tikzpicture}
	\caption{Simulation pipeline. The processing streams begin with a $256 \times 256$ ImageNet image. Then, after random flipping and transposing, random phase and random ellipsoid cropping are applied. Then, the network input $\mathbf{x}_d$ has Gibbs ringing simulated along with noise and partial Fourier masking, followed by normalization. The target $\mathbf{m}_d$ images have the absolute value operation applied followed by spline resizing to the target image matrix size.}
	\label{fig:simpipeline}
\end{figure}
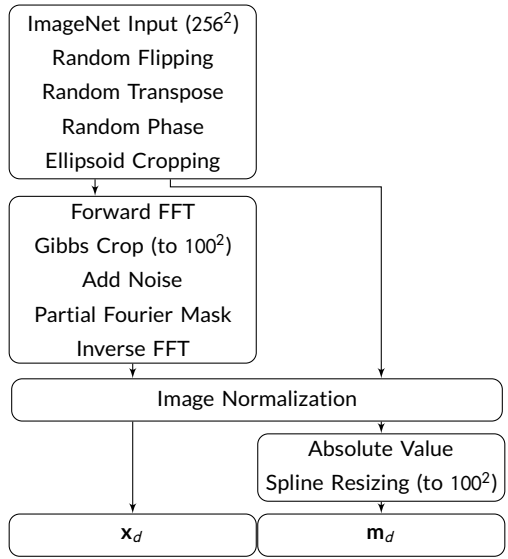
Figure \ref{fig:simpipeline} shows our simulation pipeline. We describe the details of the steps in Section \ref{subsec:simmeth}. Section \ref{subsec:modelmeth} describes the model architecture for $f_{\theta}(\cdot)$. In Section \ref{subsec:expmeth} we describe test experiments that were conducted on data sets completely separate from that used for training, including real-world diffusion validation experiments in which we test the performance of two different networks and compare to state-of-the-art methods across partial Fourier factors. We show results from these experiments in Section \ref{sec:results}.

\subsection{Model Architecture}
\label{subsec:modelmeth}
There are many potential model architectures for accomplishing deGibbsing. Architectures such as variational \cite{hammernik2018learning} and cascaded \cite{schlemper2018deep} networks offer the most flexibility for explicitly modeling system physics. However, these models often require access to the raw data and system parameters, e.g., for sensitivity map estimation. Use of such models may require accessing proprietary information or signing research agreements with vendors. Furthermore, the use of models that rely on raw data can incur substantial data storage requirements. Finally, they preclude the use of the model on previously-acquired data.

\begin{figure*}[htb]
    \captionsetup[subfigure]{justification=centering}
    \centering
    \begin{subfigure}{1\textwidth}
		\centering
        \begin{tikzpicture}
            [
                node distance=0.3cm, >=stealth', very thick,
                cross/.style={path picture={ \draw[black]
                                (path picture bounding box.south east) -- (path picture bounding box.north west)
                                (path picture bounding box.south west) -- (path picture bounding box.north east);
                            }},
                BigRectangle/.style={rectangle,draw=black,fill=white,thick,
                        inner sep=0pt,minimum width=1.2cm, minimum height=0.5cm, rounded corners},
                SmallRectangle/.style={rectangle,draw=black,fill=white,thick,
                        inner sep=0pt,minimum width=0.5cm, minimum height=1cm},
                Circle/.style={circle,draw=black,fill=white,thick,
                        inner sep=0pt,minimum size=6mm},
            ]
            \node[SmallRectangle] (A) [minimum width=0.01cm]       {};
            \node[above of=A,yshift=0.4cm,align=center]           {3};
            \node[SmallRectangle] (B) [right=of A]                 {};
            \node[above of=B,yshift=0.4cm,align=center]           {64};
            \node[BigRectangle]   (C) [right=of B]                 {Dn(64)};
            \node[BigRectangle]   (D) [below=of C,xshift=0.5cm]    {Dn(128)};
            \node[BigRectangle]   (E) [below=of D,xshift=0.5cm]    {Dn(256)};
            \node[BigRectangle]   (F) [below=of E,xshift=0.5cm]    {Dn(512)};
            \node[BigRectangle]   (P) [right=of F, yshift=-0.5cm]    {Int};
            \node[BigRectangle]   (I) [right=of P,yshift=0.5cm]    {Up(256)};
            \node[BigRectangle]   (J) [above=of I,xshift=0.5cm]    {Up(128)};
            \node[BigRectangle]   (K) [above=of J,xshift=0.5cm]    {Up(64)};
            \node[BigRectangle]   (L) [above=of K,xshift=0.5cm]    {Up(64)};
            \node[SmallRectangle]   (M) [right=of L]               {};
            \node[above of=M,yshift=0.4cm,align=center]           {64};
            \node[SmallRectangle] (N) [right=of M,minimum width=0.01cm]  {};
            \node[above of=N,yshift=0.4cm,align=center]           {2};
            \node[Circle] (O) [right=of N] {$+$};
    
            \draw[->,black] ([xshift=-0.4cm]A.west) -- (A.west);
            \draw[->,red] (A.east) -- (B.west);
            \draw[->,blue] (B.east) -- (C.west);
            \draw[->,purple] ([yshift=.16cm]C.east) -- ([yshift=0.16cm]L.west);
            \draw[->,gray] ([yshift=.08cm]C.east)   -- ([yshift=0.08cm]L.west);
            \draw[->,gray] ([yshift=0cm]C.east)     -- ([yshift=0cm]L.west);
            \draw[->,gray] ([yshift=-.08cm]C.east)  -- ([yshift=-0.08cm]L.west);
            \draw[->,blue] ([xshift=0.25cm]C.south) -- ([xshift=-0.25cm]D.north);
            \draw[->,purple] ([yshift=.16cm]D.east) -- ([yshift=0.16cm]K.west);
            \draw[->,gray] ([yshift=.08cm]D.east)   -- ([yshift=0.08cm]K.west);
            \draw[->,gray] ([yshift=0cm]D.east)     -- ([yshift=0cm]K.west);
            \draw[->,gray] ([yshift=-.08cm]D.east)  -- ([yshift=-0.08cm]K.west);
            \draw[->,blue] ([xshift=0.25cm]D.south) -- ([xshift=-0.25cm]E.north);
            \draw[->,purple] ([yshift=.16cm]E.east) -- ([yshift=0.16cm]J.west);
            \draw[->,gray] ([yshift=.08cm]E.east)   -- ([yshift=0.08cm]J.west);
            \draw[->,gray] ([yshift=0cm]E.east)     -- ([yshift=0cm]J.west);
            \draw[->,gray] ([yshift=-.08cm]E.east)  -- ([yshift=-0.08cm]J.west);
            \draw[->,blue] ([xshift=0.25cm]E.south) -- ([xshift=-0.25cm]F.north);
            \draw[->,purple] ([yshift=.16cm]F.east) -- ([yshift=0.16cm]I.west);
            \draw[->,gray] ([yshift=.08cm]F.east)   -- ([yshift=0.08cm]I.west);
            \draw[->,gray] ([yshift=0cm]F.east)     -- ([yshift=0cm]I.west);
            \draw[->,gray] ([yshift=-.08cm]F.east)  -- ([yshift=-0.08cm]I.west);
            \draw[->,blue] ([yshift=-.16cm]F.east)  -- ([xshift=0.1cm,yshift=-0.16cm]F.east) -- ([xshift=-0.2cm]P.west) -- (P.west);
            \draw[->,blue] (P.east)  -- ([xshift=0.1cm]P.east) -- ([xshift=-0.2cm,yshift=-.16cm]I.west) -- ([yshift=-.16cm]I.west);
            \draw[->,blue] ([xshift=.25cm]I.north)  -- ([xshift=-0.25cm]J.south);
            \draw[->,blue] ([xshift=.25cm]J.north)  -- ([xshift=-0.25cm]K.south);
            \draw[->,blue] ([xshift=.25cm]K.north)  -- ([xshift=-0.25cm]L.south);
            \draw[->,blue] (L.east) -- (M.west);
            \draw[->,red] (M.east) -- (N.west);
            \draw[->,black] (N.east) -- (O.west);
            \draw[->,black] (O.east) -- ([xshift=0.4cm]O.east);
            \draw[->,black] ([yshift=0.2cm]A.east) -- ([yshift=0.2cm,xshift=0.2cm]A.east) -- ([yshift=1cm,xshift=0.2cm]A.east) -- ([yshift=0.7cm]O.north) -- (O.north);
    
        \end{tikzpicture}
        \caption{Overall Architecture}
		\label{subfig:overallarchitecture}
    \end{subfigure}
    
    \begin{subfigure}{0.35\textwidth}
		\centering
		\begin{tikzpicture}
			[
				node distance=0.5cm, >=stealth', very thick,
				cross/.style={path picture={ \draw[black]
									(path picture bounding box.south east) -- (path picture bounding box.north west)
									(path picture bounding box.south west) -- (path picture bounding box.north east);
							}},
				BigRectangle/.style={rectangle,draw=black,fill=white,thick,
						inner sep=0pt,minimum width=0.3cm, minimum height=.6cm},
				SmallRectangle/.style={rectangle,draw=black,fill=white,thick,
						inner sep=0pt,minimum height=0.3cm, minimum width=0.3cm},
				Circle/.style={circle,draw=black,fill=white,thick,
						inner sep=0pt,minimum size=6mm},
            ]
            \node[BigRectangle]   (A) [minimum width=0.15cm]       {};
            \node[above of=A,yshift=0.1cm,align=center]           {$N$};
            \node[BigRectangle]   (B) [right=of A]                 {};
            \node[above of=B,yshift=0.1cm,align=center]          {$P$};
            \node[BigRectangle]   (C) [right=of B]                 {};
            \node[above of=C,yshift=0.1cm,align=center]          {$P$};
            \node[SmallRectangle] (D) [right=of C, yshift=-0.5cm]  {};
            \node[above of=D,yshift=-0.1cm,align=center]          {$P$};
			\node[SmallRectangle] (E) [below=of D, yshift=0.4cm]  {};
			\node[SmallRectangle] (F) [below=of E, yshift=0.4cm]  {};
            \node[SmallRectangle] (G) [below=of F, yshift=0.4cm]  {};

			\draw[->,blue] ([xshift=-0.5cm]A.west) -- (A.west);
			\draw[->,red] (A.east) -- (B.west);
			\draw[->,red] (B.east) -- (C.west);
			\draw[->,purple] ([yshift=0.25cm]C.east) -- ([xshift=1.5cm, yshift=0.25cm]C.east);
			\draw[->,orange] (C.east) -- ([xshift=0.25cm]C.east) |- (D.west);
			\draw[->,orange] (C.east) -- ([xshift=0.25cm]C.east) |- (E.west);
			\draw[->,orange] (C.east) -- ([xshift=0.25cm]C.east) |- (F.west);
			\draw[->,orange] (C.east) -- ([xshift=0.25cm]C.east) |- (G.west);
			\draw[->,gray] (D.east) -- ([xshift=.5cm]D.east);
			\draw[->,gray] (E.east) -- ([xshift=.5cm]E.east);
			\draw[->,gray] (F.east) -- ([xshift=.5cm]F.east);
            \draw[->,blue] (G.east) -- ([xshift=.5cm]G.east);

            \node[BigRectangle] (L1) [below of=A,yshift=-1.1cm,xshift=.7cm, align=left, draw=white, minimum height=2.5cm, minimum width=2.5cm] {Pass Images \\ Conv,BN,ReLU \\ Wave Decomp. \\ To Concat. \\ To Wave Trans.};
            
            \draw[->,blue] ([yshift=-.4cm,xshift=-0.6cm]A.south) -- ([yshift=-.4cm,xshift=-0.3cm]A.south);
            \draw[->,red] ([yshift=-.85cm,xshift=-0.6cm]A.south) -- ([yshift=-.85cm,xshift=-0.3cm]A.south);
            \draw[->,orange] ([yshift=-1.3cm,xshift=-0.6cm]A.south) -- ([yshift=-1.3cm,xshift=-0.3cm]A.south);
            \draw[->,purple] ([yshift=-1.75cm,xshift=-0.6cm]A.south) -- ([yshift=-1.75cm,xshift=-0.3cm]A.south);
            \draw[->,gray] ([yshift=-2.2cm,xshift=-0.6cm]A.south) -- ([yshift=-2.2cm,xshift=-0.3cm]A.south);

        \end{tikzpicture}
        \caption{Downsampling, Dn($P$)}
		\label{subfig:downblock}
	\end{subfigure}
	\begin{subfigure}{0.25\textwidth}
		\centering
		\begin{tikzpicture}
			[
            node distance=0.5cm, >=stealth', very thick,
            cross/.style={path picture={ \draw[black]
                                (path picture bounding box.south east) -- (path picture bounding box.north west)
                                (path picture bounding box.south west) -- (path picture bounding box.north east);
                        }},
            BigRectangle/.style={rectangle,draw=black,fill=white,thick,
                    inner sep=0pt,minimum width=0.3cm, minimum height=.6cm},
            SmallRectangle/.style={rectangle,draw=black,fill=white,thick,
                    inner sep=0pt,minimum height=0.3cm, minimum width=0.3cm},
            Circle/.style={circle,draw=black,fill=white,thick,
                    inner sep=0pt,minimum size=6mm},
            ]
            \node[SmallRectangle]   (A) []       {};
            \node[above of=A,yshift=0.1cm,align=center]           {$N$};
            \node[SmallRectangle]   (B) [minimum width=.6cm,right=of A]                 {};
            \node[above of=B,yshift=0.1cm,align=center]          {$2N$};
            \node[SmallRectangle]   (C) [right=of B] {};
            \node[above of=C,yshift=0.1cm,align=center]          {$N$};

            \draw[->,blue] ([xshift=-0.5cm]A.west) -- (A.west);
            \draw[->,red] (A.east) -- (B.west);
            \draw[->,red] (B.east) -- (C.west);
            \draw[->,blue] (C.east) -- ([xshift=0.5cm]C.east);
        \end{tikzpicture}
        \caption{Intermediate, Int}
        \label{subfig:int}
    \end{subfigure}
	\begin{subfigure}{0.39\textwidth}
		\centering
		\begin{tikzpicture}
			[
				node distance=0.5cm, >=stealth', very thick,
				cross/.style={path picture={ \draw[black]
								(path picture bounding box.south east) -- (path picture bounding box.north west)
								(path picture bounding box.south west) -- (path picture bounding box.north east);
							}},
				BigRectangle/.style={rectangle,draw=black,fill=white,thick,
						inner sep=0pt,minimum width=0.3cm, minimum height=.6cm},
				SmallRectangle/.style={rectangle,draw=black,fill=white,thick,
						inner sep=0pt,minimum size=0.3cm},
				Circle/.style={circle,draw=black,fill=white,thick,
						inner sep=0pt,minimum size=6mm},
			]
            \node[SmallRectangle] (A)                              {};
            \node[above of=A,yshift=-0.1cm,align=center]           {$N$};
			\node[SmallRectangle] (B) [below=of A, yshift=0.4cm]  {};
			\node[SmallRectangle] (C) [below=of B, yshift=0.4cm]  {};
			\node[SmallRectangle] (D) [below=of C, yshift=0.4cm]  {};
			\node[BigRectangle]   (E) [right=of A, yshift=0.5cm]   {};
			\node[BigRectangle]   (F) [right=of E, xshift=-0.5cm]  {};
            \node[BigRectangle]   (G) [right=of F, xshift=-0.5cm]  {};
            \node[above of=G,yshift=0.2cm,align=center]           {$5N$};
			\node[BigRectangle]   (H) [right=of G, xshift=-0.5cm]  {};
			\node[BigRectangle]   (I) [right=of H, xshift=-0.5cm]  {};
            \node[BigRectangle]   (J) [right=of I, minimum width=0.6cm]                 {};
            \node[above of=J,yshift=0.2cm,align=center]           {$2P$};
            \node[BigRectangle]   (K) [right=of J, minimum width=0.25cm] {};
            \node[above of=K,yshift=0.2cm,align=center]           {$P$};

			\draw[->,purple] ([xshift=-1.3cm,yshift=0.25cm]E.west) -- ([yshift=0.25cm]E.west);
			\draw[->,gray] ([xshift=-0.4cm]A.west) -- (A.west);
			\draw[->,gray] ([xshift=-0.4cm]B.west) -- (B.west);
			\draw[->,gray] ([xshift=-0.4cm]C.west) -- (C.west);
			\draw[->,blue] ([xshift=-0.4cm]D.west) -- (D.west);
			\draw[->,green] (A.east) -- ([xshift=0.25cm]A.east) |- (E.west);
			\draw[->,green] (B.east) -- ([xshift=0.25cm]B.east) |- (E.west);
			\draw[->,green] (C.east) -- ([xshift=0.25cm]C.east) |- (E.west);
			\draw[->,green] (D.east) -- ([xshift=0.25cm]D.east) |- (E.west);
			\draw[->,red] (I.east) -- (J.west);
			\draw[->,red] (J.east) -- (K.west);
            \draw[->,blue] (K.east) -- ([xshift=0.5cm]K.east);
            
            \node[BigRectangle] (L1) [below of=H,yshift=-1.1cm,xshift=.8cm, align=left, draw=white, minimum height=2.5cm, minimum width=2.5cm] {From Conv. \\ From Wave Decomp. \\ Pass Images \\ Wave Trans.,Concat. \\ Conv,BN,ReLU};
            
            \draw[->,purple] ([yshift=-.35cm,xshift=-0.5cm]G.south) -- ([yshift=-.35cm,xshift=-0.2cm]G.south);
            \draw[->,gray] ([yshift=-.8cm,xshift=-0.5cm]G.south) -- ([yshift=-.8cm,xshift=-0.2cm]G.south);
            \draw[->,blue] ([yshift=-1.25cm,xshift=-0.5cm]G.south) -- ([yshift=-1.25cm,xshift=-0.2cm]G.south);
            \draw[->,green] ([yshift=-1.7cm,xshift=-0.5cm]G.south) -- ([yshift=-1.7cm,xshift=-0.2cm]G.south);
            \draw[->,red] ([yshift=-2.15cm,xshift=-0.5cm]G.south) -- ([yshift=-2.15cm,xshift=-0.2cm]G.south);

        \end{tikzpicture}
        \caption{Upsampling, Up($P$)}
		\label{subfig:upblock}
    \end{subfigure}
    \caption{Model architecture (\protect\subref{subfig:overallarchitecture}). Boxes with sharp boundaries denote images with the number of channels displayed above the box. Boxes with rounded boundaries are submodules. Also shown are example downsampling (\protect\subref{subfig:downblock}), upsampling (\protect\subref{subfig:upblock}), and intermediate (\protect\subref{subfig:int}) submodules at a constant network depth. The model as shown here has 31,646,338 parameters.}
    \label{fig:architecture}
\end{figure*}
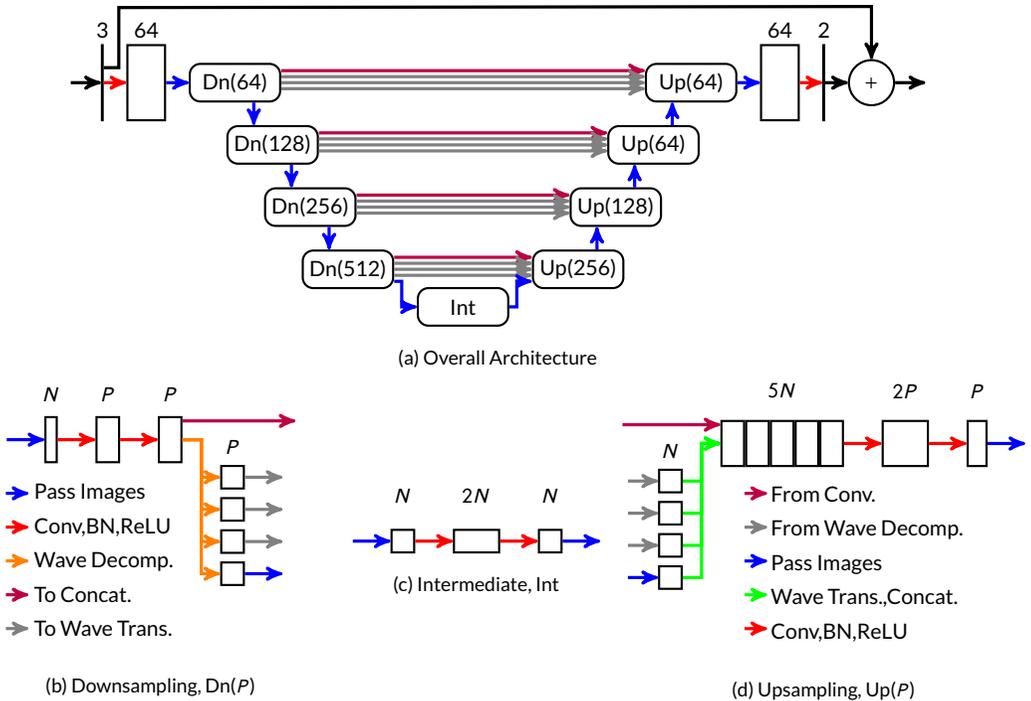

Due to these considerations, we address the Gibbs phenomenon with an image-to-image architecture. Our model comes from the class of U-Net \cite{ronneberger2015u} architectures. Figure \ref{fig:architecture} shows a diagram of the architecture. At the midpoint of the U-Net (shown as the bottom of the U in Figure \ref{fig:architecture}), convolution operations affect almost the entire image, enabling the network to learn global features such as conjugate k-space symmetry.
We process complex images by inputting the real and imaginary components into the network as two separate channels \cite{schlemper2018deep}. After each convolution layer we apply batch normalization \cite{ioffe2015batch}. After batch normalization, we apply rectified linear unit (ReLU) activation functions, as is standard practice \cite{glorot2011deep}. Unlike the standard U-Net, our model includes wavelet transforms for downsampling and upsampling. The theoretical properties of these modifications were analyzed previously \cite{ye2018deep} and demonstrated to better-preserve high-resolution features. Based on the intuition that it is easier to learn residuals rather than signals \cite{he2016deep}, we use a skipped connection over the entire architecture for the real and imaginary channels. For cases with complex inputs, we apply the magnitude operation as a final step after the skipped connection.

\subsection{Training}
\label{sec:training}
\matt{Code for training our models is available on Github at <link>.} We trained the network by simulating MRI acquisitions on photographs from the ImageNet data base \cite{deng2009imagenet}, using resized, noise-free versions of the original photographs as target images $\mathbf{m}_d$. The data set contained 1,281,167 images in the training set and 50,000 images in the validation set. The ``best'' model was selected based on the minimal loss over the validation data set during the training epochs. Use of ImageNet for training requires specification of 1) an encoding simulation pipeline (Section \ref{subsec:simmeth}) and 2) an optimization procedure (Section \ref{subsec:optim}). 

Although the images in ImageNet are not true MR images, we take this approach due to the fact that it is impossible to acquire large volumes of high-resolution diffusion images for the purpose of simulating Gibbs artifacts. An alternative would be to train from high-resolution anatomical data, but diffusion-weighted images can have substantially different features from their high-resolution anatomical counterparts. Thus, training a network on simulated Gibbs artifacts from high-resolution anatomical data runs the risk of having the network reconstruct signals that would not be present in the corresponding diffusion data. Furthermore, an anatomical data set would be highly homogeneous, limiting the generalizability of the resulting network. Conversely, ImageNet is a highly heterogeneous data set, enabling the network to learn more generic features and potentially enabling application of this approach in other MRI modalities that suffer from Gibbs artifacts. A key determinant of the feasibility of this approach is the simulation of encoding operations as they occur in vivo, which we discuss below.

\subsubsection{Simulation}
\label{subsec:simmeth}
\begin{figure*}[ht]
    \centering
    \begin{subfigure}{0.25\textwidth}
        \centering
        {Simulated Phase}
    \end{subfigure}
    \begin{subfigure}{0.25\textwidth}
        \centering
        {In Vivo Phase}
    \end{subfigure}
    \begin{subfigure}{0.07\textwidth}
        {\hspace{1mm}}
    \end{subfigure}
    \begin{subfigure}{0.02\textwidth}
        \raisebox{0in}{\rotatebox[origin=t]{-90}{\hspace{1mm}}}
    \end{subfigure}

    \begin{subfigure}{0.25\textwidth}
        \centering
        \includegraphics[width=1\linewidth]{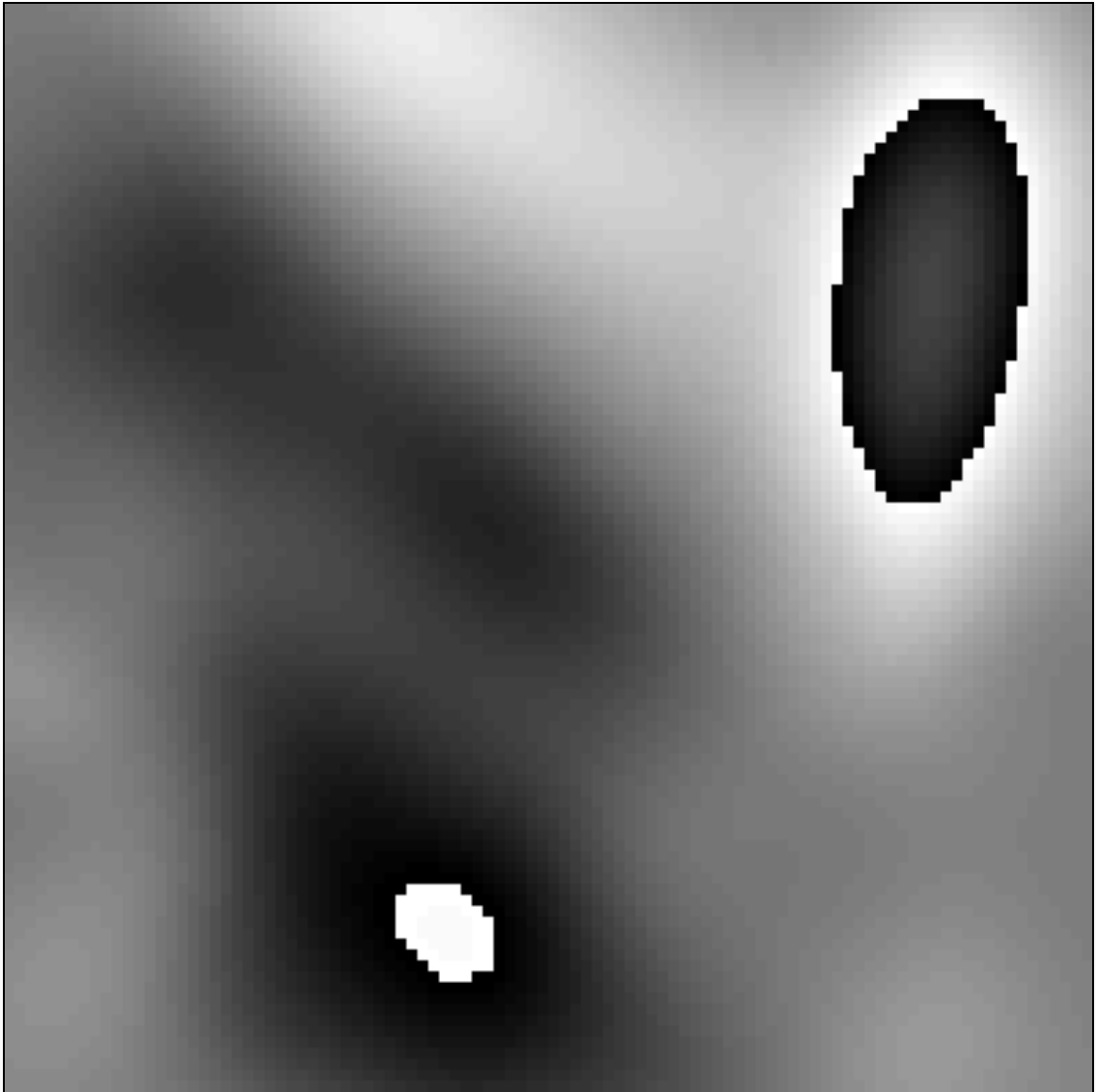}
    \end{subfigure}
    \begin{subfigure}{0.25\textwidth}
        \centering
        \includegraphics[width=1\linewidth]{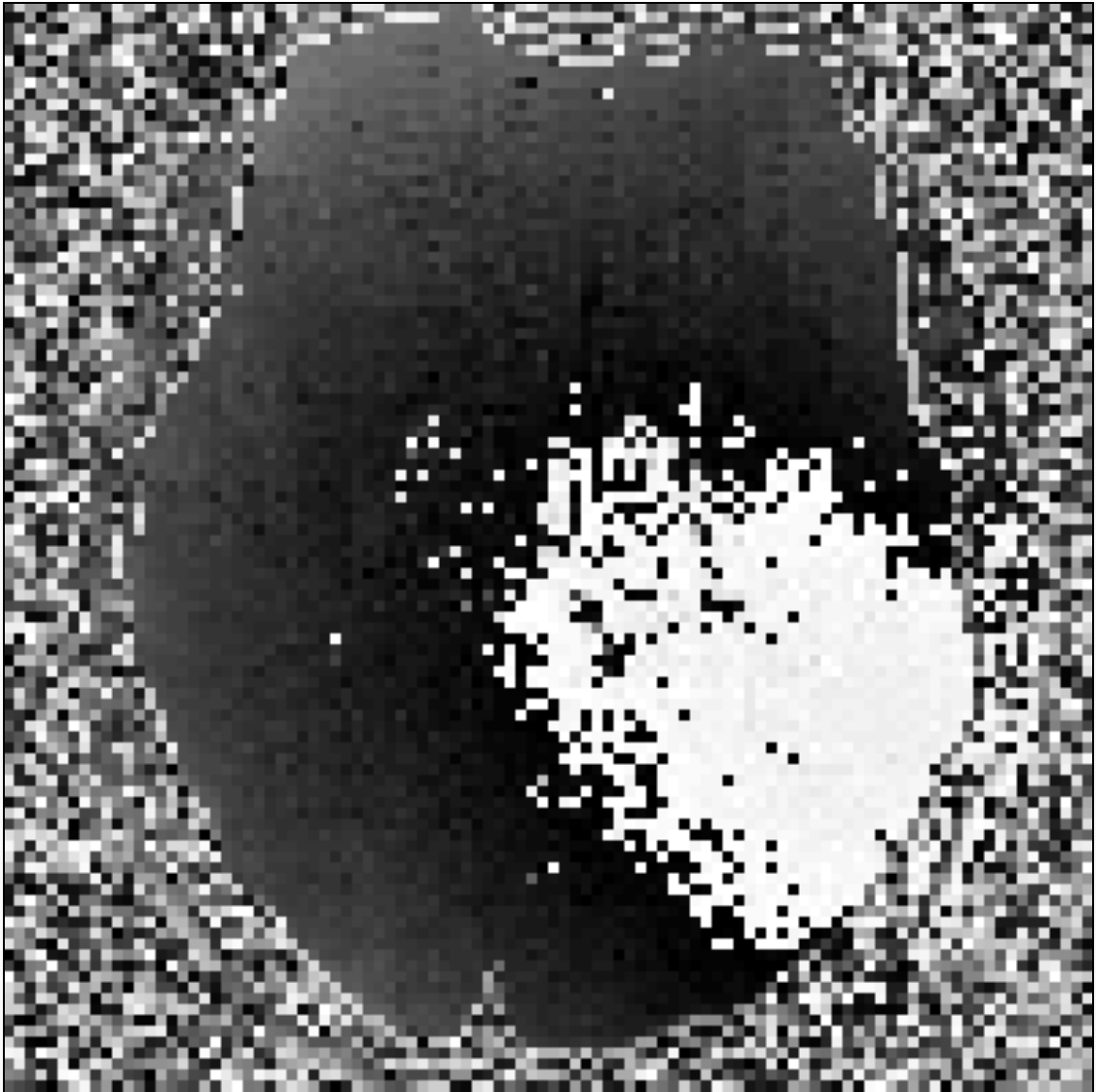}
    \end{subfigure}
    \begin{subfigure}{0.07\textwidth}
        \includegraphics[height=1.4in]{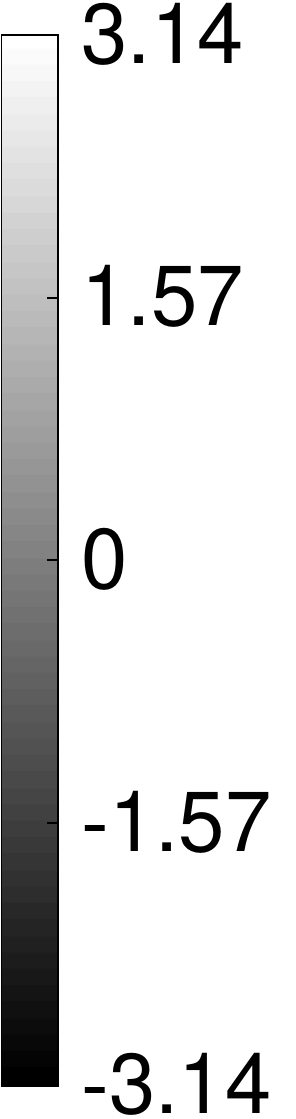}
    \end{subfigure}
    \begin{subfigure}{0.02\textwidth}
        \raisebox{0in}{\rotatebox[origin=t]{-90}{Phase Angle (Radians)}}
    \end{subfigure}

    \caption{Examples of phase from in vivo diffusion-weighted data and the simulations used for network training. Phase maps can have substantial variations that must be considered when applying partial Fourier imaging.}
    \label{fig:phase_example}
\end{figure*}

The simplest Gibbs simulation could be accomplished by cropping the k-space representation of a high-resolution image. However, in practice MR acquisitions are affected by many factors other than the Gibbs phenomenon. We simulate a subset of these effects.
Figure \ref{fig:simpipeline} shows a flowchart of the simulation pipeline. The simulation begins with raw grayscale ImageNet image (converted via BT.601 color standards), resized to a standard size of $256 \times 256$ using bilinear interpolation \cite{paszke2017automatic}. The simulation then randomly flips and/or transposes the image, each with a 50\% probability. After flipping and transposing, the simulation proceeds to the random phase step. The random phase simulation generates a random set of Gaussian radial basis functions for each simulated image - this is described in detail in Appendix \ref{app:phasesim}. An example of simulated phase compared to a true in vivo phase case is shown in Figure \ref{fig:phase_example}.
The random phase simulation requires a number of probabilistic parameters for specifying the distribution of the simulated phase - these are reported in the appendix.

Since most images have a no-signal background, following random phase simulation we apply random ellipsoid cropping to the images. We also include a 10\% probability that no ellipsoid cropping occurs to handle cases that may not have a no-signal background. After image-domain cropping, the processing pipelines split. For the data, $\mathbf{x}_d$, we apply the FFT and crop the image from a $256 \times 256$ grid to a $100 \times 100$ grid in k-space to simulate the Fourier series truncation effect. The simulation adds noise at levels from 1 to 32 times the mean absolute k-space value, varying the noise level within this range on a uniform base-2 logarithmic scale (average noise ratio of 10). 

After the inverse FFT, both processing pipelines are merged into an image normalization layer where both images are divided by the maximum absolute value in $\mathbf{x}_d$. The absolute value operation is then applied to $\mathbf{m}_d$. To simulate an artifact-free target, $\mathbf{m}_d$, we interpolate $\mathbf{m}_d$ from a $256 \times 256$ grid to a $100 \times 100$ grid using cubic splines. Finally, to provide the complex-valued CNN with more information, we concatenate a standard partial Fourier reconstruction method \cite{margosian1986faster} to the complex-valued $\mathbf{x}_d$ output in the channel dimension prior to input to the neural network.

When magnitude images are used, the inputs have Rician noise rather than Gaussian. We maintain most of the above processing stream aside from three modifications. 1) The absolute value operation is applied after inverse FFT in the $\mathbf{x}_d$ stream, and 2) followed by dividing by the maximum value of the $\mathbf{x}_d$ image for both the $\mathbf{m}_d$ and $\mathbf{x}_d$ final images. This implicitly builds Rician bias correction into the network training.
 
\subsubsection{Optimization}
\label{subsec:optim}
Our optimization routine followed standard practice. We used the ImageNet data set \cite{deng2009imagenet} with standard training and validation splits. Our convolution weights were initialized with a uniform distribution \cite{he2015delving}. ReLUs were initialized with zero parameters. We used the Adam algorithm \cite{kingma2014adam} with a learning rate of $1 \times 10^{-3}$ over 10 epochs to minimize the mean-squared error cost function in (\ref{eq:optim}). At each epoch, we calculated the loss over the entire validation data set and saved the model with the best validation loss for the in vivo test experiments. Each model was trained independently on each partial Fourier factor. We implemented our models in the PyTorch deep learning framework \cite{paszke2017automatic}. Training was performed on an IBM GPU computing cluster with Power9 8335-GTH compute nodes. Each node in the compute cluster had four Nvidia V100 GPUs with 16 GB of memory. Our specific trainings utilized one GPU with a batch size of 55. Training time took approximately 10 days.

\subsection{Testing}
In the test phase we evaluated the proposed CNN methods for their ability to process diffusion-weighted images (DWIs) on different data sets. In all our test experiments, we define ``SNR'' to be the mean absolute image value divided by the complex Gaussian standard deviation. Note that this is different from the k-space ratio used during training. We conducted canonical signal processing experiments and in vivo experiments. To provide context for the performance of the CNN methods, we also processed the DWIs with a state-of-the-art (SoA) method \cite{ades2018evaluation}, as well as a method designed for partial Fourier imaging \cite{margosian1986faster}. In all cases with diffusion parameter maps, the raw DWIs were first processed by the designated method prior to parameter map estimation. To clearly show that our test data sets are distinct from our training data sets, we index test signals with $t$ instead of $d$.

\textbf{Raw:} These parameter maps were calculated from the raw diffusion-weighted images with no extra processing before the parameter estimation stage.

\textbf{State-of-the-Art (SoA) \cite{ades2018evaluation}:} Prior to parameter map estimation, the images were denoised using random matrix theory-based MP-PCA method \cite{veraart2016denoising} using a 5 x 5 x 5 voxel kernel extent with Rician bias correction \cite{veraart2016denoising}, followed by Gibbs removal via subvoxel shifting \cite{kellner2016gibbs} where each voxel was discretized into 20 subvoxel elements. Unlike the CNN methods, the SoA method is able to use information from low-$b$ diffusion-weighted images to denoise images from high-$b$ diffusion gradients.

\textbf{Magnitude-input CNN (MCNN):} The Raw diffusion images after the magnitude operation, $\mathbf{x}_t$, are processed by a neural network trained with the simulation pipeline in Figure \ref{fig:simpipeline} with the target image $\mathbf{m}_t$ being a magnitude, deGibbsed image without PF. This particular network attempts to restore the missing partial Fourier information. This network has 31,646,338 parameters.

\textbf{Standard Partial Fourier (Standard PF):} The Raw diffusion images are processed by using a standard partial Fourier reconstruction method that utilizes the phase maps from the symmetric region of k-space \cite{margosian1986faster} prior to parameter estimation.

\textbf{Complex-input CNN (CCNN):} The Raw complex-valued diffusion images, $\mathbf{x}_t$, are processed by a neural network trained with the simulation pipeline in Figure \ref{fig:simpipeline} with the target image $\mathbf{m}_t$ being a magnitude, deGibbsed image without PF. This particular network attempts to restore the missing partial Fourier information. This network has 31,646,338 parameters.

\subsubsection{Canonical Experiments}
\label{subsec:resexpmeth}
To characterize the deGibbsing and denoising performance of the network, we performed a set of canonical signal processing experiments with two data sets. The first data set consisted of a simple edge phantom, while the second data set consisted of the test split of the 2013 ImageNet challenge \cite{russakovsky2015imagenet}.

\textbf{Edge-spread function}: The edge phantom, meant to study the response of the networks to an ideal unit-step, consisted of a 2D digital image with half of the image filled with 1s and the other half filled with 0s. This image was generated at a 1024 $\times$ 1024 matrix size (cropped to a 100 $\times$ 100 matrix size during simulation). The location of the edge was varied by rotating it. Figure \ref{fig:cnrresults} includes a composite image of the phantom generated from $f_{\hat{\theta}}(\mathbf{x}_t)$ at many different SNR levels. We performed two sets of experiments with the edge phantom: contrast-to-noise (CNR) and deGibbsing experiments. In a previous study \cite{kellner2016gibbs}, it was noted that deGibbsing performance could vary depending on the angle of the edge. To assess the performance of the method on removing Gibbs artifacts at different angles, we generated the phantom at a set of angles beetween 0 and 45 degrees at a 1024 SNR level and computed the CCNN deGibbsed image.

We also performed contrast-to-noise ratio (CNR) experiments. The goal of the CNR experiments was to quantify the full-width-half-maximum (FWHM) of the network's line-spread function (analogous to a point-spread function in 1D) across a variety of noise levels. We accomplished this by applying the simulation pipeline in Figure \ref{fig:simpipeline} without random flipping and transposing at a variety of CNRs between 0 and 10, resulting in images at 100 $\times$ 100 matrix sizes. For each CNR, we applied both the CCNN model (generating $f_{\hat{\theta}}(\mathbf{x}_t)$ realizations for each one). This resulted in a series of edge-spread function images. To compensate for noise in calculating the FWHM values, for each row of the resulting image, we fit a summation of logistic functions \cite{li2009comparison}:
\begin{equation}
    l(s) = c + \mathlarger{\mathlarger{\sum}_{i=1}^3} \frac{\alpha_i}{1 + \text{exp}\left ( \frac{s-\beta_i}{\gamma_i} \right )},
\end{equation}
where $c$, $\alpha_i$, $\beta_i$, and $\gamma_i$, $i \in [1, 2, 3]$ are fitting parameters. We computed the derivative of this function and used it to calculate a line-spread function for each row of the image. From the line-spread functions, we calculated the FWHMs and averaged across the rows. For each CNR level, we repeated and averaged this experiment 50 times.

\textbf{ImageNet Test Data:} The goal of the tests on the ImageNet data was to assess the network's performance across a broad variety of signals. For this we used the test split from the 2013 ImageNet challenge \cite{russakovsky2015imagenet}. This data was not used at all in training and is distinct from the validation data set. The test set included 40,152 images. We applied the simulation pipeline in Figure \ref{fig:simpipeline} and computed the power spectral ratio on the resulting images \cite{veraart2016gibbs}. Let $\mathcal{S}(\cdot)$ be the operator that computes the power spectral density of the image. For each image-target pair in the validation data set, $(\mathbf{x}_t, \mathbf{m}_t)$, we computed
\begin{equation}
    h_t(\boldsymbol{f}) = \sqrt{\frac{\mathcal{S}\left ( f_{\hat{\theta}}\left ( \mathbf{x}_t \right ) \right )}{\mathcal{S} \left ( \mathbf{m}_t \right )}}.
\label{def-hf}
\end{equation}
Then, we averaged over all $t \in [1, ..., n_{\textit{test}}]$ to compute a mean $\bar{h}(\boldsymbol{f})$ for each partial Fourier factor and SNR level and computed profiles across $\boldsymbol{f}$ in the partial Fourier direction to assess resolution performance.

\subsubsection{In Vivo Test Experiments}
\label{subsec:expmeth}
We tested the method in vivo by scanning a volunteer with a diffusion imaging protocol under IRB approval, saving the raw data. We scanned the volunteer after all model and training parameters were complete, so that these data qualify as a true prospective test set. The volunteer was scanned on a 3T Prisma scanner with a 16-channel head coil (Siemens Healthineers, Erlangen, Germany). Diffusion-weighted images were acquired with a spin echo EPI sequence at 100 $\times$ 100 matrix size over 38 slices (2 mm isotropic resolution) and along 66 isotropically-distributed directions at $b$-values of 0, 1,000, and 1,500 s/mm$^2$. We applied the diffusion-weighted acquisition four times: once for each PF factor of 5/8ths, 6/8ths, 7/8ths, and fully-sampled, with the intention that the fully-sampled DWIs with current state-of-the-art processing serve as a gold standard. The acquisition parameters were set based on the shortest values we could achieve for the fully-sampled acquisition, leading to a TR of 9 s and a TE of 105 ms. The partial Fourier acquisitions used a standard protocol.

The images were reconstructed with adaptive coil combination \cite{walsh2000adaptive}. Parameter map estimation was performed on the processed image using a constrained weighted linear least squares approach \cite{veraart2013weighted}. Since the complex CNN can take into account smooth variations in the phase of the image to recover high-resolution information, we also included as a comparison a reconstruction using a standard partial Fourier method that estimates smooth phase from the symmetric region of k-space \cite{margosian1986faster} applied with our own custom software to the raw data.

\section{Results}
\label{sec:results}
\subsection{Canonical Experimental Results}
\label{subsec:resexpresults}
\begin{figure*}[htb]
    \centering
    \begin{subfigure}{0.02\textwidth}
        \raisebox{0in}{\rotatebox[origin=t]{90}{\hspace{1mm}}}
    \end{subfigure}
    \begin{subfigure}{0.38\textwidth}
        \centering
        {Angle = $0\degree$}
    \end{subfigure}
    \begin{subfigure}{0.02\textwidth}
        \raisebox{0in}{\rotatebox[origin=t]{90}{\hspace{1mm}}}
    \end{subfigure}
    \begin{subfigure}{0.38\textwidth}
        \centering
        {Angle = $45\degree$}
    \end{subfigure}

    \begin{subfigure}{0.02\textwidth}
        \raisebox{0in}{\rotatebox[origin=t]{90}{Signal}}
    \end{subfigure}
    \begin{subfigure}{0.38\textwidth}
        \centering
        \includegraphics[width=1\linewidth]{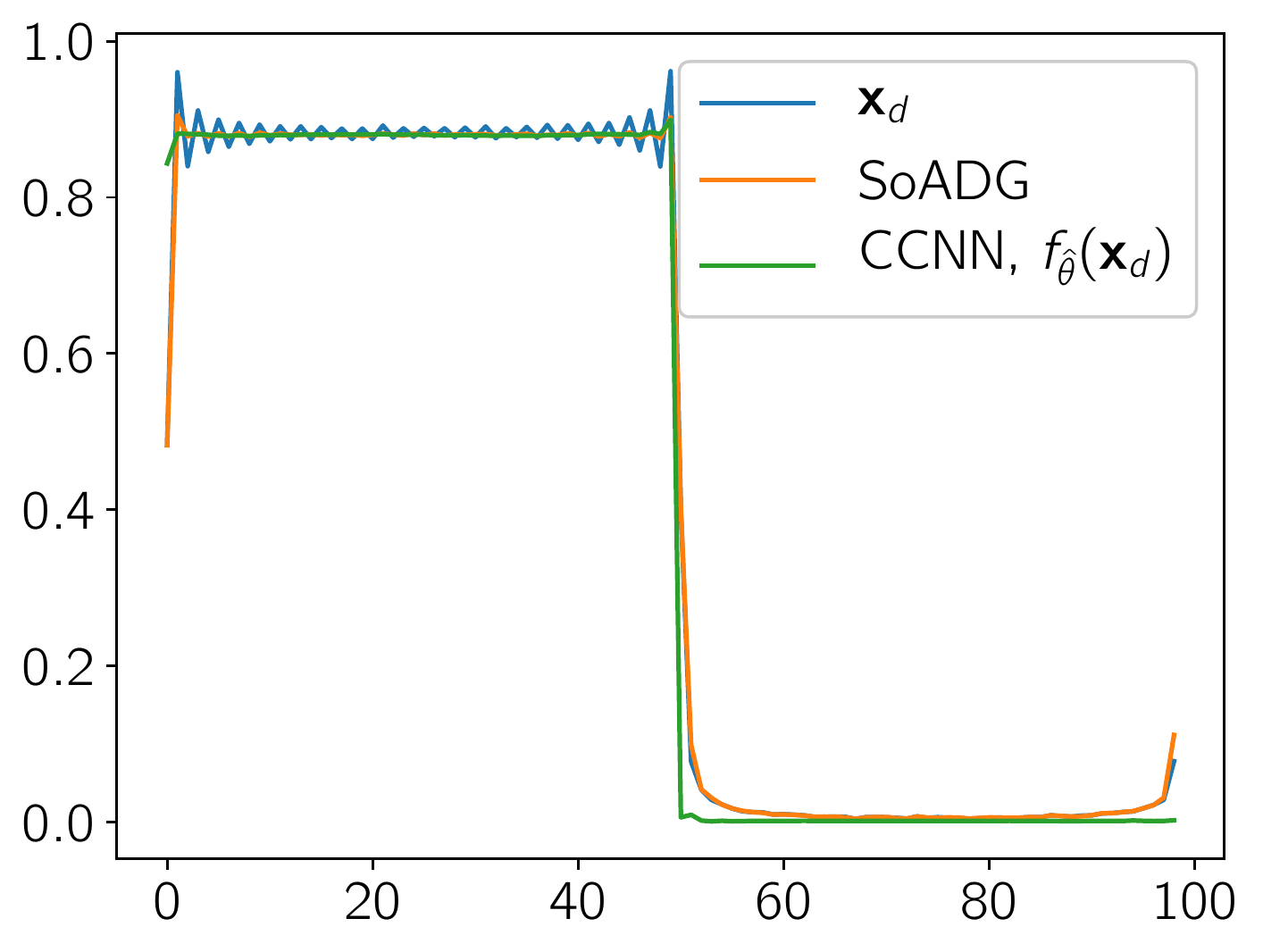}
    \end{subfigure}
    \begin{subfigure}{0.02\textwidth}
        \raisebox{0in}{\rotatebox[origin=t]{90}{Signal}}
    \end{subfigure}
    \begin{subfigure}{0.38\textwidth}
        \centering
        \includegraphics[width=1\linewidth]{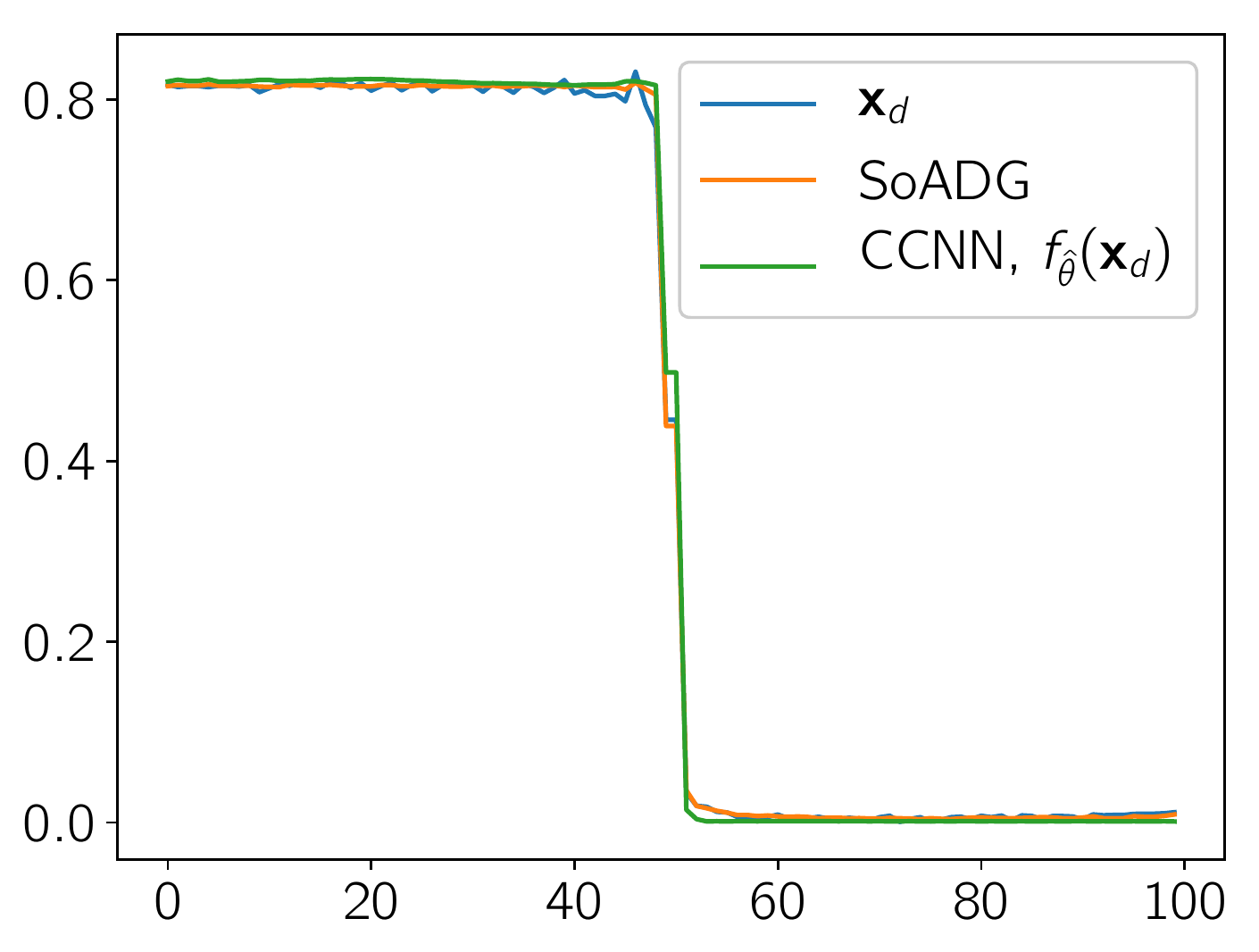}
    \end{subfigure}
    
    \begin{subfigure}{0.02\textwidth}
        \raisebox{0in}{\rotatebox[origin=t]{90}{\hspace{1mm}}}
    \end{subfigure}
    \begin{subfigure}{0.38\textwidth}
        \centering
        {Image Pixel Index}
    \end{subfigure}
    \begin{subfigure}{0.38\textwidth}
        \centering
        {Image Pixel Index}
    \end{subfigure}
    \caption{Profiles across the edge phantom for the raw, Gibbs-corrupted data, $\mathbf{x}_t$, the CCNN model, $f_{\hat{\theta}}(\mathbf{x}_t)$, and the MCNN model for edges at different angles. Both the CCNN and the MCNN method reduce the Gibbs artifacts. The MCNN method has residual artifacts at the image boundaries and some rounding of the edge. Both plots are done at the same length scale, so the 45-degree plot domain does not reach the image boundary.}
    \label{fig:gibbsedge}
\end{figure*}

Figure \ref{fig:gibbsedge} shows the performance of the network in a canonical deGibbsing experiment with a hard edge with the edge oriented at 0 and 45 degrees.
Both the CCNN and the MCNN methods performed similarly at both angles, with the main difference being the sharpness of the edges with the MCNN method both at the simulated edge and the image boundaries. These image boundary residual artifacts do not appear in the $45\degree$ plot since this profile does not reach the image boundaries.

\begin{figure*}[htb]
    \centering
    \begin{subfigure}{0.02\textwidth}
        \raisebox{0in}{\rotatebox[origin=t]{90}{Input, $\mathbf{x}_t$}}
    \end{subfigure}
    \begin{subfigure}{0.6\textwidth}
        \centering
        \includegraphics[width=1\linewidth]{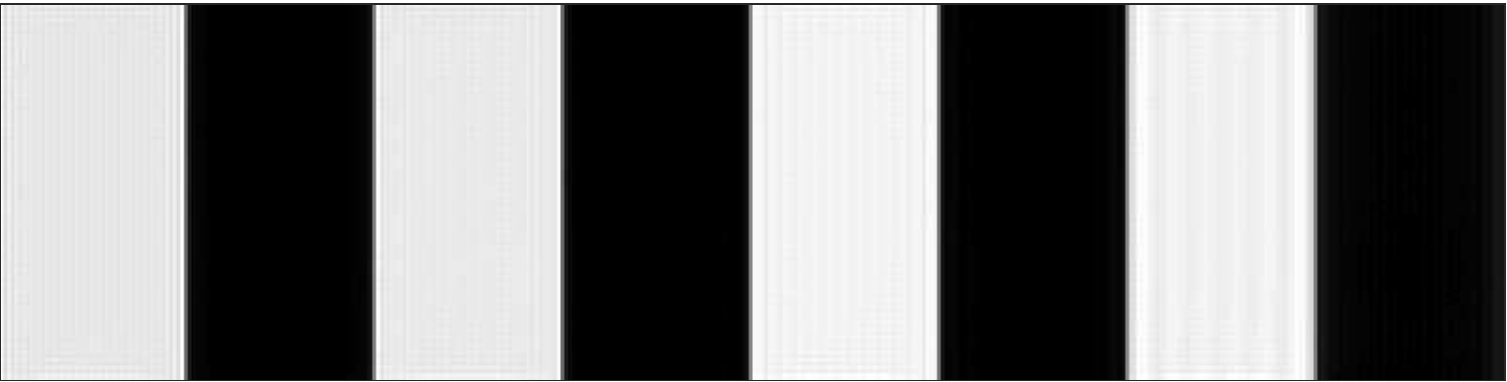}
    \end{subfigure}

    \begin{subfigure}{0.02\textwidth}
        \raisebox{0in}{\rotatebox[origin=t]{90}{Activations}}
    \end{subfigure}
    \begin{subfigure}{0.6\textwidth}
        \centering
        \includegraphics[width=1\linewidth]{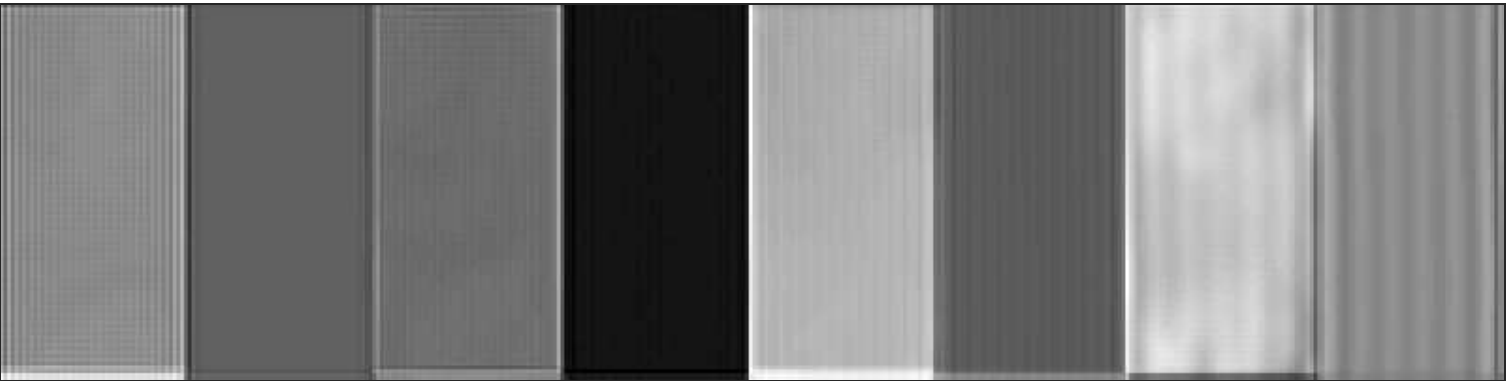}
    \end{subfigure}

    \begin{subfigure}{0.02\textwidth}
        \raisebox{0in}{\rotatebox[origin=t]{90}{\hspace{1mm}}}
    \end{subfigure}
    \begin{subfigure}{0.15\textwidth}
        \centering
        {No PF}
    \end{subfigure}
    \begin{subfigure}{0.15\textwidth}
        \centering
        {7/8 PF}
    \end{subfigure}
    \begin{subfigure}{0.15\textwidth}
        \centering
        {6/8 PF}
    \end{subfigure}
    \begin{subfigure}{0.15\textwidth}
        \centering
        {5/8 PF}
    \end{subfigure}
    \caption{Activations across PF factors. ({\it top}) Original $\mathbf{x}_t$ input to the neural network at an SNR of 1024. ({\it bottom}) Activations from one channel prior to first wavelet transform. The activations in early layers enhance the Gibbs artifact, which is then subtracted from the image in the final step.}
    \label{fig:gibbsact}
\end{figure*}

We also examined the CCNN activations (i.e., images from each channel after a CNN layer) in the case of the edge phantom, which are shown in Figure \ref{fig:gibbsact}.
Figure \ref{fig:gibbsact} shows one channel (out of 64) from the activations after three convolutional steps and prior to the first wavelet transform. In early layers, we found that the networks tended to enhance Gibbs artifacts for removal in the final layers. Other features of the image are detected in the other network channels, as shown in Figure 1 of the supplementary material. Some channels detect changes in the phase of the image, while other channels detect flat regions. The overall properties of the network depend on integrations of all of these channels. We also examined activations in other layers after wavelet downsampling operations, but the activations in these layers were less interpretable.

\begin{figure*}
    \centering
    \begin{subfigure}{0.02\textwidth}
        \raisebox{0in}{\rotatebox[origin=t]{90}{\hspace{1mm}}}
    \end{subfigure}
    \begin{subfigure}{0.4\textwidth}
        \centering
        {Composite of $f_{\hat{\theta}} \left ( \mathbf{x}_t \right )$}
    \end{subfigure}
    \begin{subfigure}{0.44\textwidth}
        \centering
        {FWHM Across CNR}
    \end{subfigure}

    \begin{subfigure}{0.02\textwidth}
        \raisebox{0in}{\rotatebox[origin=t]{90}{Image Row}}
    \end{subfigure}
    \begin{subfigure}{0.4\textwidth}
        \centering
        \includegraphics[width=1\linewidth]{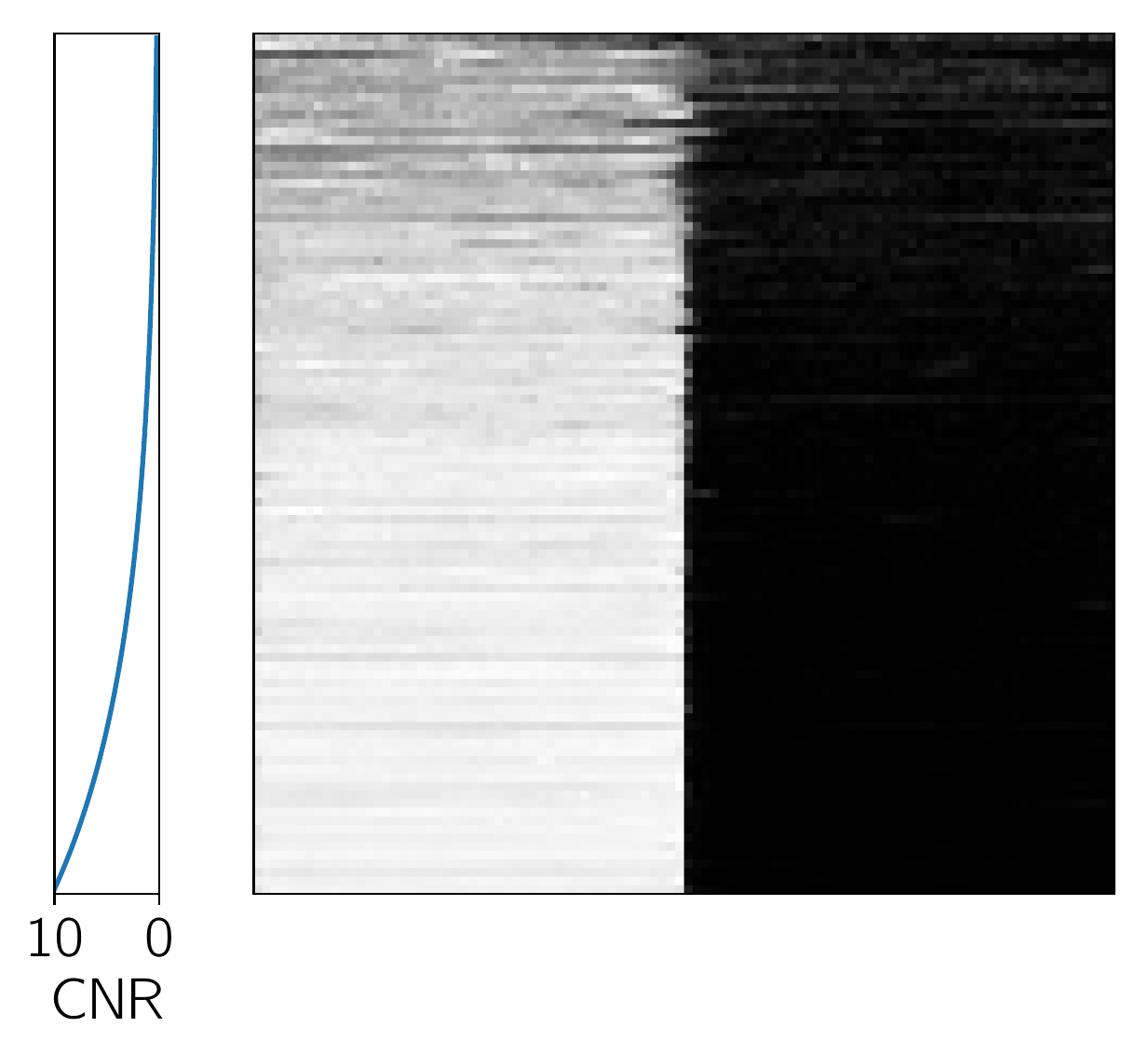}
    \end{subfigure}
    \begin{subfigure}{0.44\textwidth}
        \centering
        \includegraphics[width=1\linewidth]{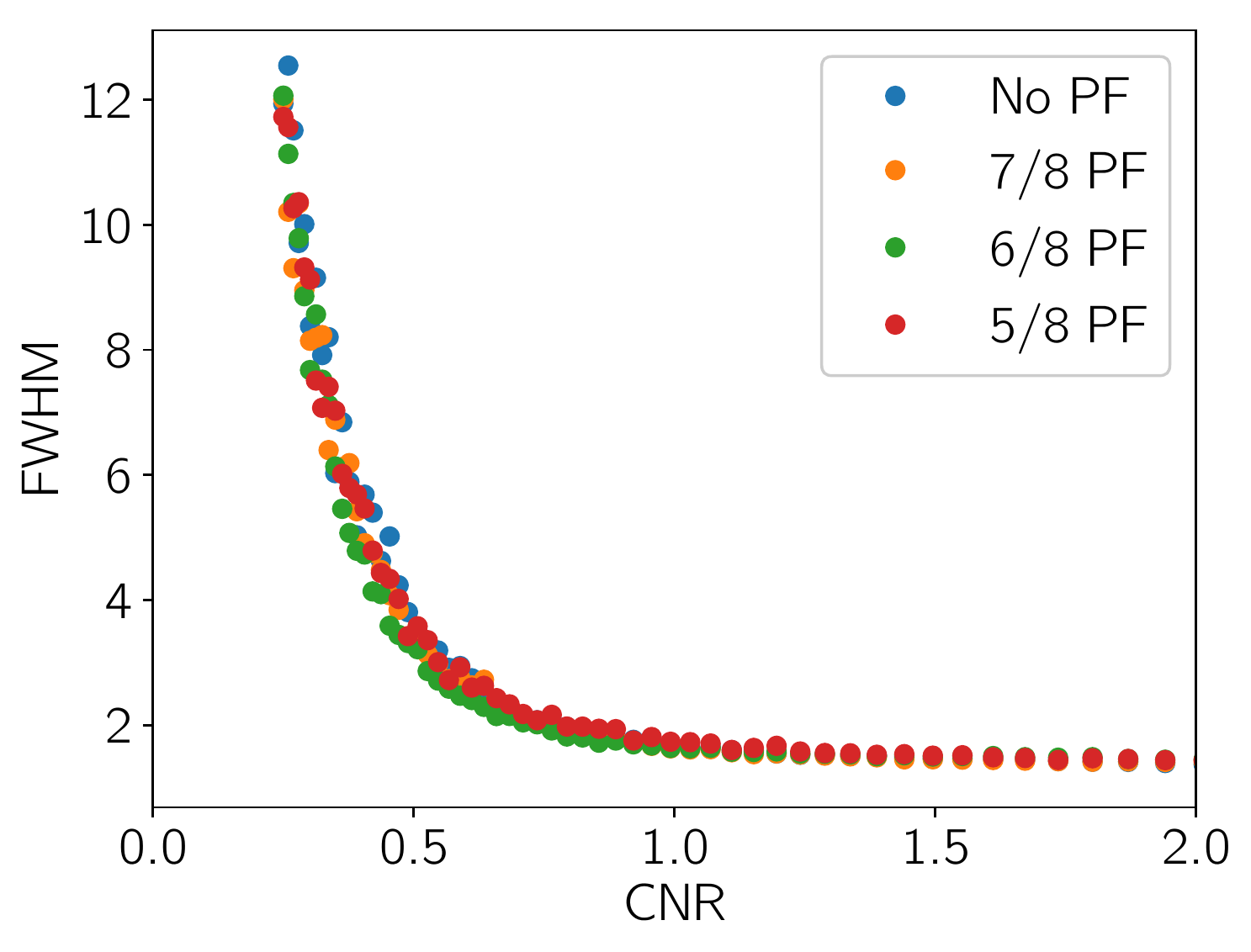}
    \end{subfigure}
    \caption{Responses of the CCNN model across CNR levels. (\textit{left}) shows a composite image of network estimates across a variety of CNRs from 0.25 to 10, while (\textit{right}) shows the change in estimated FWHM as a function of CNR. Above a CNR of 1, CNN performance is fairly stable. Below a CNR of 1, the FWHM increases as the CNR approaches 0.}
    \label{fig:cnrresults}
\end{figure*}

Figure \ref{fig:cnrresults} shows the response of the CCNN in the edge experiments across a range of contrast-to-noise ratios.
Above a CNR of 1, the CNN FWHM rapidly converges towards 1. Below a CNR ratio of 1, the network's FWHM increases as the CNR approaches 0. Figure \ref{fig:cnrresults} also shows a composite image of CNN image estimates, $f_{\hat{\theta}} \left ( \mathbf{x}_t \right )$, across CNR factors. Figure \ref{fig:cnrresults} shows visually how as the CNR decreases, the edge begins to blur.

\begin{figure*}[htb]
    \centering
    \begin{subfigure}{0.03\textwidth}
        \raisebox{0in}{\rotatebox[origin=t]{90}{\hspace{1mm}}}
    \end{subfigure}
    \begin{subfigure}{0.40\textwidth}
        \centering
        {No PF}
    \end{subfigure}
    \begin{subfigure}{0.40\textwidth}
        \centering
        {7/8 PF}
    \end{subfigure}

    \begin{subfigure}{0.02\textwidth}
        \raisebox{0in}{\rotatebox[origin=t]{90}{Spectral Response, $\bar{h}(\boldsymbol{f})$}}
    \end{subfigure}
    \begin{subfigure}{0.40\textwidth}
        \centering
        \includegraphics[width=1\linewidth]{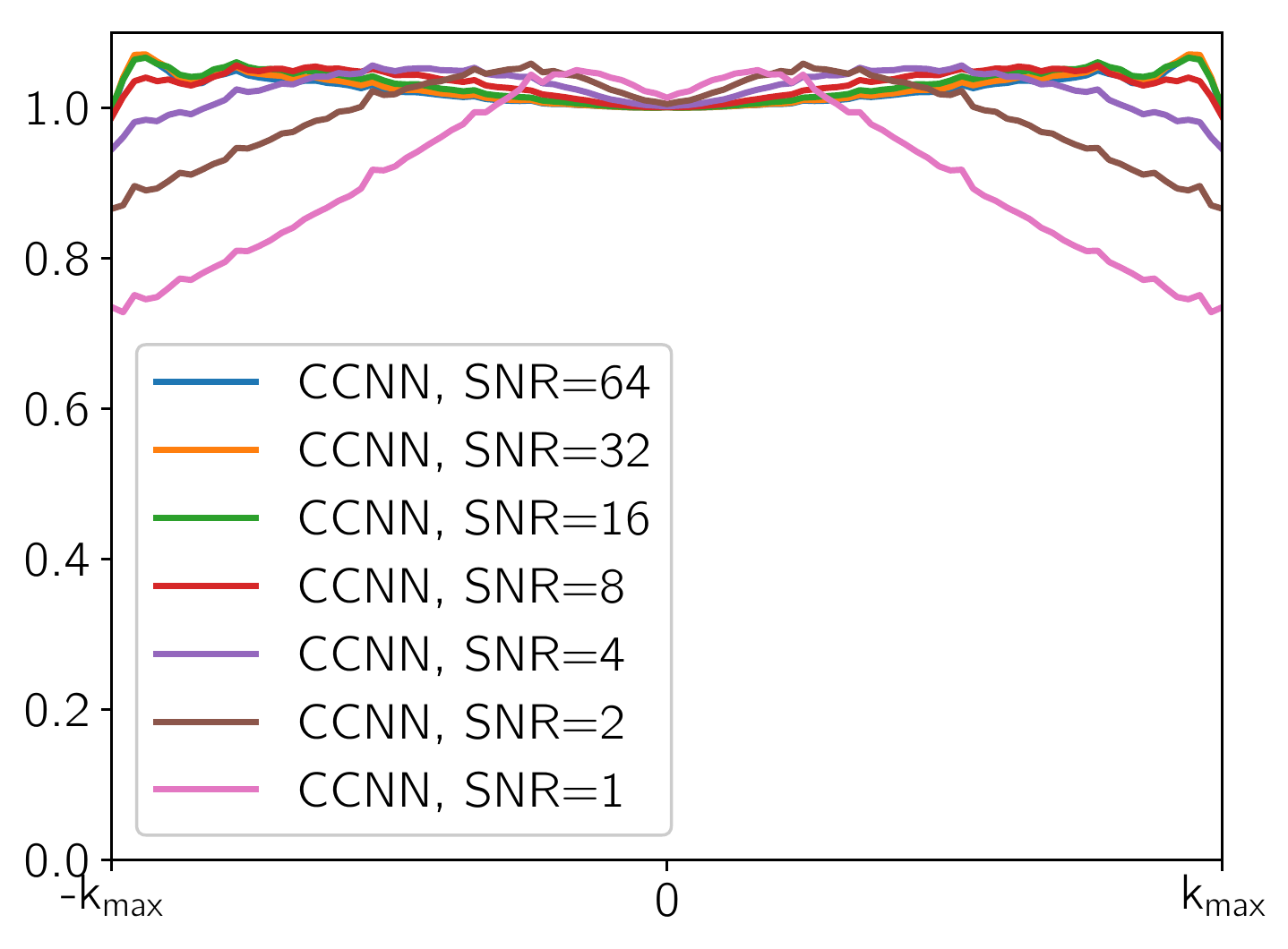}
    \end{subfigure}
    \begin{subfigure}{0.40\textwidth}
        \centering
        \includegraphics[width=1\linewidth]{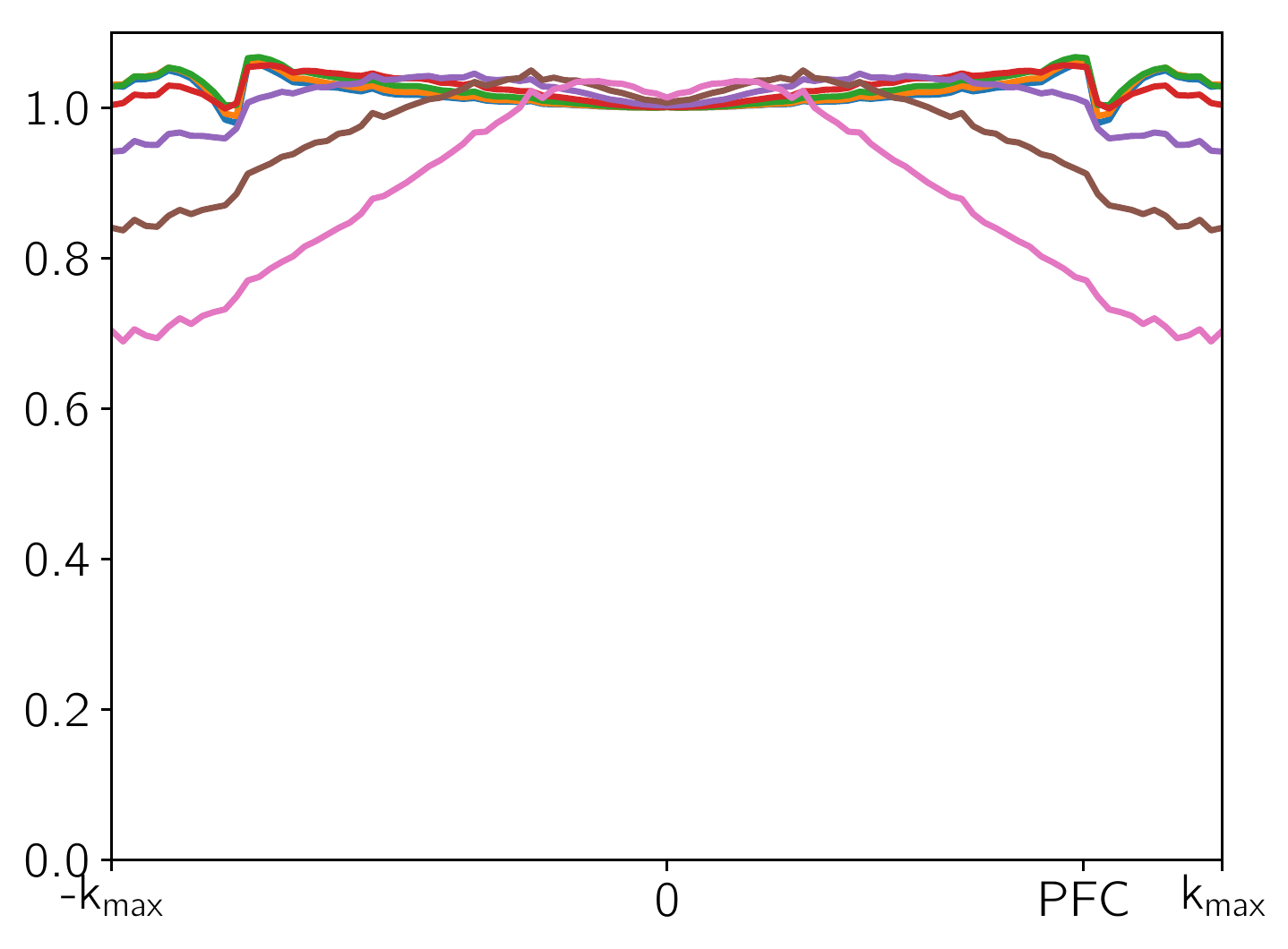}
    \end{subfigure}
    
    \begin{subfigure}{0.03\textwidth}
        \raisebox{0in}{\rotatebox[origin=t]{90}{\hspace{1mm}}}
    \end{subfigure}
    \begin{subfigure}{0.40\textwidth}
        \centering
        {6/8 PF}
    \end{subfigure}
    \begin{subfigure}{0.40\textwidth}
        \centering
        {5/8 PF}
    \end{subfigure}

    \begin{subfigure}{0.02\textwidth}
        \raisebox{0in}{\rotatebox[origin=t]{90}{Spectral Response, $\bar{h}(\boldsymbol{f})$}}
    \end{subfigure}
    \begin{subfigure}{0.40\textwidth}
        \centering
        \includegraphics[width=1\linewidth]{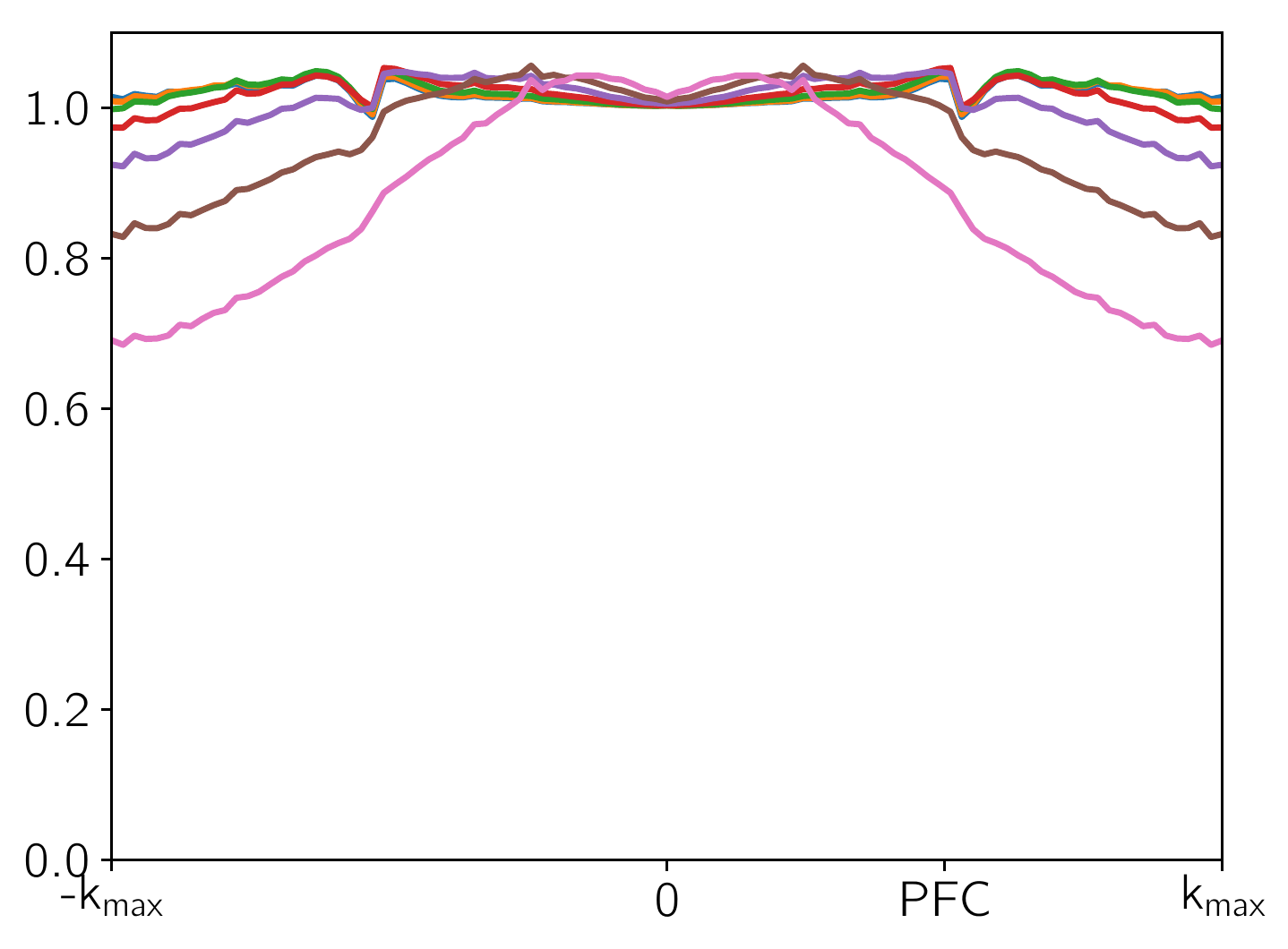}
    \end{subfigure}
    \begin{subfigure}{0.40\textwidth}
        \centering
        \includegraphics[width=1\linewidth]{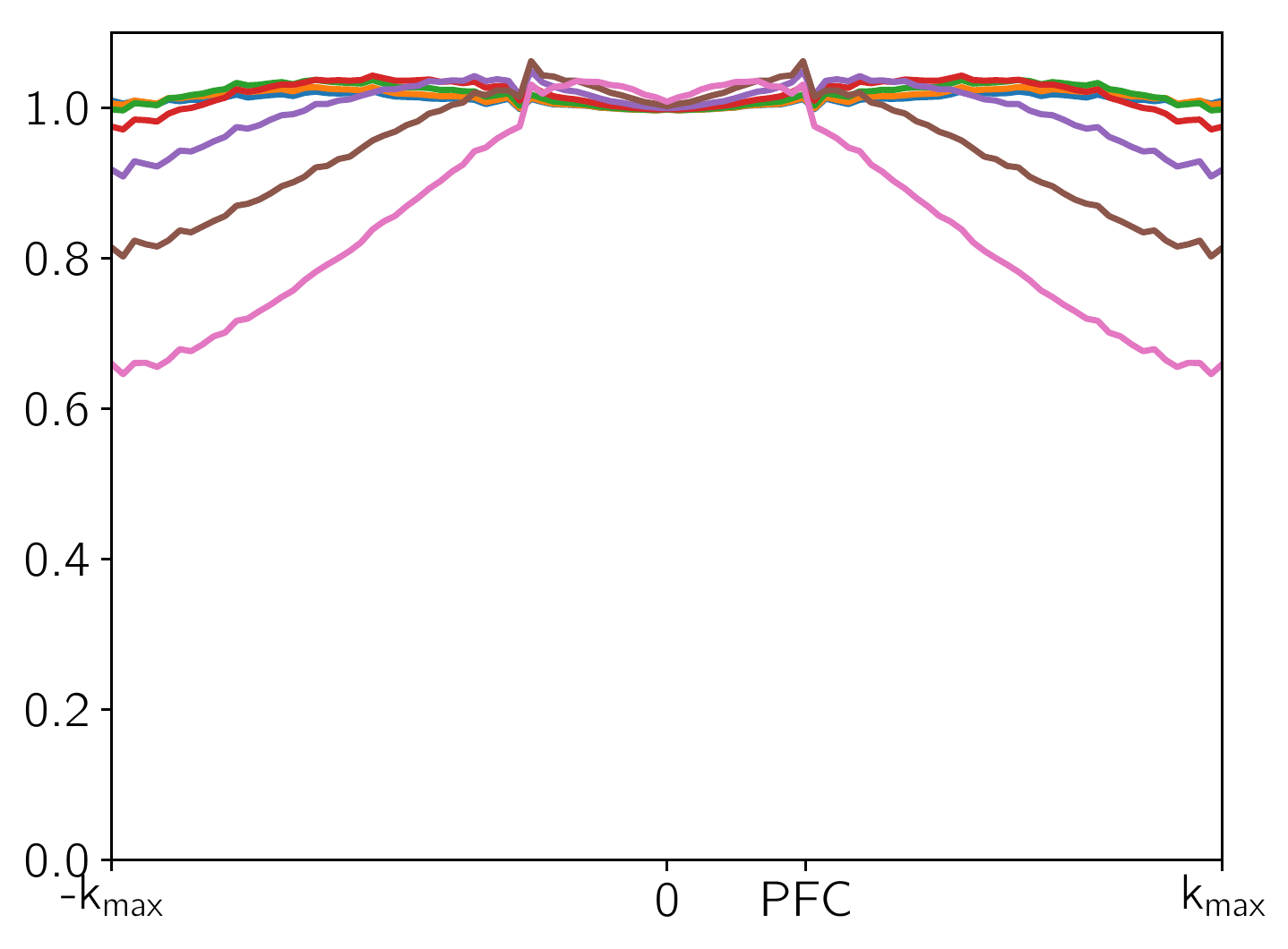}
    \end{subfigure}
    \caption{Spectral responses in the partial Fourier direction across different partial Fourier factors and SNR levels for the CCNN model. ``PFC'' indicates the partial Fourier cutoff. Ripples can be seen that are related to the Gibbs effect and the PF cutoffs. In the partial Fourier cases, the CCNN response has a cusp in the frequency response at the level of the partial Fourier factor. The response beyond the PFC characterizes the ability to restore the missing Fourier harmonics. At lower SNRs, the PFC singularity gradually disappears, and the CCNN model becomes a low-pass filter, applying more smoothing to compensate for the increased noise. Profiles for the MCNN method were similar, but required more smoothing at the lower SNR levels.}
    \label{fig:spectresponse}
\end{figure*}

Figure \ref{fig:spectresponse} shows profiles of the mean spectral response, $\bar{h}(\boldsymbol{f})$, in the partial Fourier direction of the CCNN model across different partial Fourier factors and SNR levels over the entire test set.
The spectral responses were calculated from the 2013 test split of ImageNet as described in Section \ref{subsec:resexpmeth}. As the SNR decreases, the network applies more smoothing, resulting in a loss of high-resolution features of about 30\% at an SNR value of 1. The MCNN method exhibited similar profiles, but with more smoothing at low SNR - the MCNN smoothing without partial Fourier at an SNR of 1 led to about 50\% attenuation of the high frequencies. In all cases, we observed features in the spectral responses around spectral cutoff points - this included the ripples at the edge of the sampled k-space in the No PF setting, with further cusps developing near the PF cutoff points in the PF settings.

\subsection{In Vivo Test Experiment Results}
\begin{figure*}[htb]
    \centering
    \begin{subfigure}{0.02\textwidth}
        \raisebox{0in}{\rotatebox[origin=t]{90}{}}
    \end{subfigure}
    \begin{subfigure}{0.18\textwidth}
        \centering
        {Raw $\mathbf{x}_t$}
    \end{subfigure}
    \begin{subfigure}{0.18\textwidth}
        \centering
        {MCNN, $f_{\hat{\theta}} \left( \mathbf{x}_t \right)$}
    \end{subfigure}
    \begin{subfigure}{0.18\textwidth}
        \centering
        {MCNN Resid., $\text{Ric}(\mathbf{x}_t) - f_{\hat{\theta}} \left( \mathbf{x}_t \right)$}
    \end{subfigure}
    \begin{subfigure}{0.18\textwidth}
        \centering
        {CCNN, $f_{\hat{\theta}} \left( \mathbf{x}_t \right)$}
    \end{subfigure}
    \begin{subfigure}{0.18\textwidth}
        \centering
        {CCNN Resid., $\text{Ric}(\mathbf{x}_t) - f_{\hat{\theta}} \left( \mathbf{x}_t \right)$}
    \end{subfigure}

    \begin{subfigure}{0.02\textwidth}
        \raggedright
        \raisebox{0in}{\rotatebox[origin=t]{90}{No PF}}
    \end{subfigure}
    \begin{subfigure}{0.18\textwidth}
        \includegraphics[width=1\linewidth]{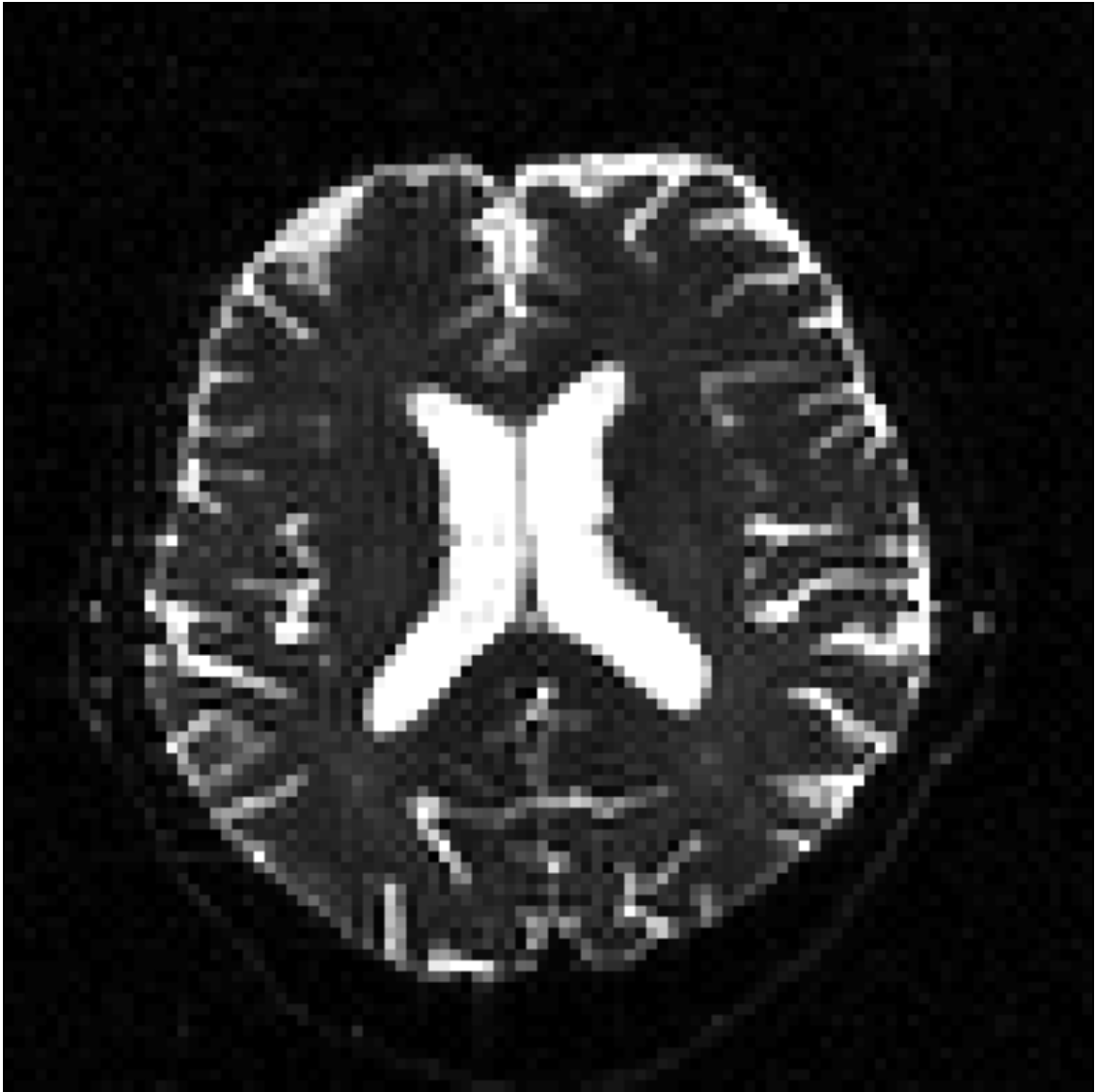}
    \end{subfigure}
    \begin{subfigure}{0.18\textwidth}
        \includegraphics[width=1\linewidth]{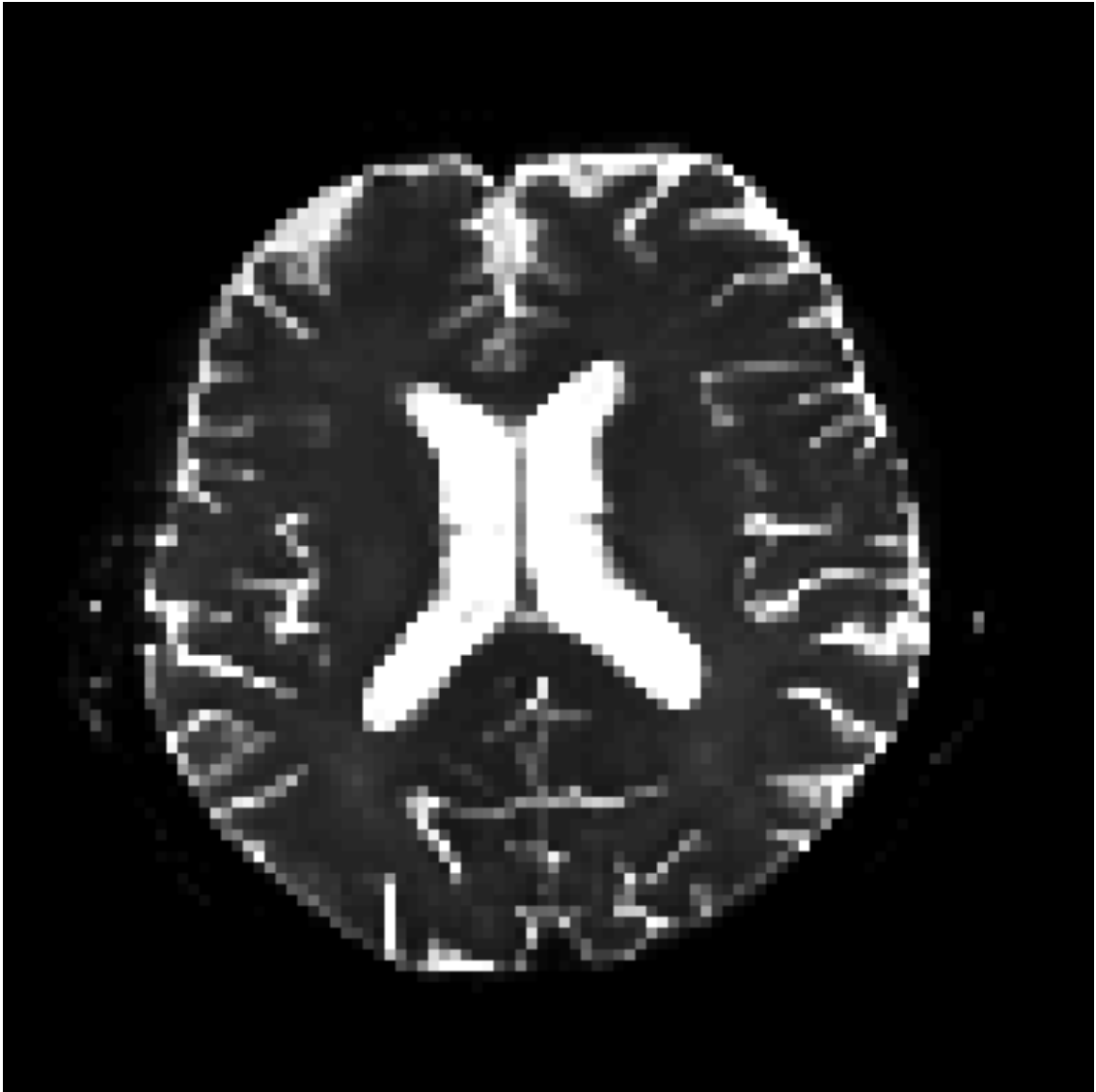}
    \end{subfigure}
    \begin{subfigure}{0.18\textwidth}
        \includegraphics[width=1\linewidth]{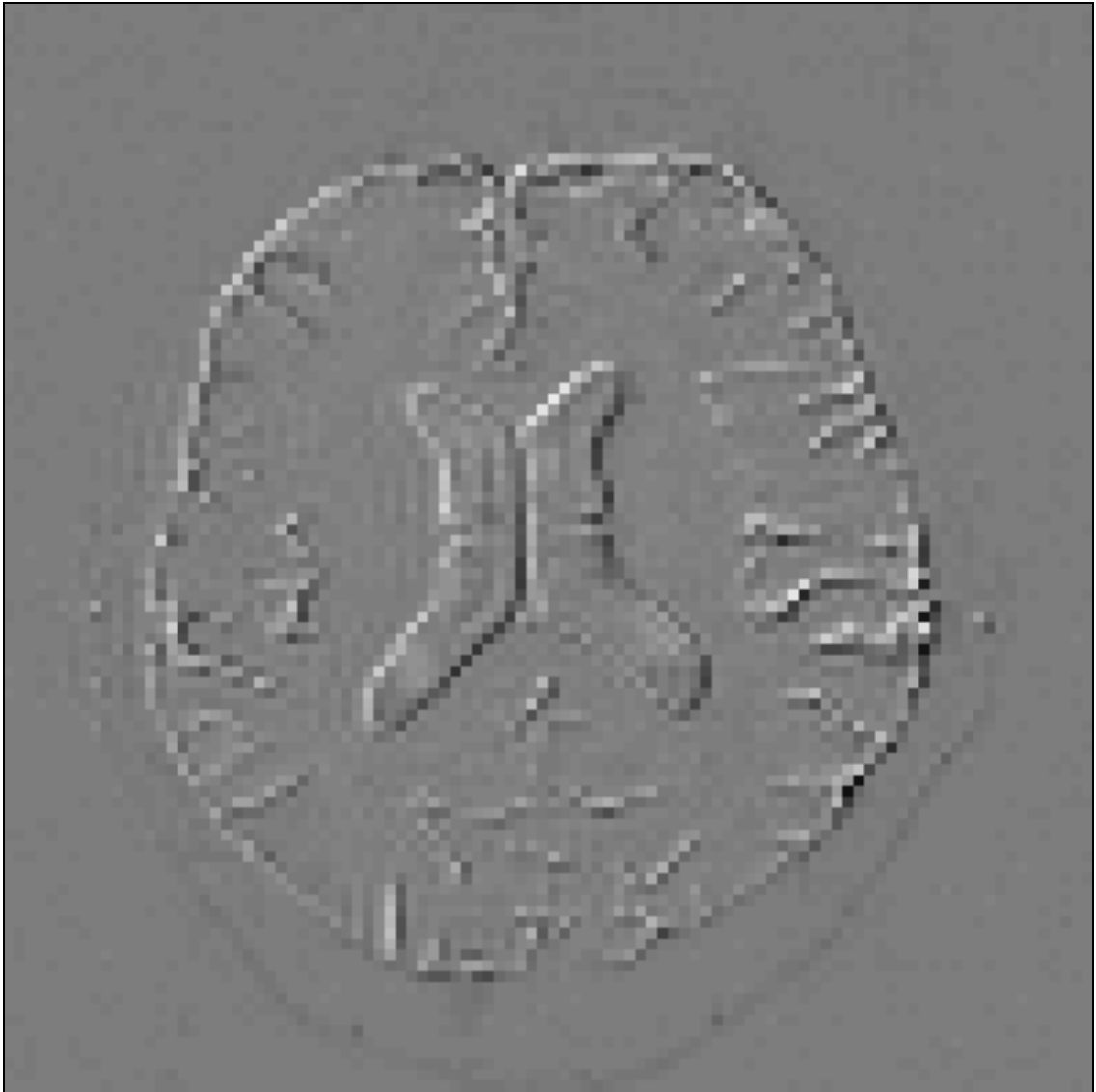}
    \end{subfigure}
    \begin{subfigure}{0.18\textwidth}
        \includegraphics[width=1\linewidth]{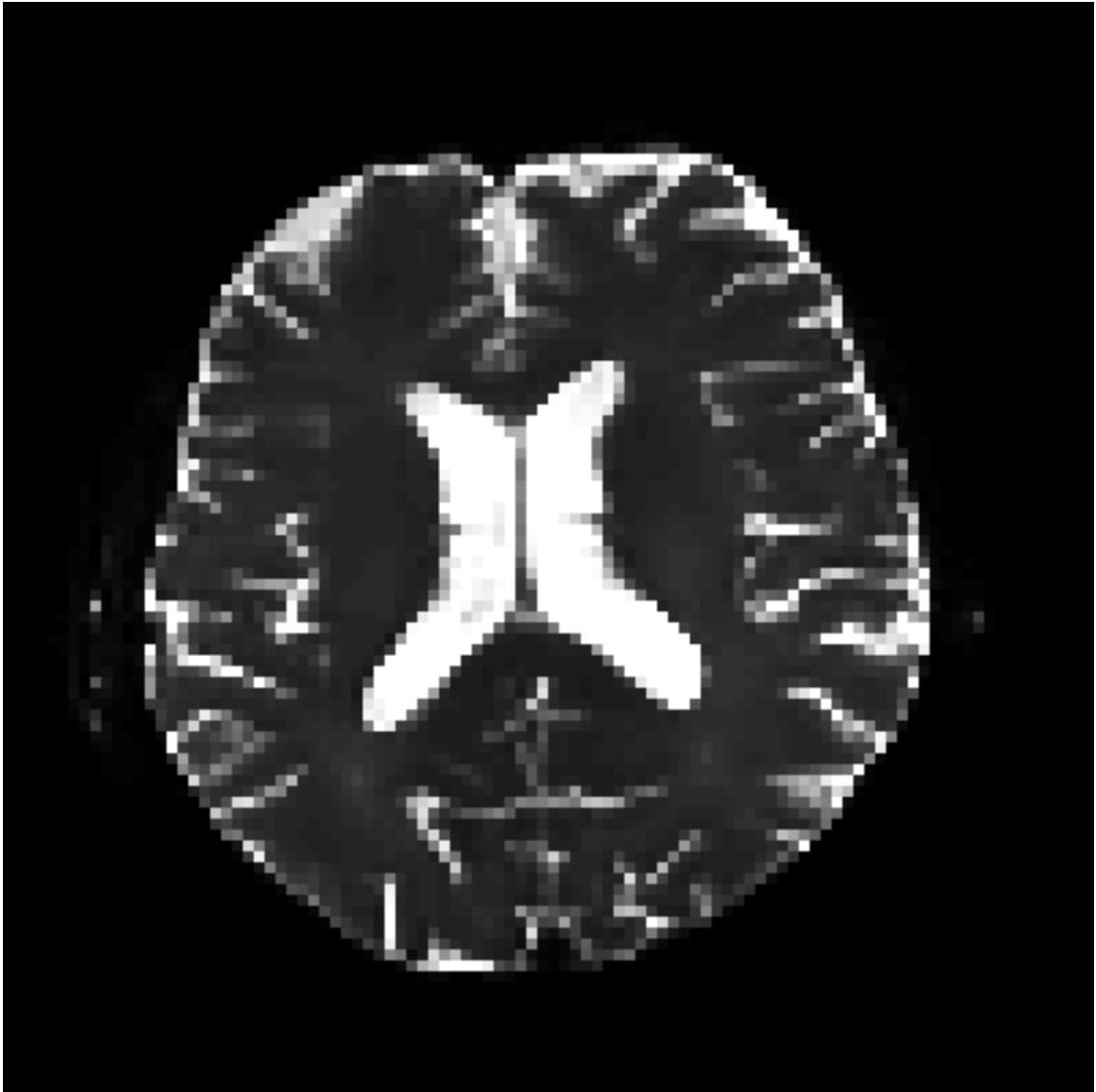}
    \end{subfigure}
    \begin{subfigure}{0.18\textwidth}
        \includegraphics[width=1\linewidth]{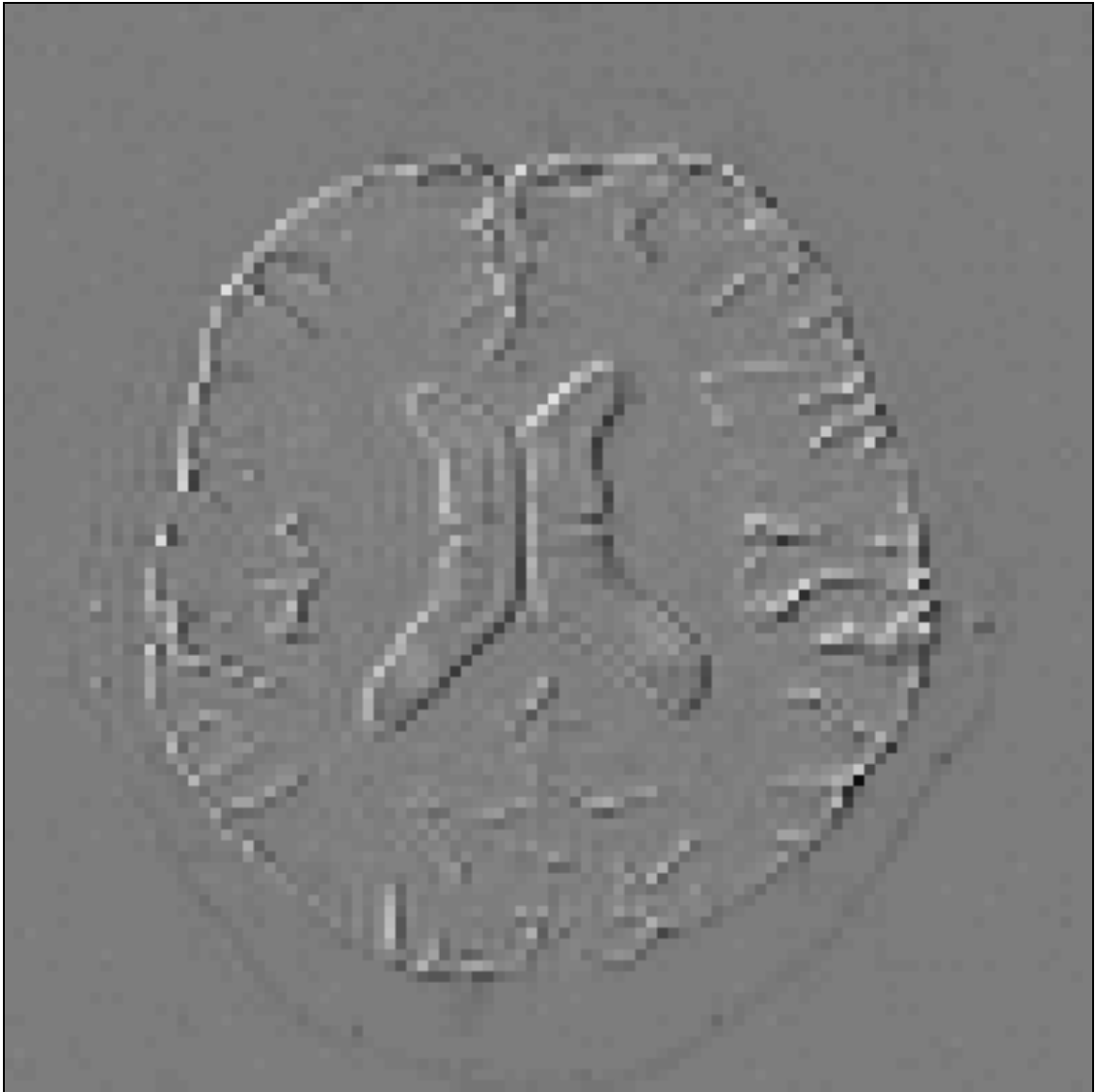}
    \end{subfigure}

    \begin{subfigure}{0.02\textwidth}
        \raggedright
        \raisebox{0in}{\rotatebox[origin=t]{90}{5/8 PF}}
    \end{subfigure}
    \begin{subfigure}{0.18\textwidth}
        \includegraphics[width=1\linewidth]{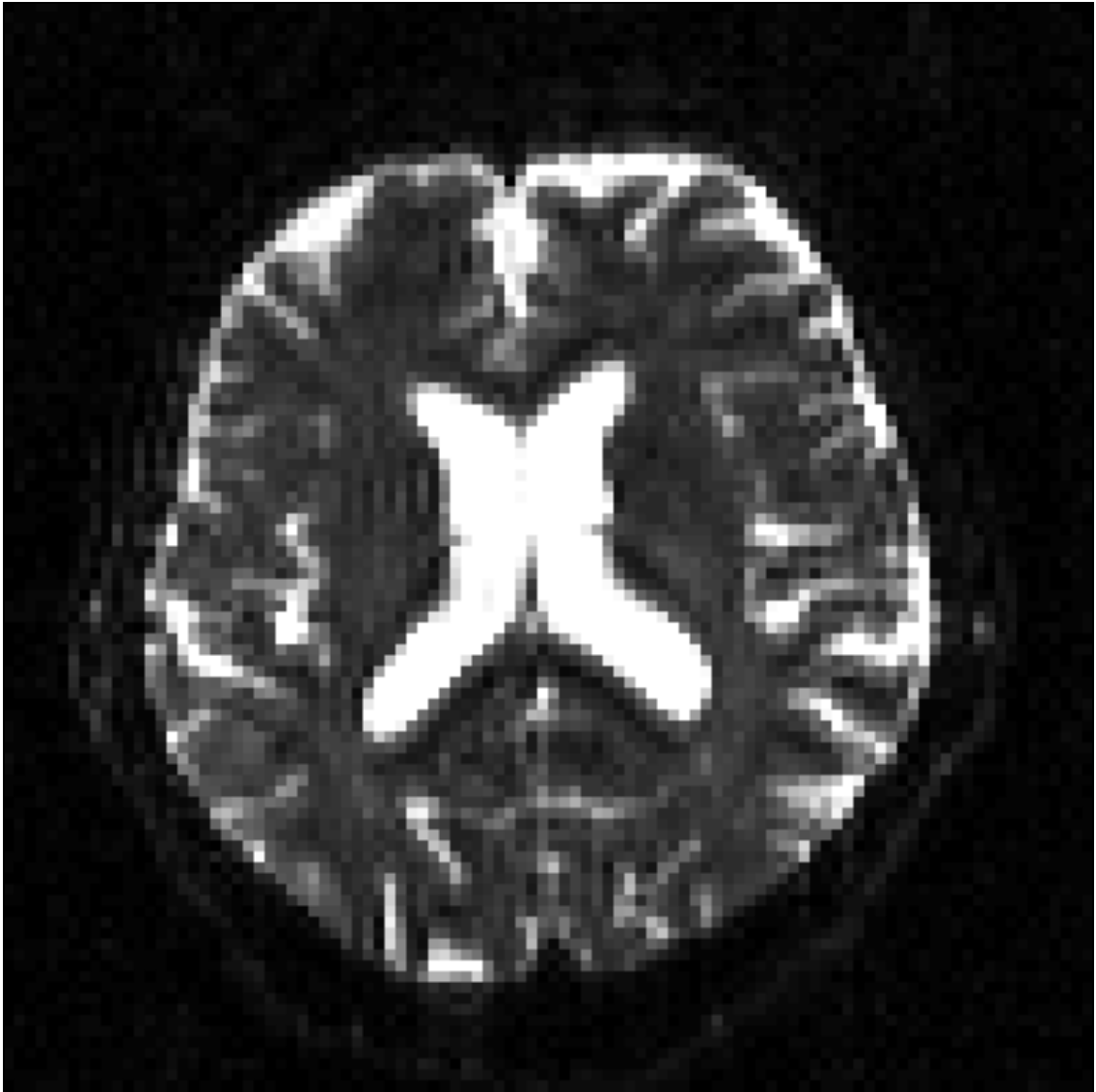}
    \end{subfigure}
    \begin{subfigure}{0.18\textwidth}
        \includegraphics[width=1\linewidth]{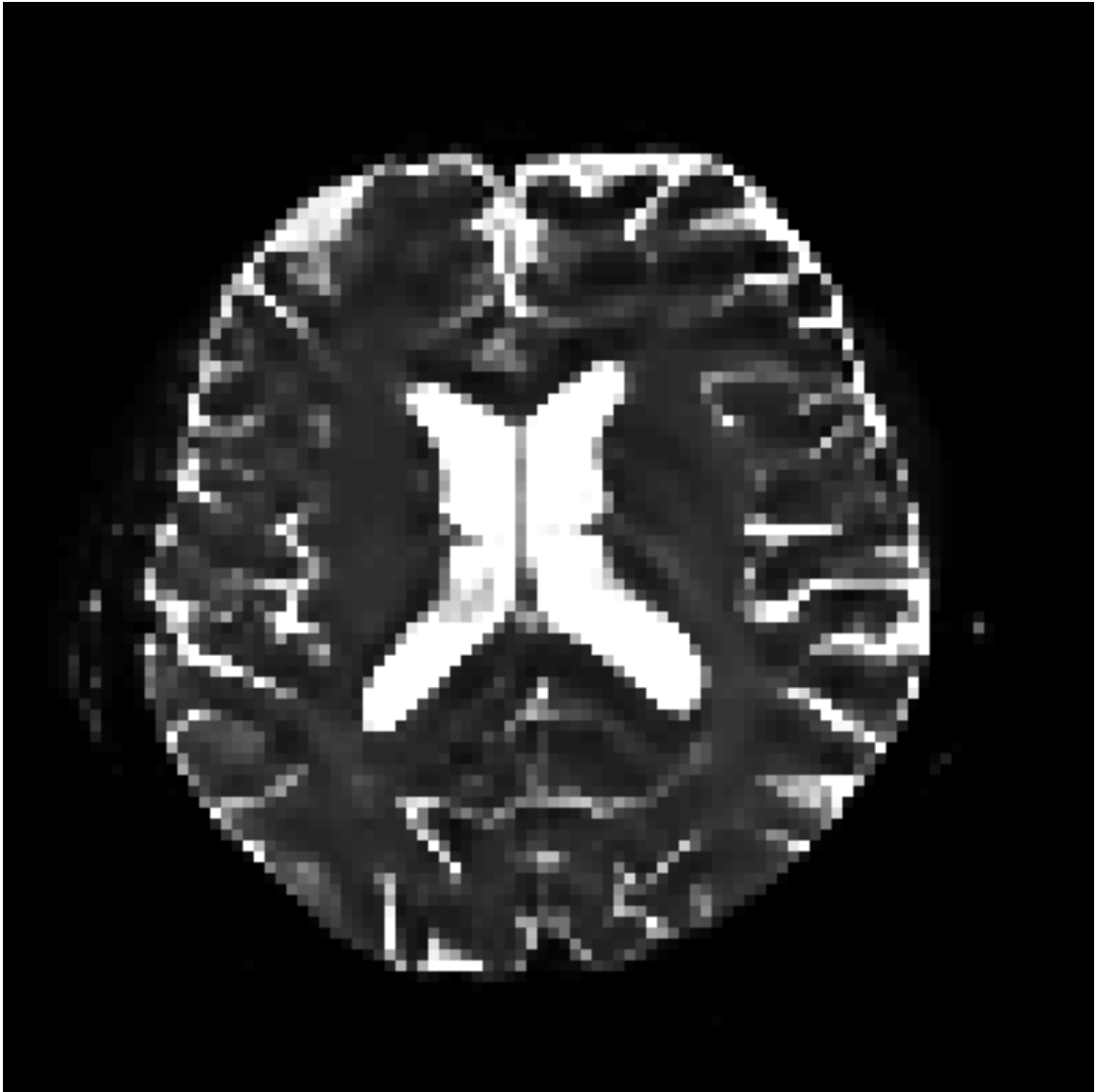}
    \end{subfigure}
    \begin{subfigure}{0.18\textwidth}
        \includegraphics[width=1\linewidth]{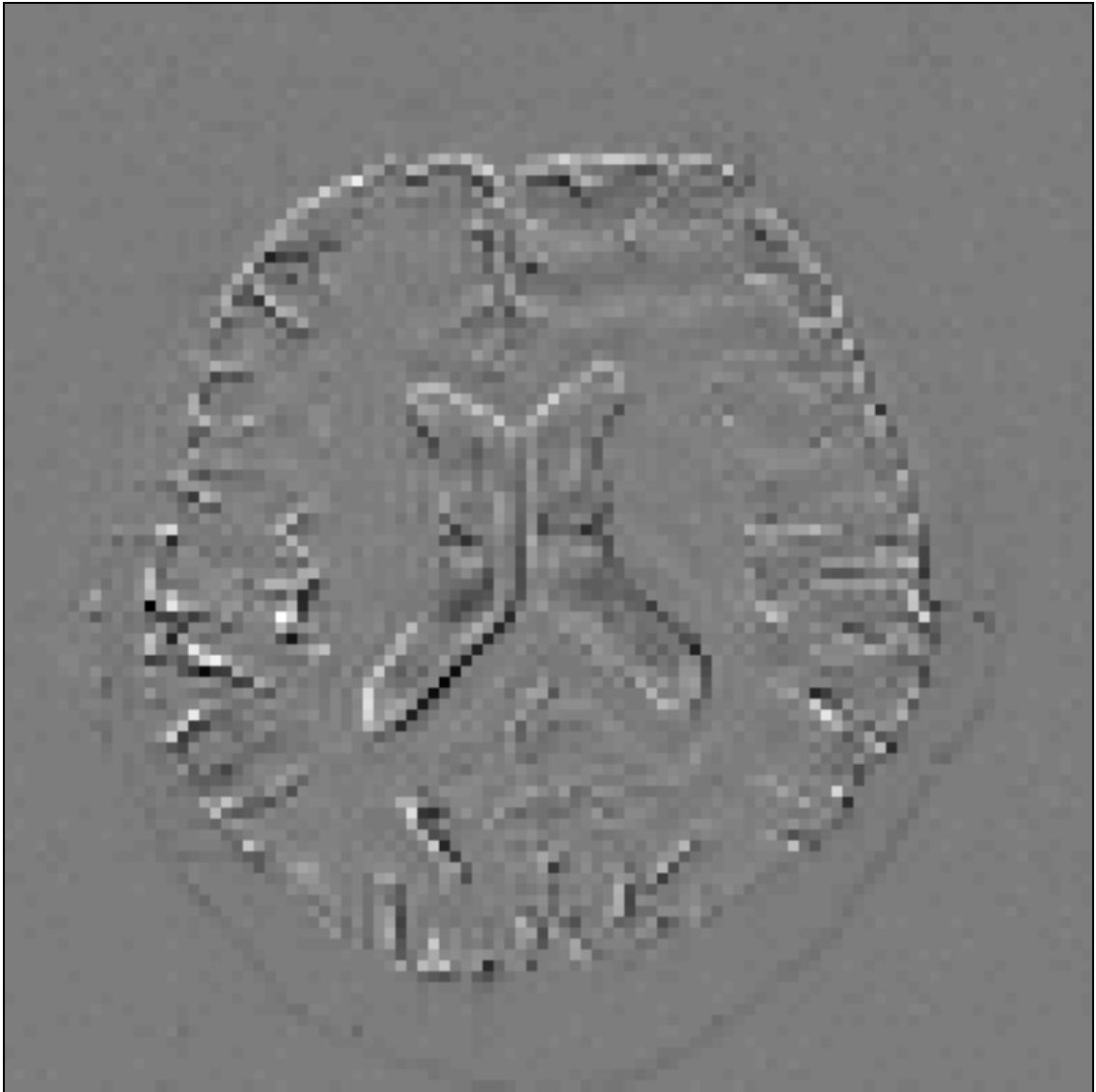}
    \end{subfigure}
    \begin{subfigure}{0.18\textwidth}
        \includegraphics[width=1\linewidth]{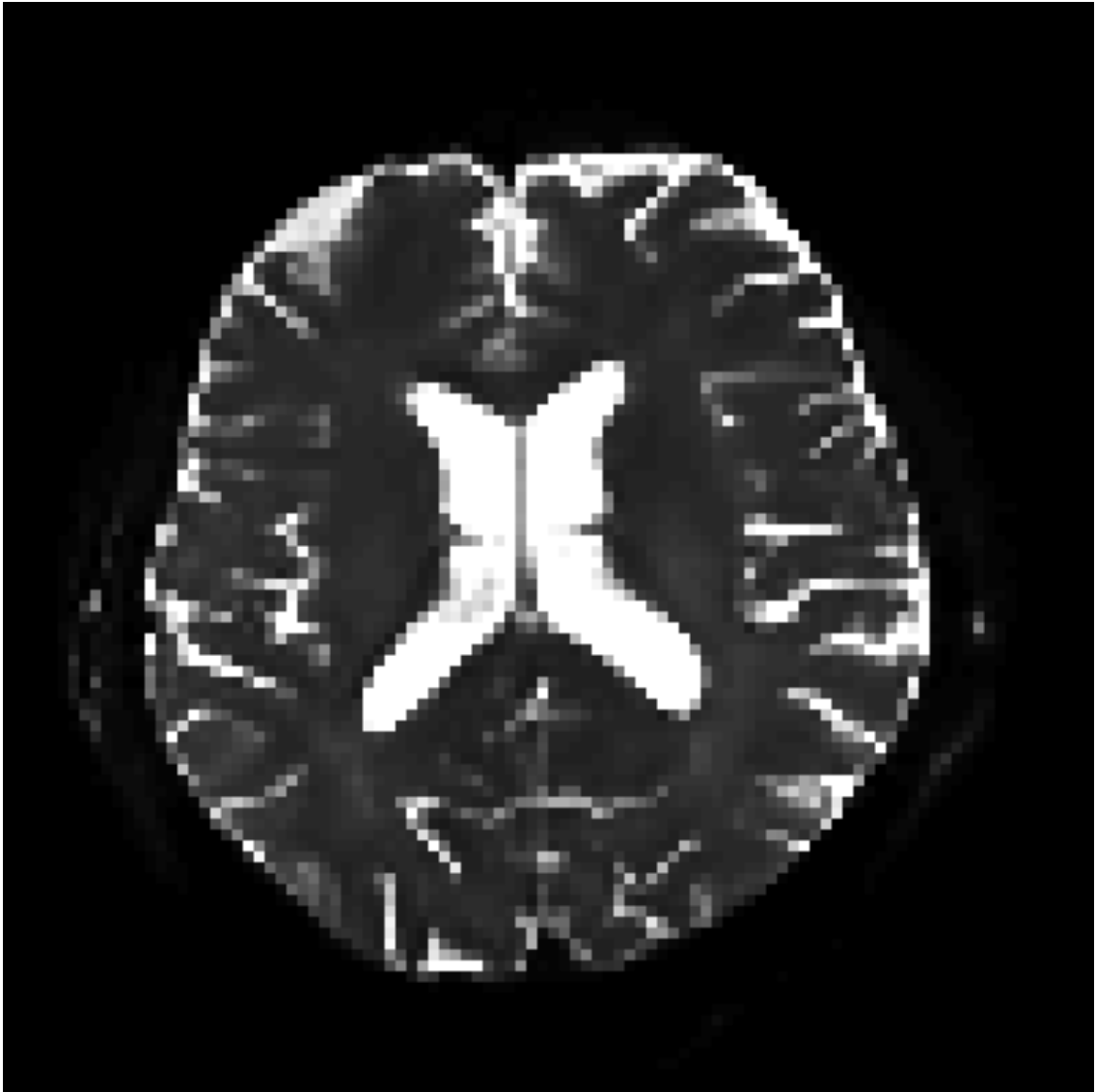}
    \end{subfigure}
    \begin{subfigure}{0.18\textwidth}
        \includegraphics[width=1\linewidth]{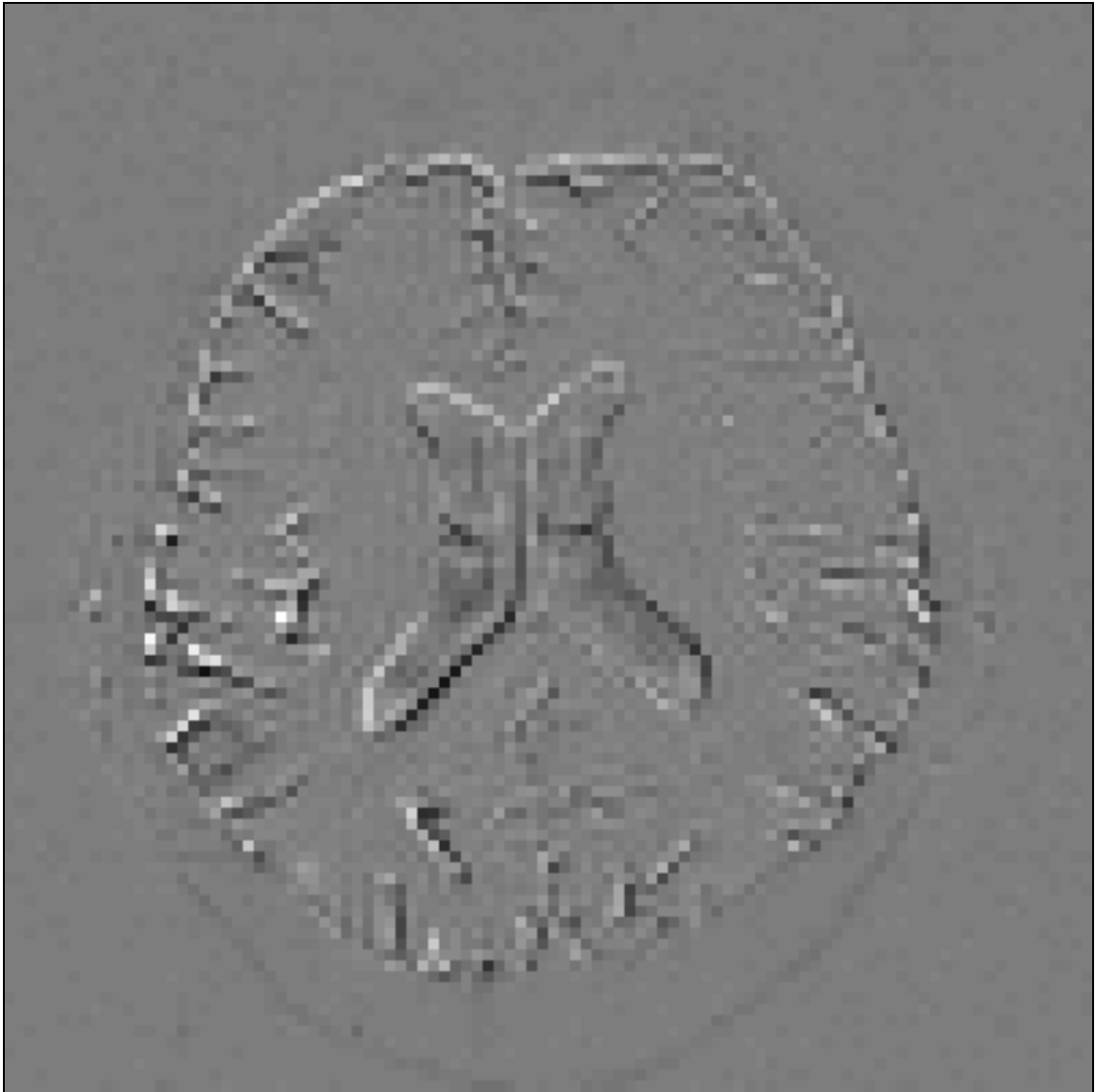}
    \end{subfigure}
    \caption{Examples of non-diffusion-weighted images from in vivo data at $b=0$ s/mm$^2$. Artifacts in the Raw image, $\mathbf{x}_t$, are corrected by the MCNN and CCNN models, $f_{\hat{\theta}}(\mathbf{x}_t)$. Also shown are the residuals between the CNN corrections and the original Raw image with Rician bias correction $\text{Ric}(\mathbf{x}_t)$ \cite{veraart2016denoising}. The Gibbs artifacts removed by the methods are observed in the residuals. The MCNN method introduces some banding artifacts at the PF 5/8ths factor that are not present in the CCNN method.}
    \label{fig:dl_meth_dwi}
\end{figure*}

Figure \ref{fig:dl_meth_dwi} shows examples of non-diffusion-weighted images before and after processing. The raw images (Raw) served as the input for the magnitude deep learning (MCNN) and complex deep learning (CCNN) methods.
Both methods remove artifacts, but the MCNN method allows residual rippling artifacts to pass through in the presence of partial Fourier. These rippling artifacts are not present in the CCNN method.

\begin{figure*}
    \centering
    \begin{subfigure}{0.02\textwidth}
        \raisebox{0in}{\rotatebox[origin=t]{90}{}}
    \end{subfigure}
    \begin{subfigure}{0.15\textwidth}
        \centering
        {Raw}
    \end{subfigure}
    \begin{subfigure}{0.15\textwidth}
        \centering
        {SoA}
    \end{subfigure}
    \begin{subfigure}{0.15\textwidth}
        \centering
        {MCNN}
    \end{subfigure}
    \begin{subfigure}{0.15\textwidth}
        \centering
        {Standard PF}
    \end{subfigure}
    \begin{subfigure}{0.15\textwidth}
        \centering
        {CCNN}
    \end{subfigure}
    \begin{subfigure}{0.05\textwidth}
        \centering
        {\hspace{1mm}}
    \end{subfigure}
    \begin{subfigure}{0.02\textwidth}
        \raisebox{0in}{\rotatebox[origin=t]{-90}{\hspace{1mm}}}
    \end{subfigure}

    \begin{subfigure}{0.02\textwidth}
        \raggedright
        \raisebox{0in}{\rotatebox[origin=t]{90}{$b=0$ s/mm$^2$}}
    \end{subfigure}
    \begin{subfigure}{0.15\textwidth}
        \includegraphics[width=1\linewidth]{figures/raw_pf_8_8_dwi-eps-converted-to.pdf}
    \end{subfigure}
    \begin{subfigure}{0.15\textwidth}
        \includegraphics[width=1\linewidth]{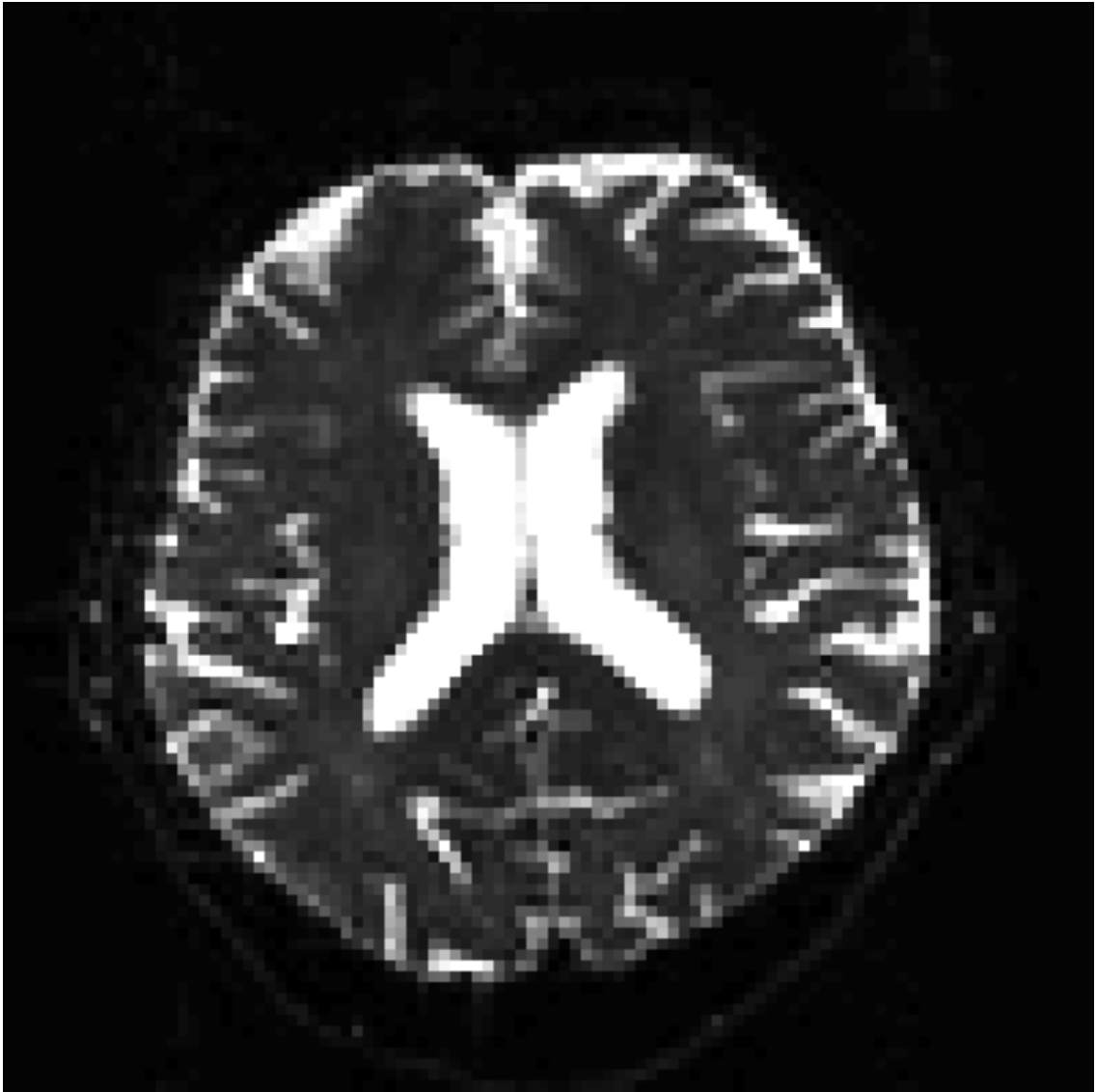}
    \end{subfigure}
    \begin{subfigure}{0.15\textwidth}
        \includegraphics[width=1\linewidth]{figures/dl_10layers_pf8_mag2mag_dwi-eps-converted-to.pdf}
    \end{subfigure}
    \begin{subfigure}{0.15\textwidth}
        \includegraphics[width=1\linewidth]{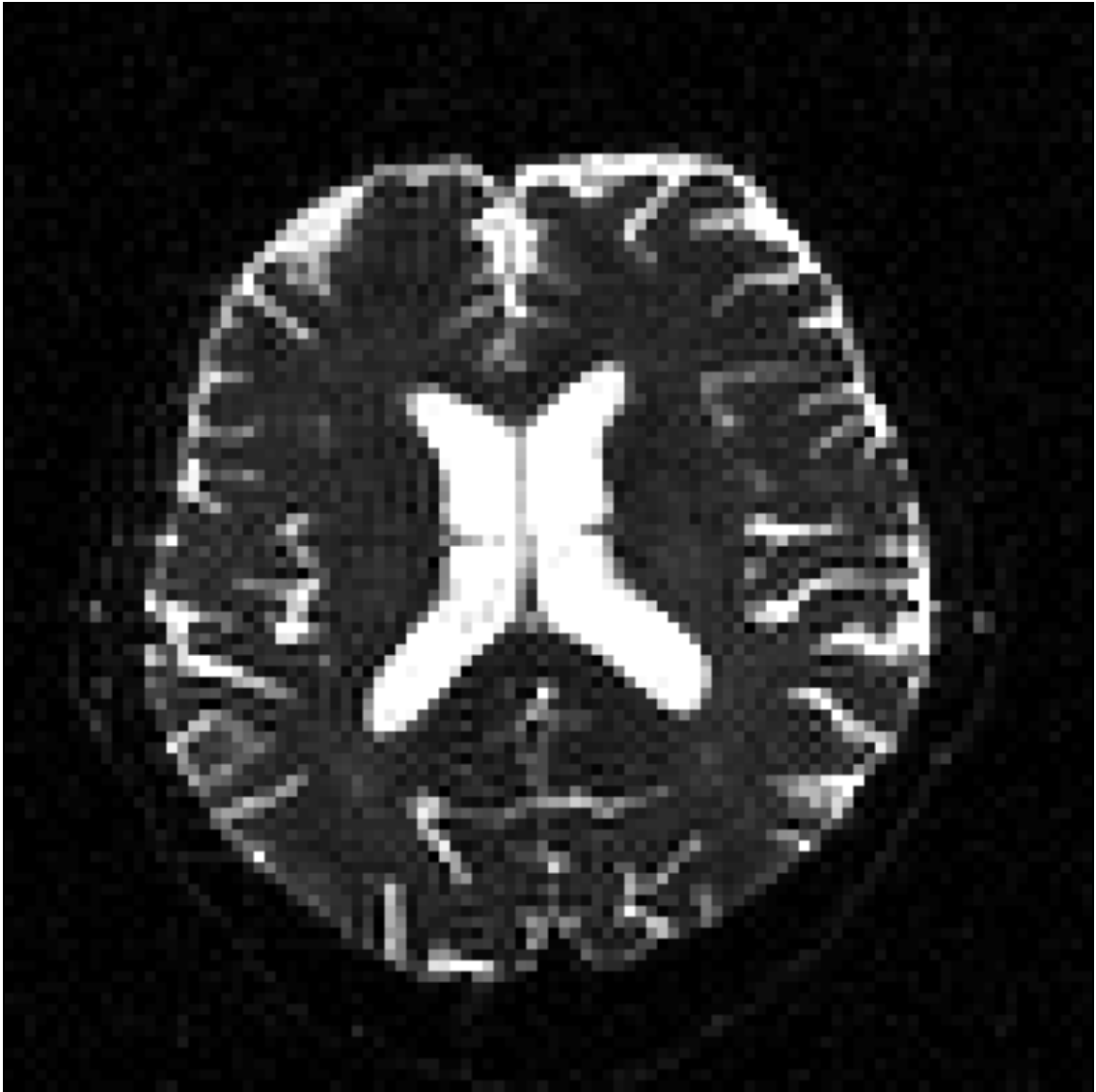}
    \end{subfigure}
    \begin{subfigure}{0.15\textwidth}
        \includegraphics[width=1\linewidth]{figures/dl_10layers_pf8_comp2mag_dwi-eps-converted-to.pdf}
    \end{subfigure}
    \begin{subfigure}{0.04\textwidth}
        \centering
        {\hspace{1mm}}
    \end{subfigure}
    \begin{subfigure}{0.02\textwidth}
        \raisebox{0in}{\rotatebox[origin=t]{-90}{\hspace{1mm}}}
    \end{subfigure}

    \begin{subfigure}{0.02\textwidth}
        \raggedright
        \raisebox{0in}{\rotatebox[origin=t]{90}{No PF}}
    \end{subfigure}
    \begin{subfigure}{0.15\textwidth}
        \includegraphics[width=1\linewidth]{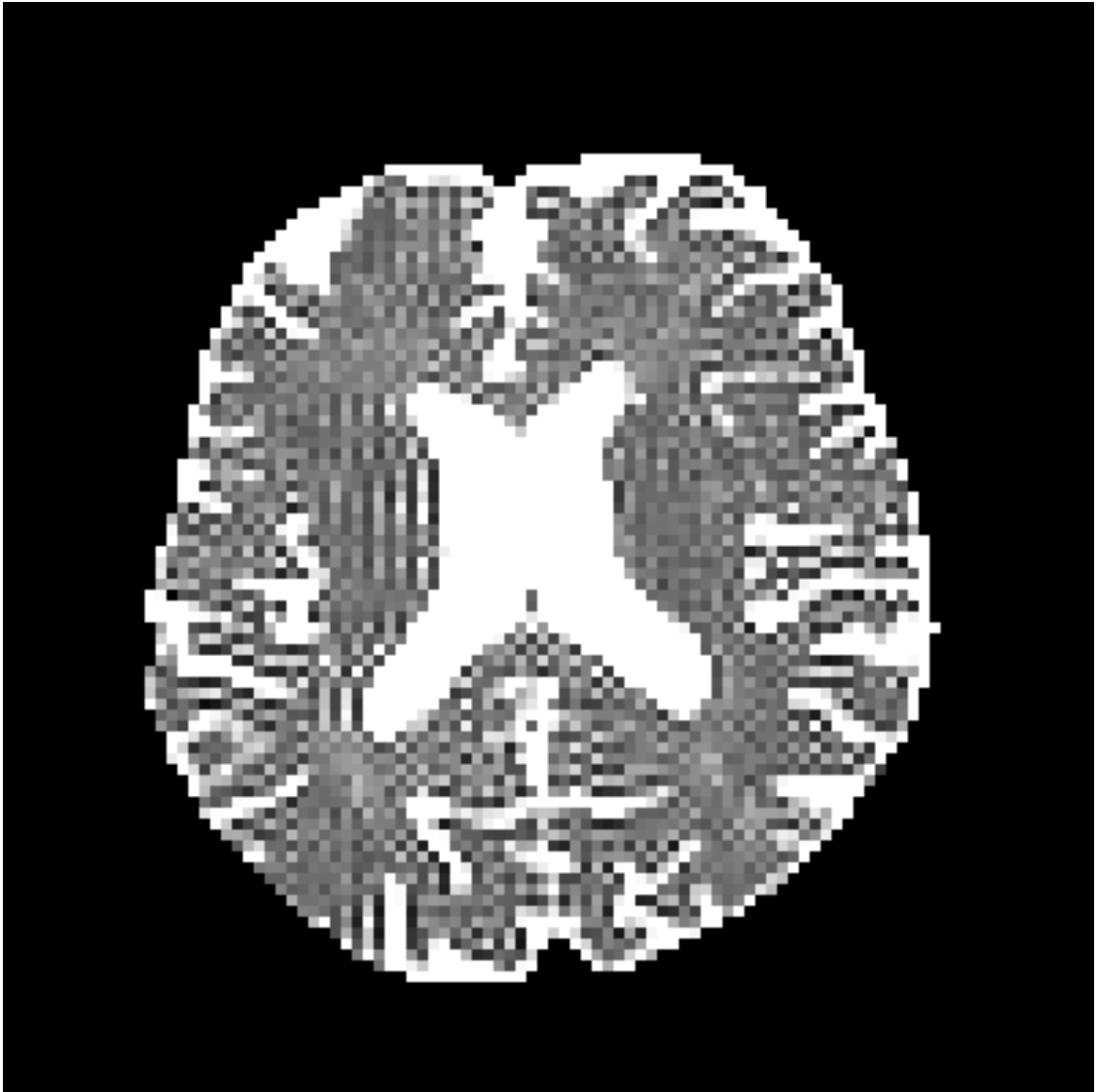}
    \end{subfigure}
    \begin{subfigure}{0.15\textwidth}
        \includegraphics[width=1\linewidth]{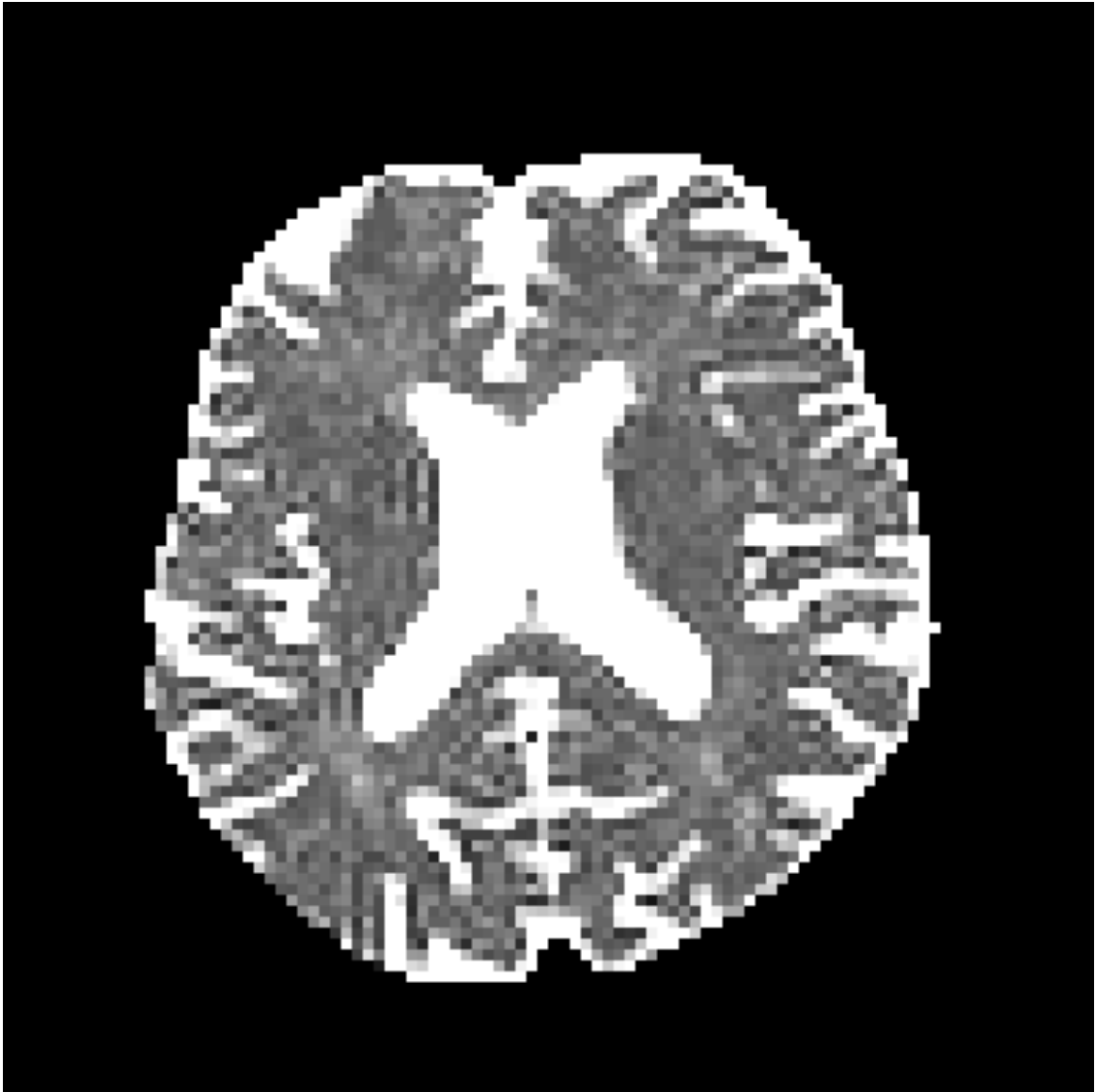}
    \end{subfigure}
    \begin{subfigure}{0.15\textwidth}
        \includegraphics[width=1\linewidth]{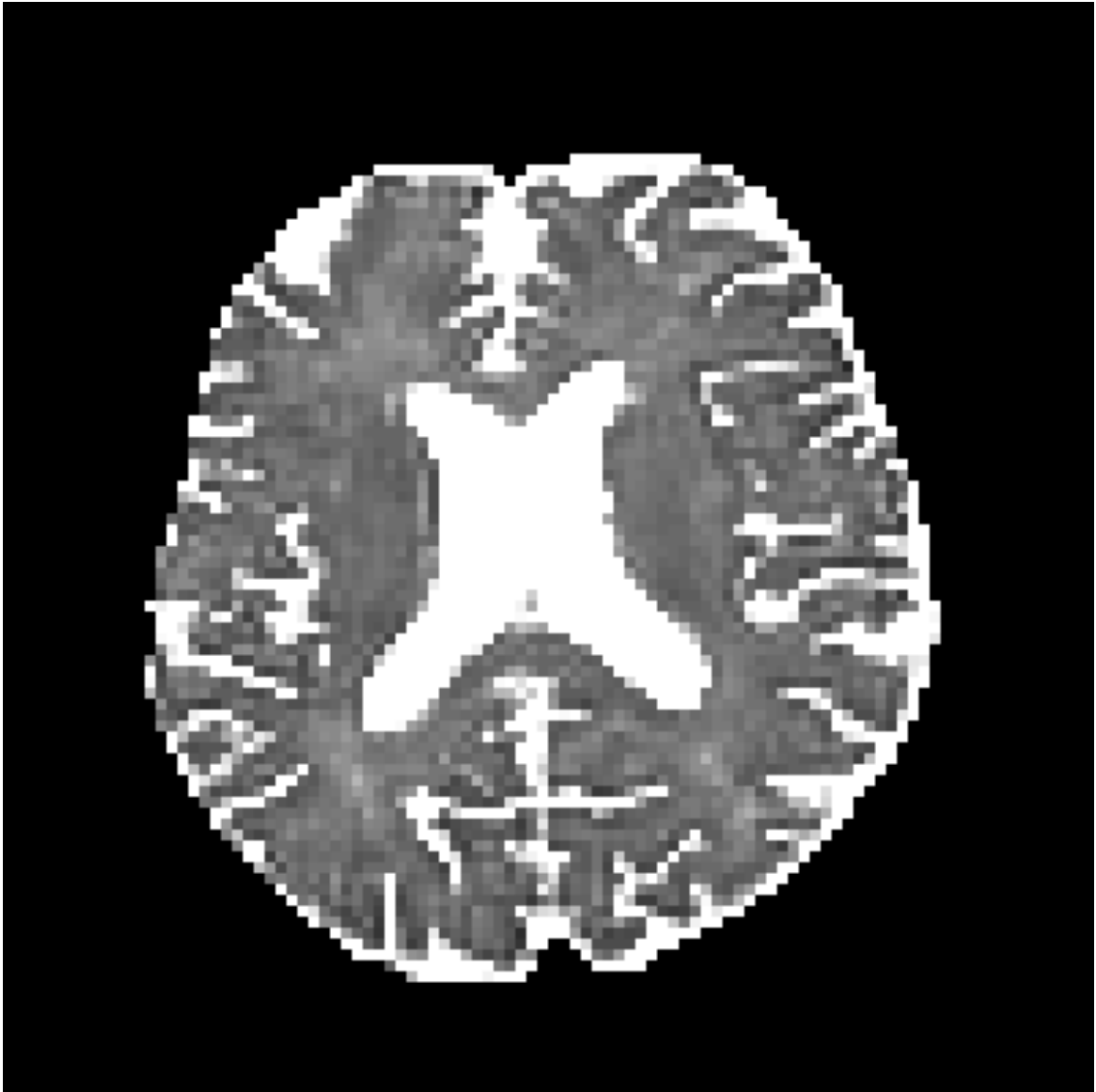}
    \end{subfigure}
    \begin{subfigure}{0.15\textwidth}
        \includegraphics[width=1\linewidth]{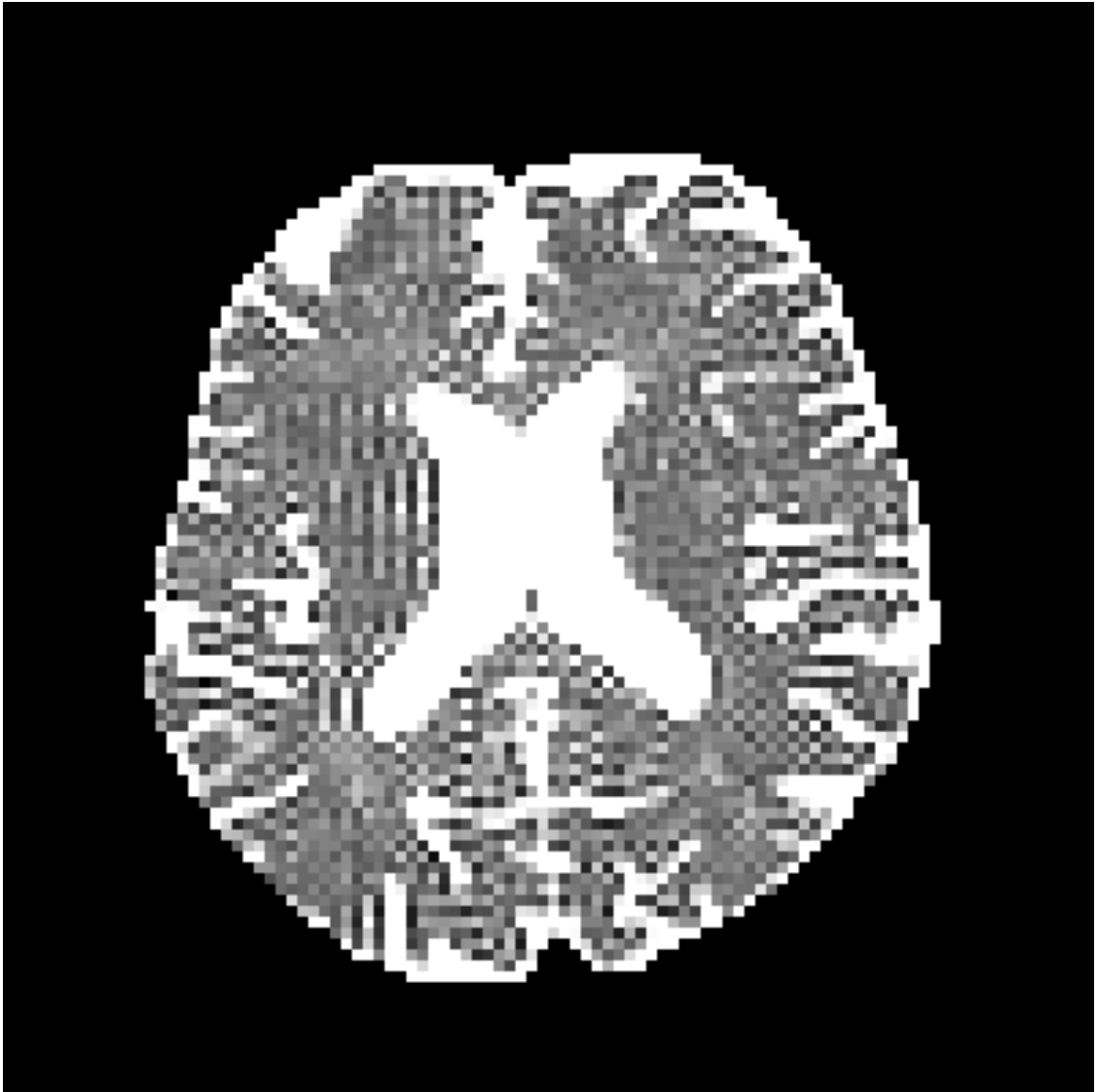}
    \end{subfigure}
    \begin{subfigure}{0.15\textwidth}
        \includegraphics[width=1\linewidth]{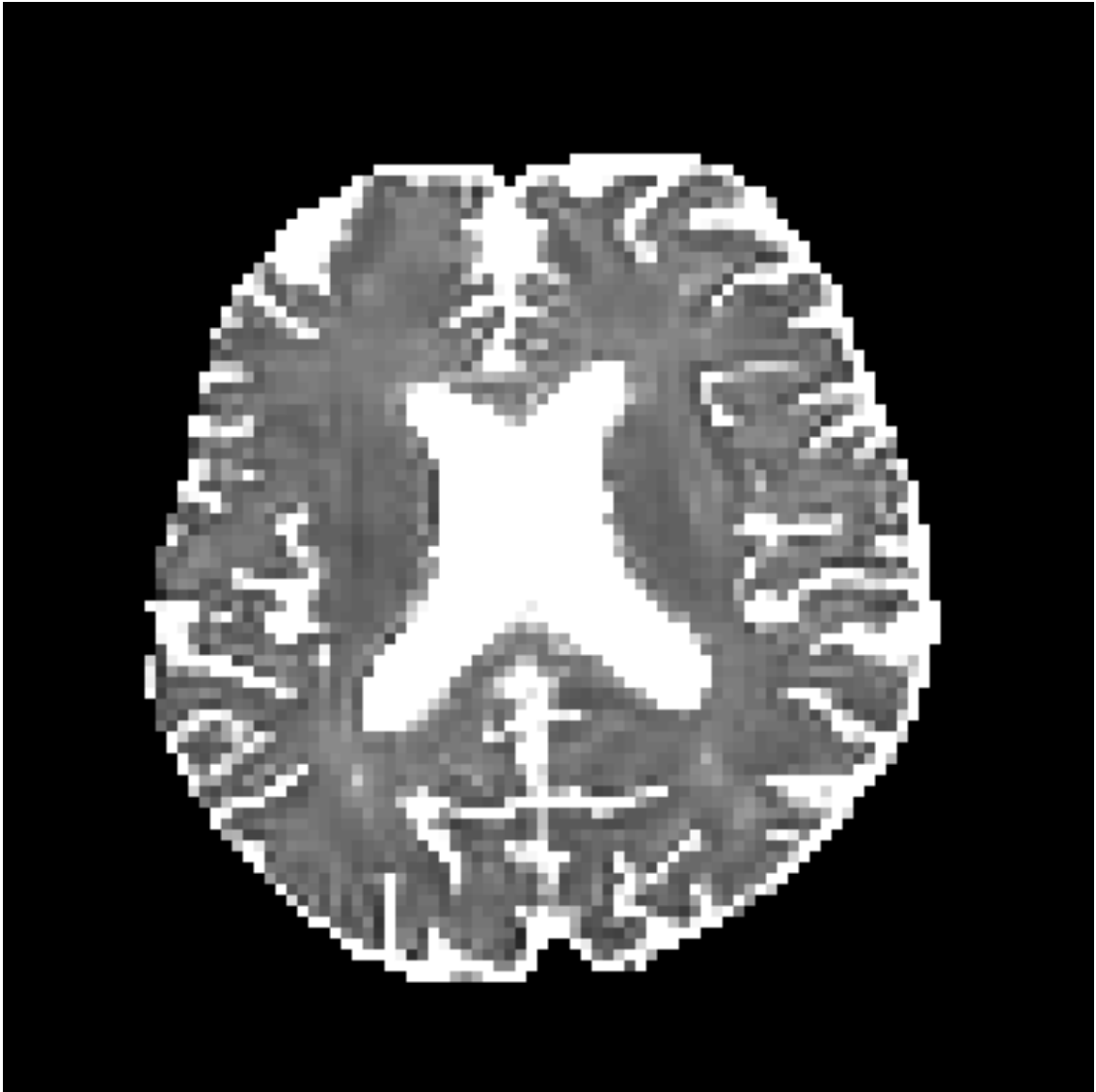}
    \end{subfigure}
    \begin{subfigure}{0.04\textwidth}
        \includegraphics[height=0.8in]{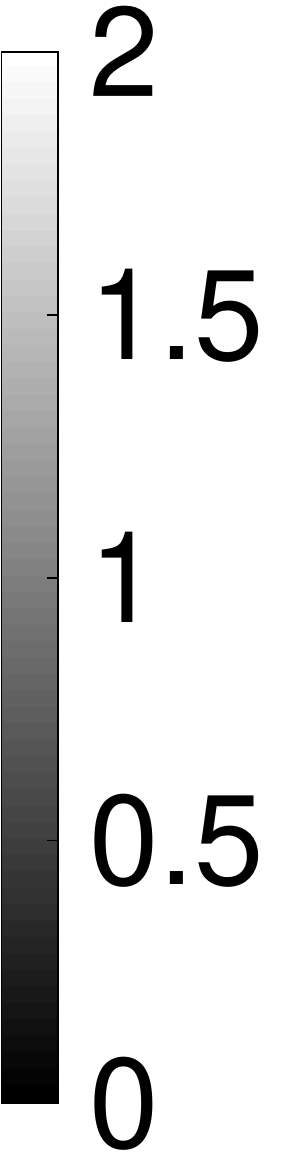}
    \end{subfigure}
    \begin{subfigure}{0.02\textwidth}
        \raisebox{0in}{\rotatebox[origin=t]{-90}{\small MD, $\mu$m$^2$/ms}}
    \end{subfigure}

    \begin{subfigure}{0.02\textwidth}
        \raggedright
        \raisebox{0in}{\rotatebox[origin=t]{90}{5/8 PF}}
    \end{subfigure}
    \begin{subfigure}{0.15\textwidth}
        \includegraphics[width=1\linewidth]{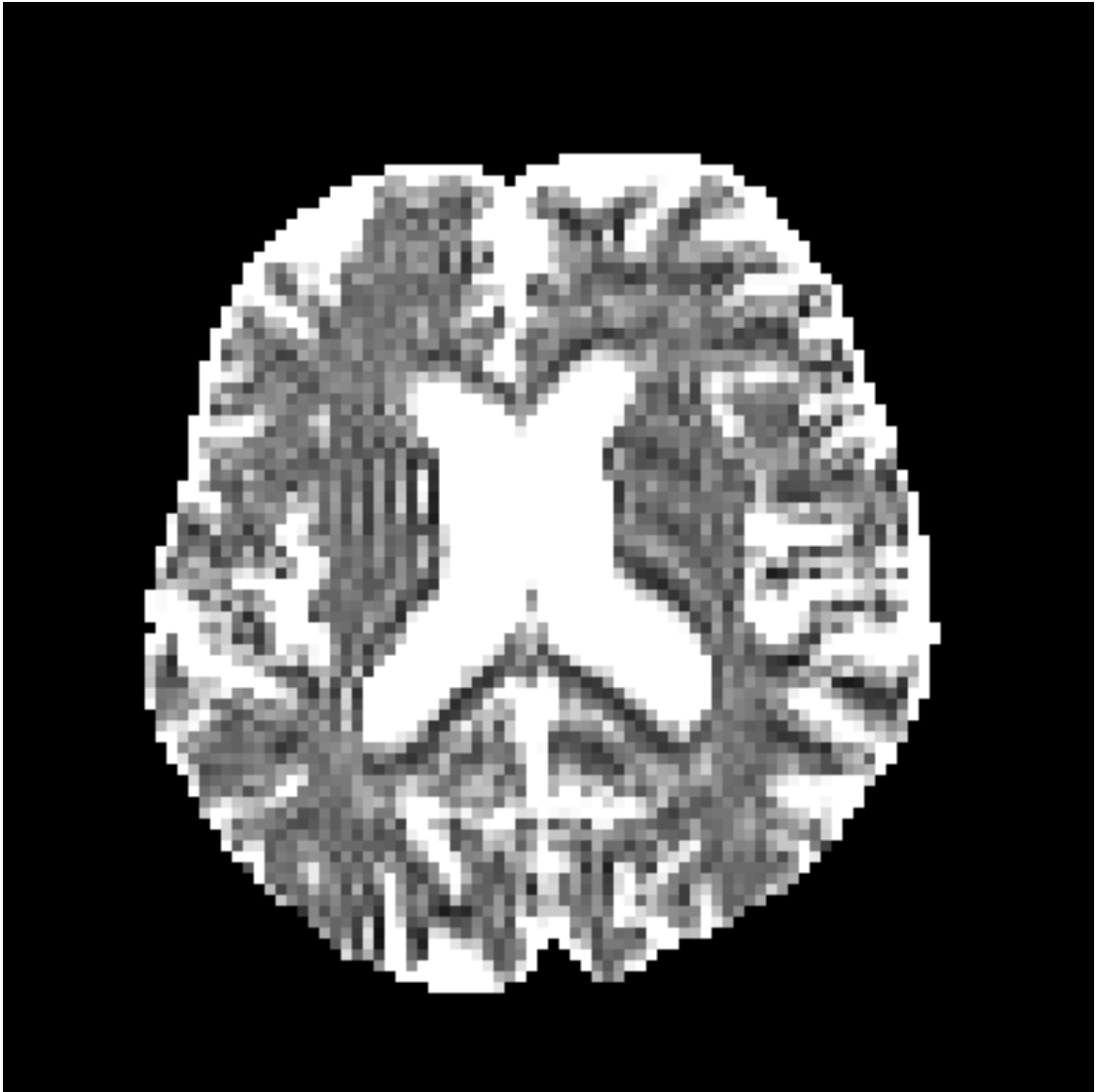}
    \end{subfigure}
    \begin{subfigure}{0.15\textwidth}
        \includegraphics[width=1\linewidth]{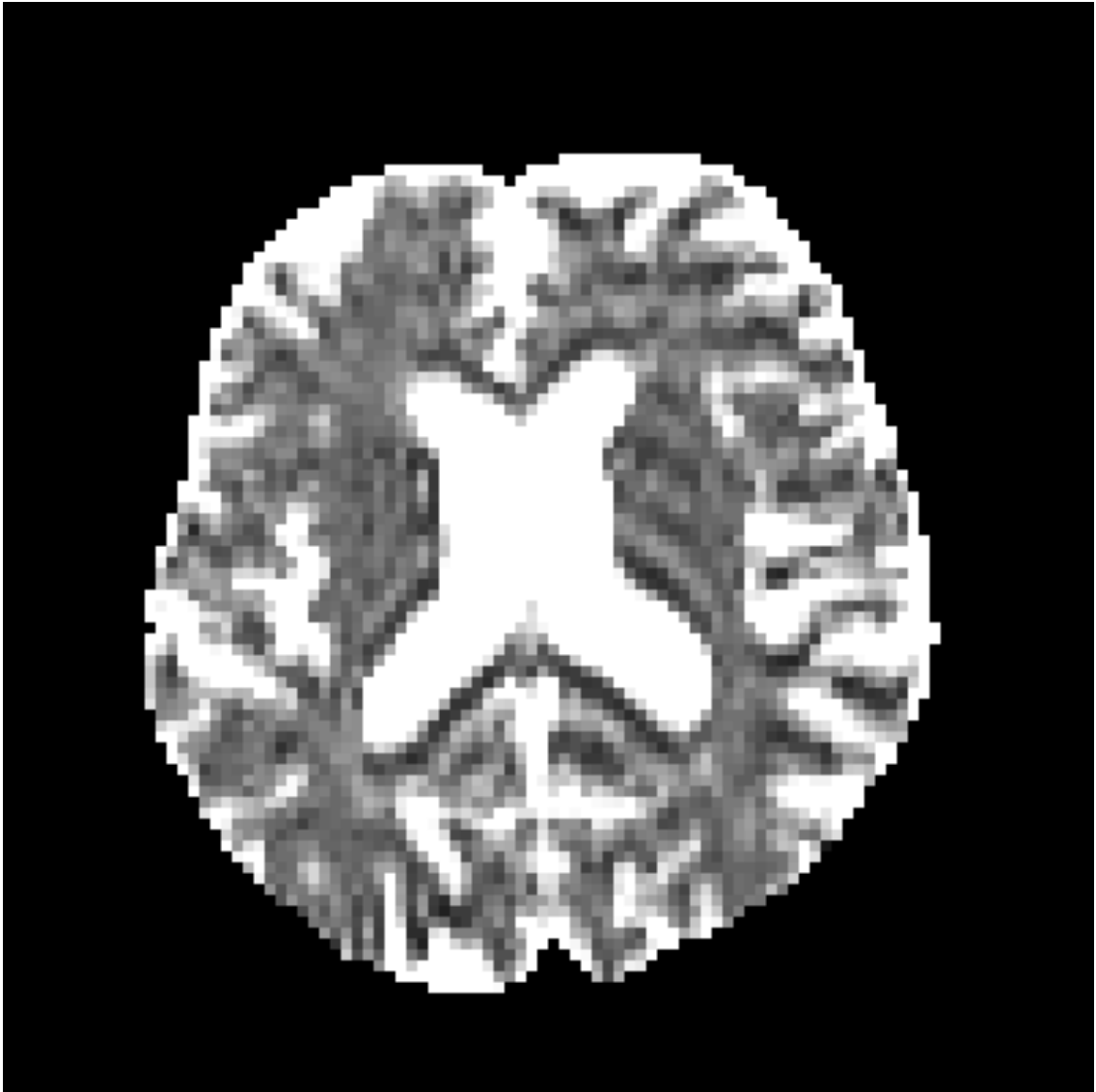}
    \end{subfigure}
    \begin{subfigure}{0.15\textwidth}
        \includegraphics[width=1\linewidth]{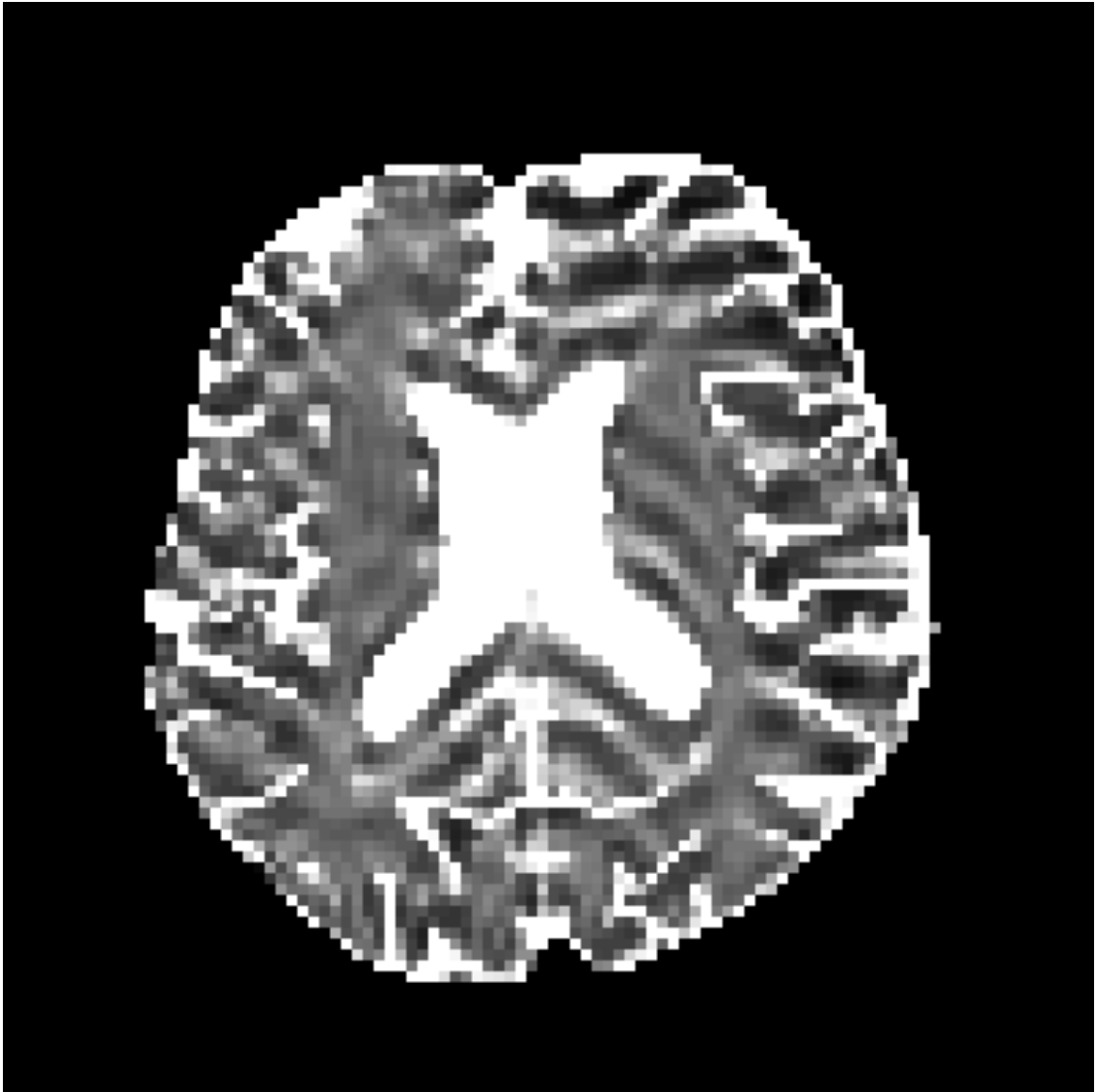}
    \end{subfigure}
    \begin{subfigure}{0.15\textwidth}
        \includegraphics[width=1\linewidth]{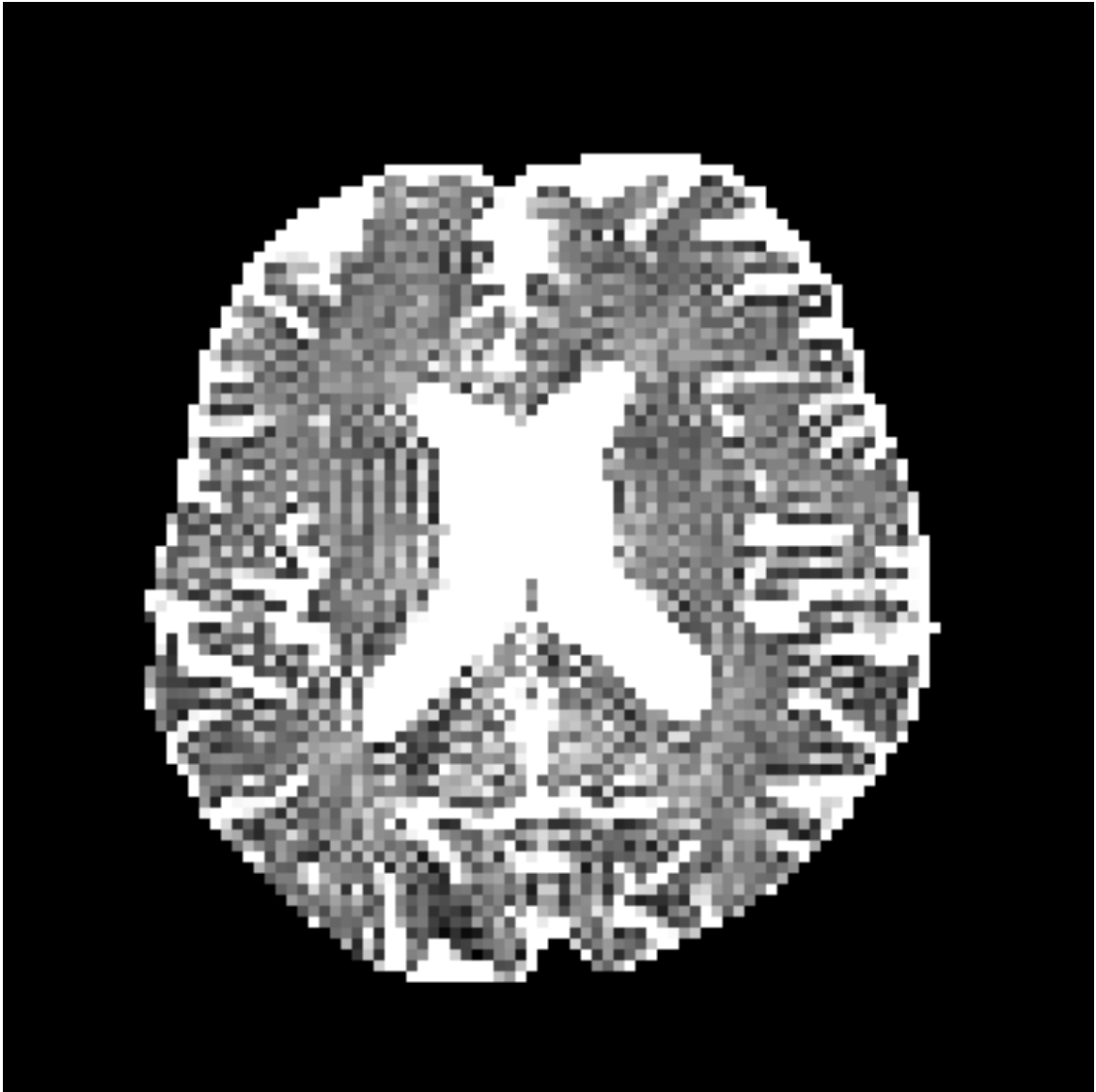}
    \end{subfigure}
    \begin{subfigure}{0.15\textwidth}
        \includegraphics[width=1\linewidth]{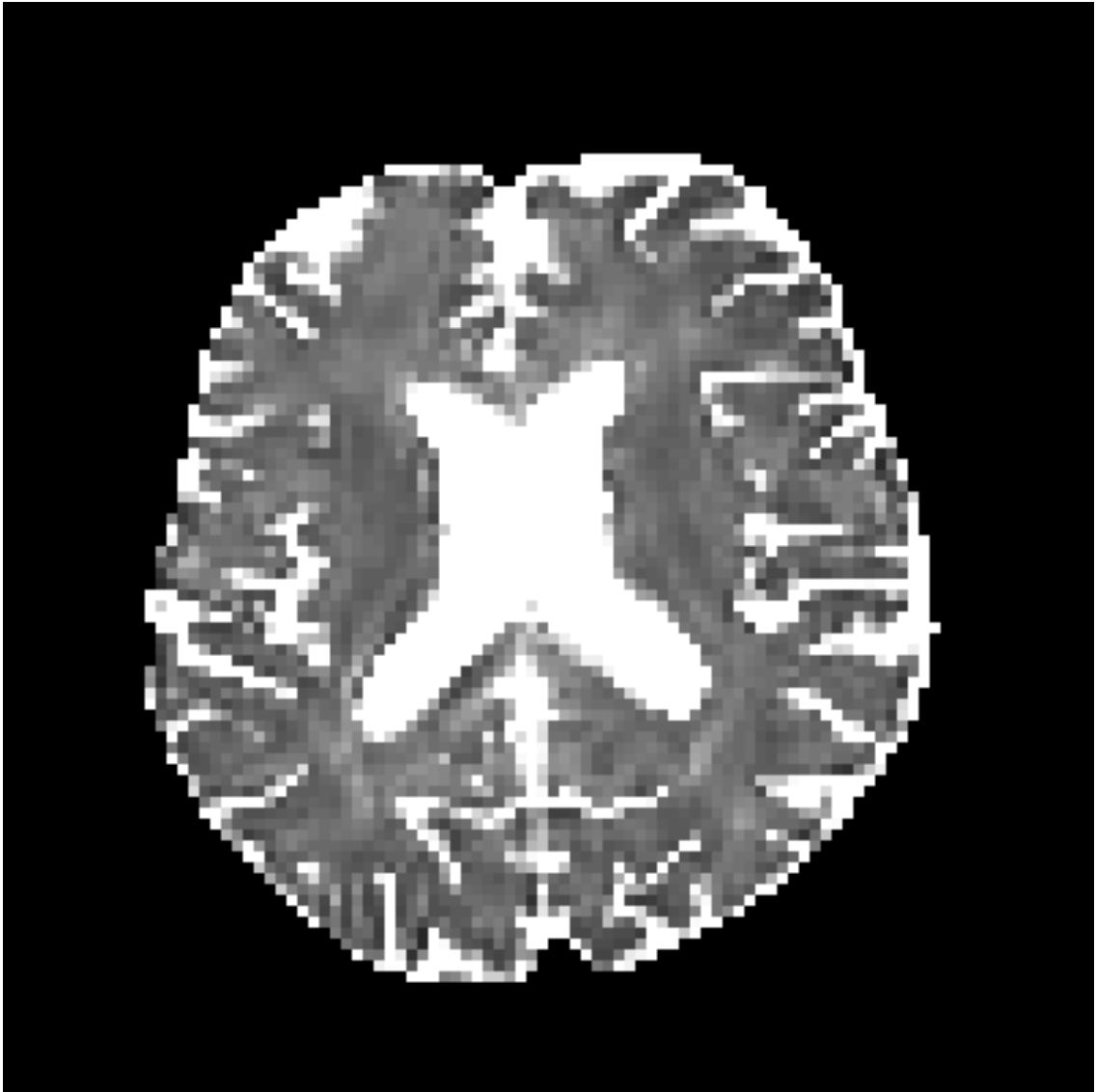}
    \end{subfigure}
    \begin{subfigure}{0.04\textwidth}
        \includegraphics[height=0.8in]{figures/md_colorbar-eps-converted-to.pdf}
    \end{subfigure}
    \begin{subfigure}{0.02\textwidth}
        \raisebox{0in}{\rotatebox[origin=t]{-90}{\small MD, $\mu$m$^2$/ms}}
    \end{subfigure}

    \begin{subfigure}{0.02\textwidth}
        \raggedright
        \raisebox{0in}{\rotatebox[origin=t]{90}{No PF}}
    \end{subfigure}
    \begin{subfigure}{0.15\textwidth}
        \includegraphics[width=1\linewidth]{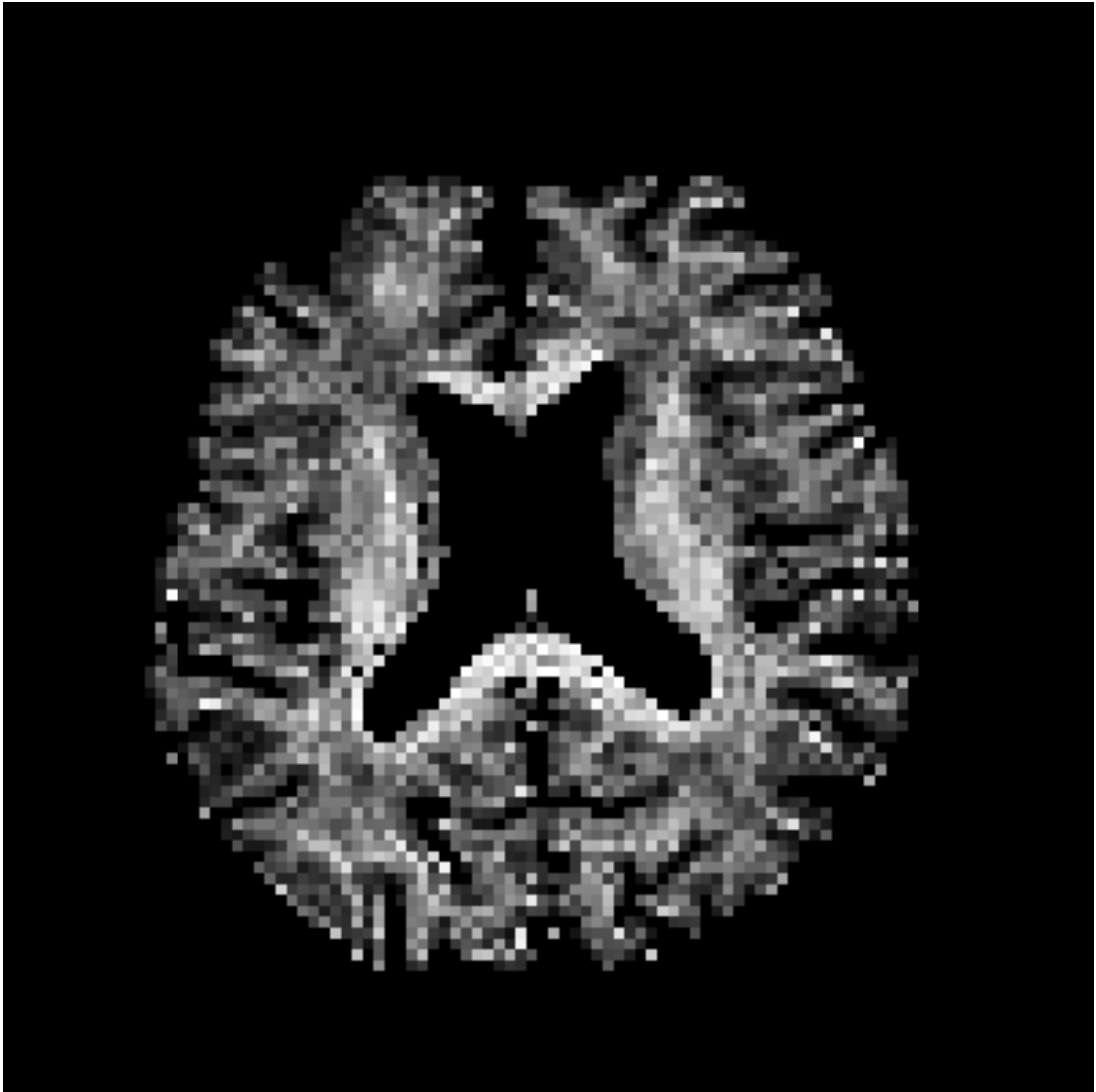}
    \end{subfigure}
    \begin{subfigure}{0.15\textwidth}
        \includegraphics[width=1\linewidth]{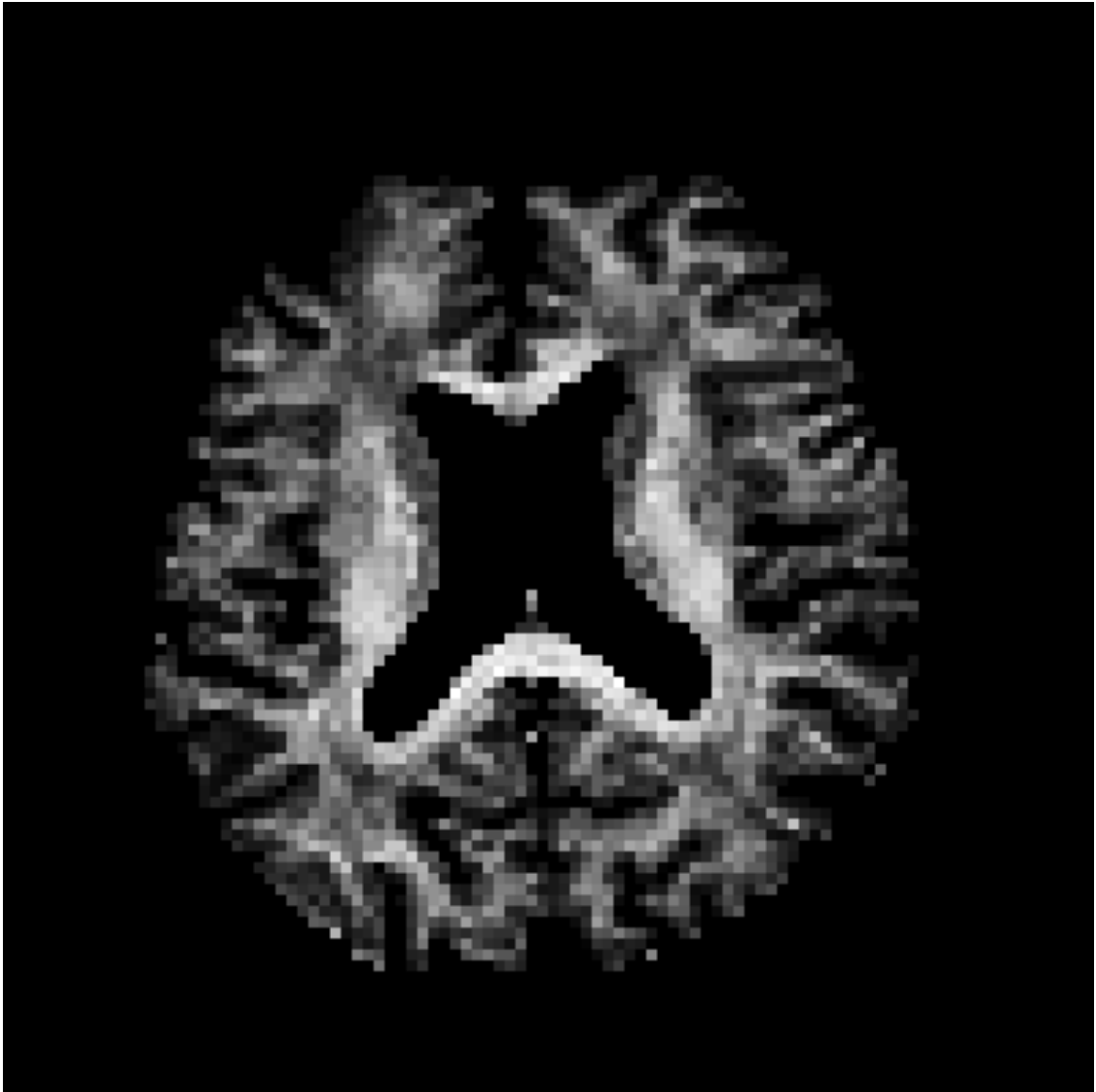}
    \end{subfigure}
    \begin{subfigure}{0.15\textwidth}
        \includegraphics[width=1\linewidth]{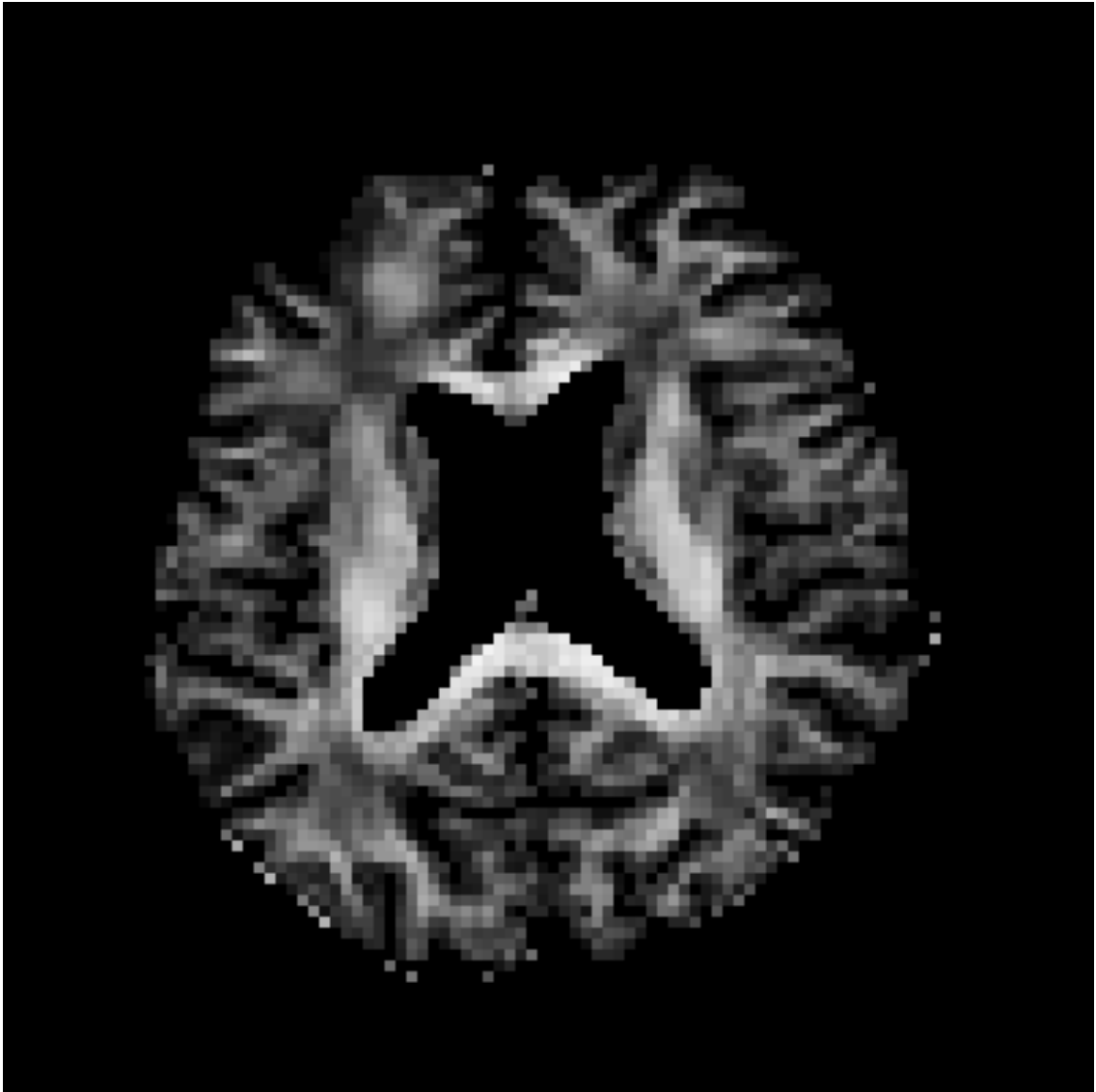}
    \end{subfigure}
    \begin{subfigure}{0.15\textwidth}
        \includegraphics[width=1\linewidth]{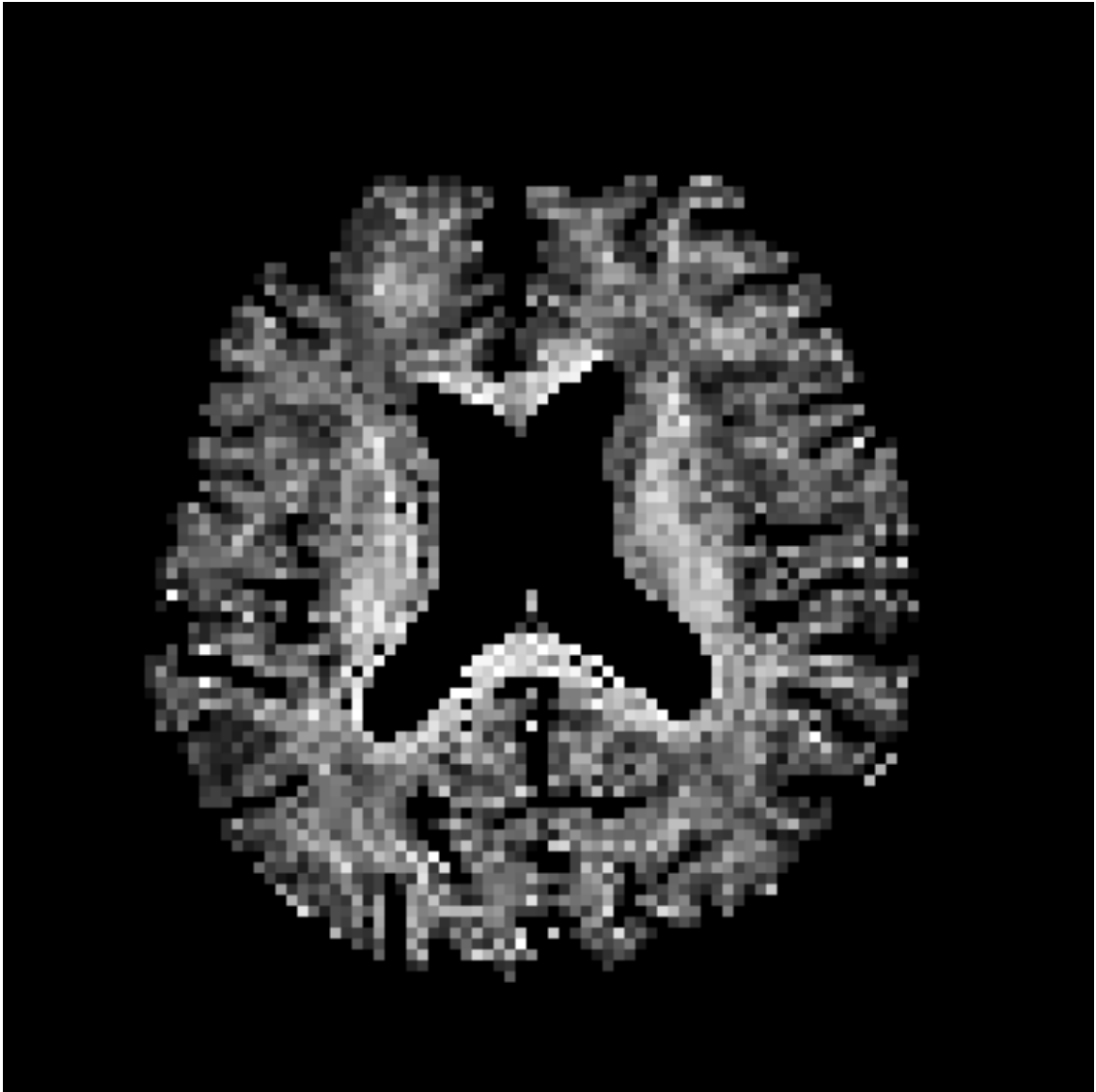}
    \end{subfigure}
    \begin{subfigure}{0.15\textwidth}
        \includegraphics[width=1\linewidth]{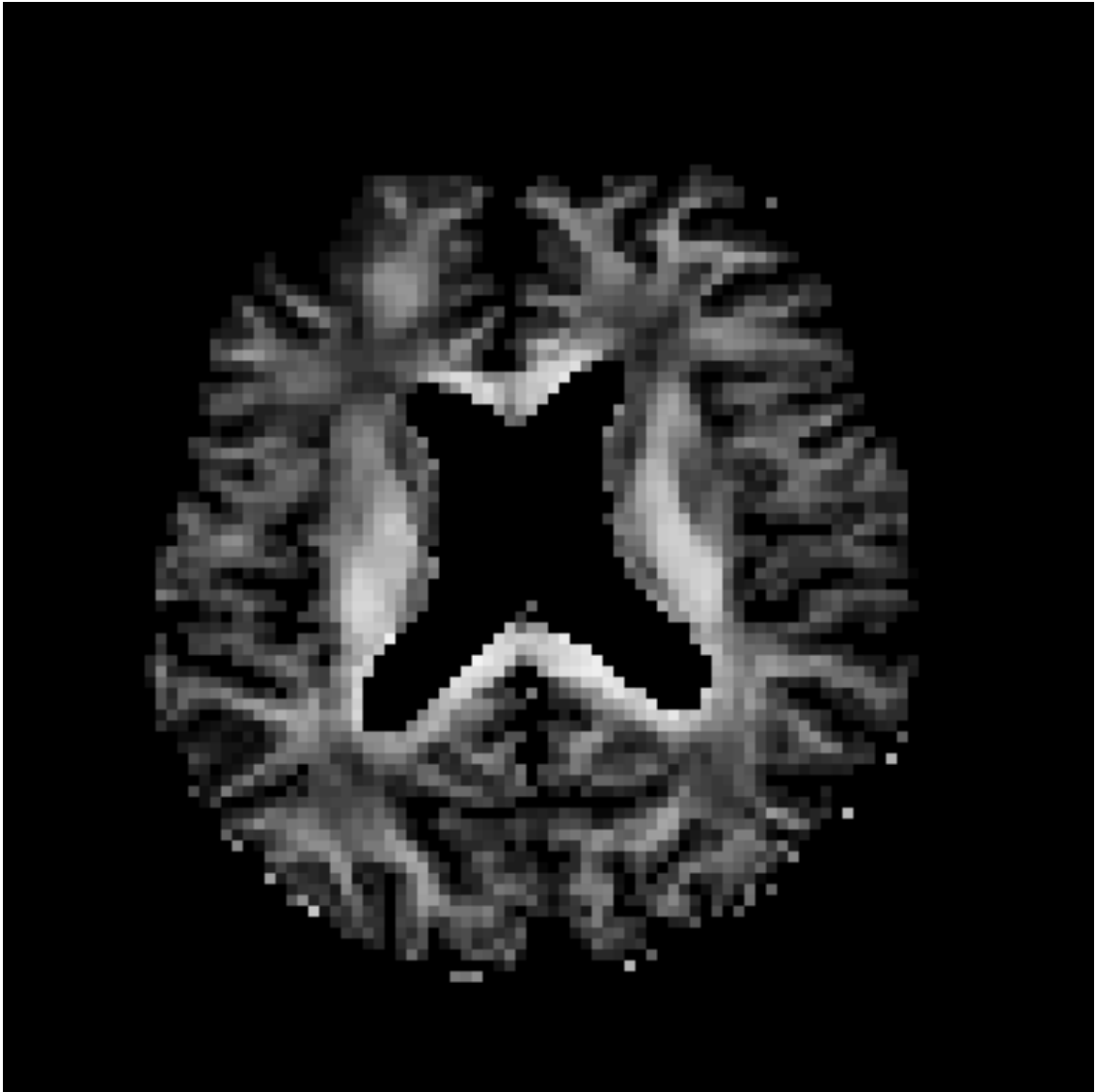}
    \end{subfigure}
    \begin{subfigure}{0.04\textwidth}
        \includegraphics[height=0.8in]{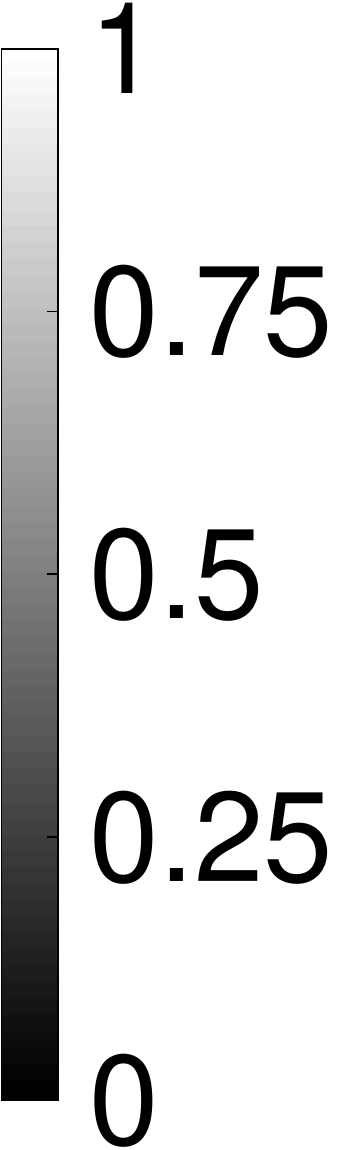}
    \end{subfigure}
    \begin{subfigure}{0.02\textwidth}
        \raggedright
        \raisebox{0in}{\rotatebox[origin=t]{-90}{\small FA}}
    \end{subfigure}

    \begin{subfigure}{0.02\textwidth}
        \raisebox{0in}{\rotatebox[origin=t]{90}{5/8 PF}}
    \end{subfigure}
    \begin{subfigure}{0.15\textwidth}
        \includegraphics[width=1\linewidth]{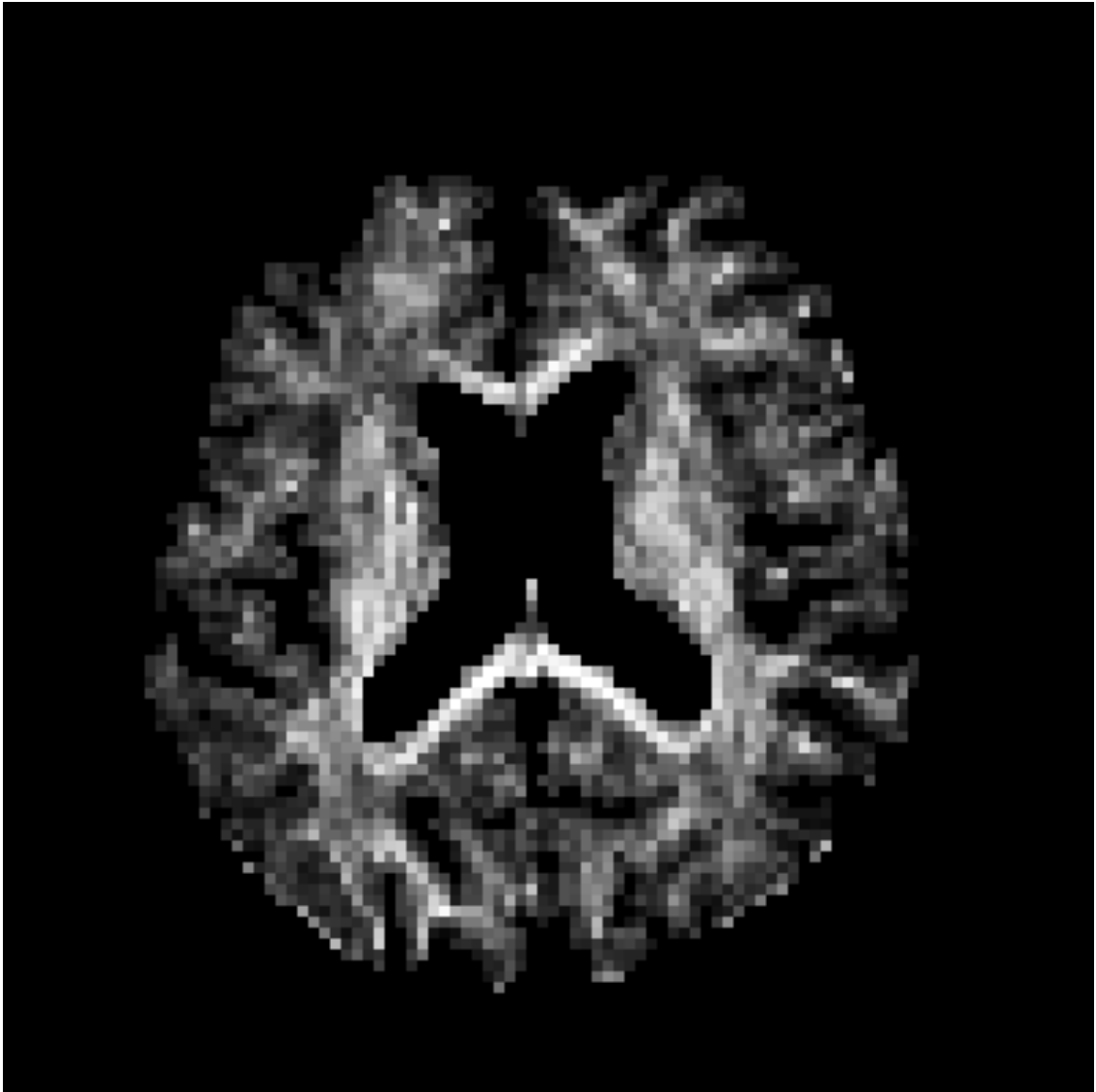}
    \end{subfigure}
    \begin{subfigure}{0.15\textwidth}
        \includegraphics[width=1\linewidth]{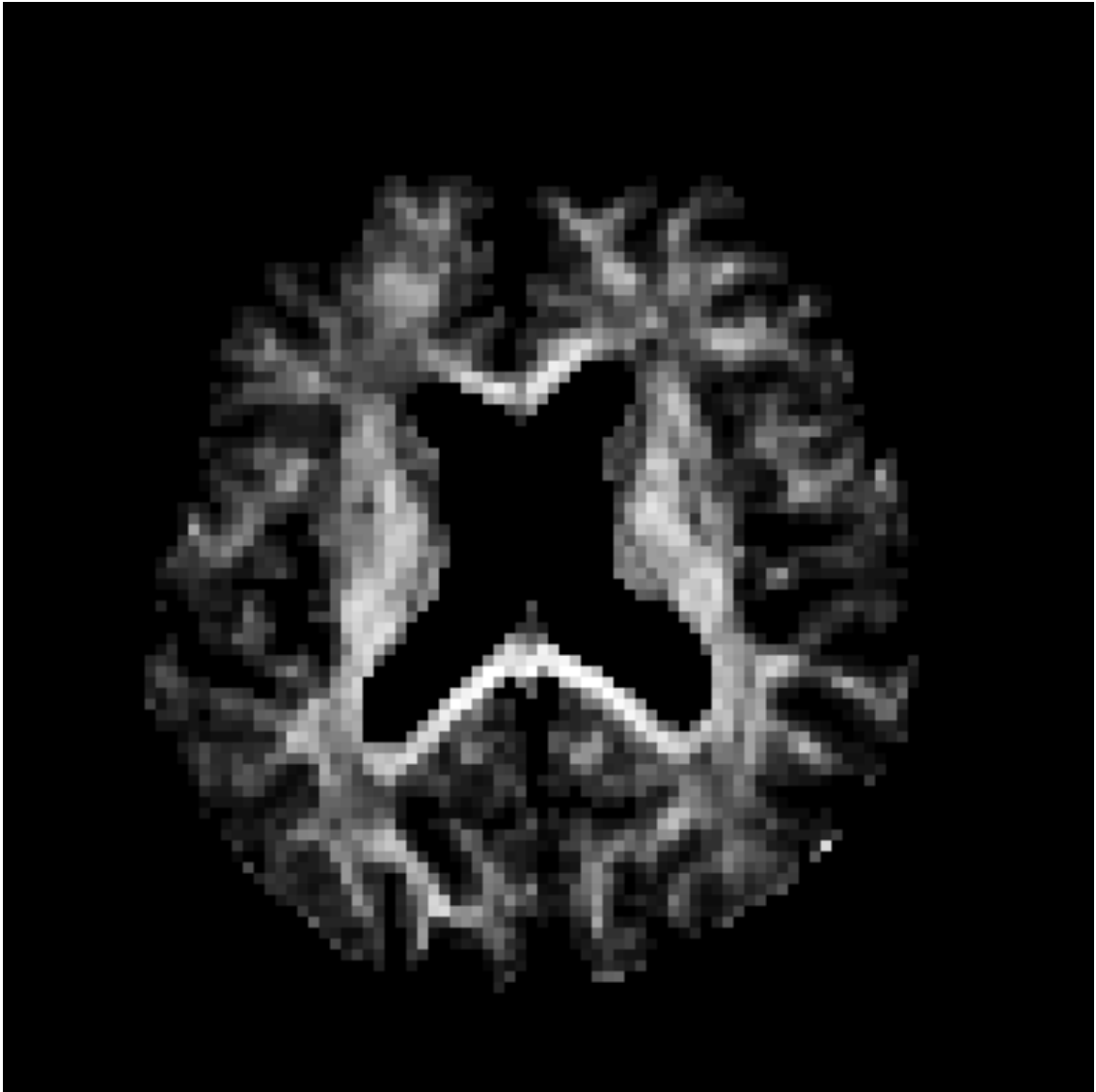}
    \end{subfigure}
    \begin{subfigure}{0.15\textwidth}
        \includegraphics[width=1\linewidth]{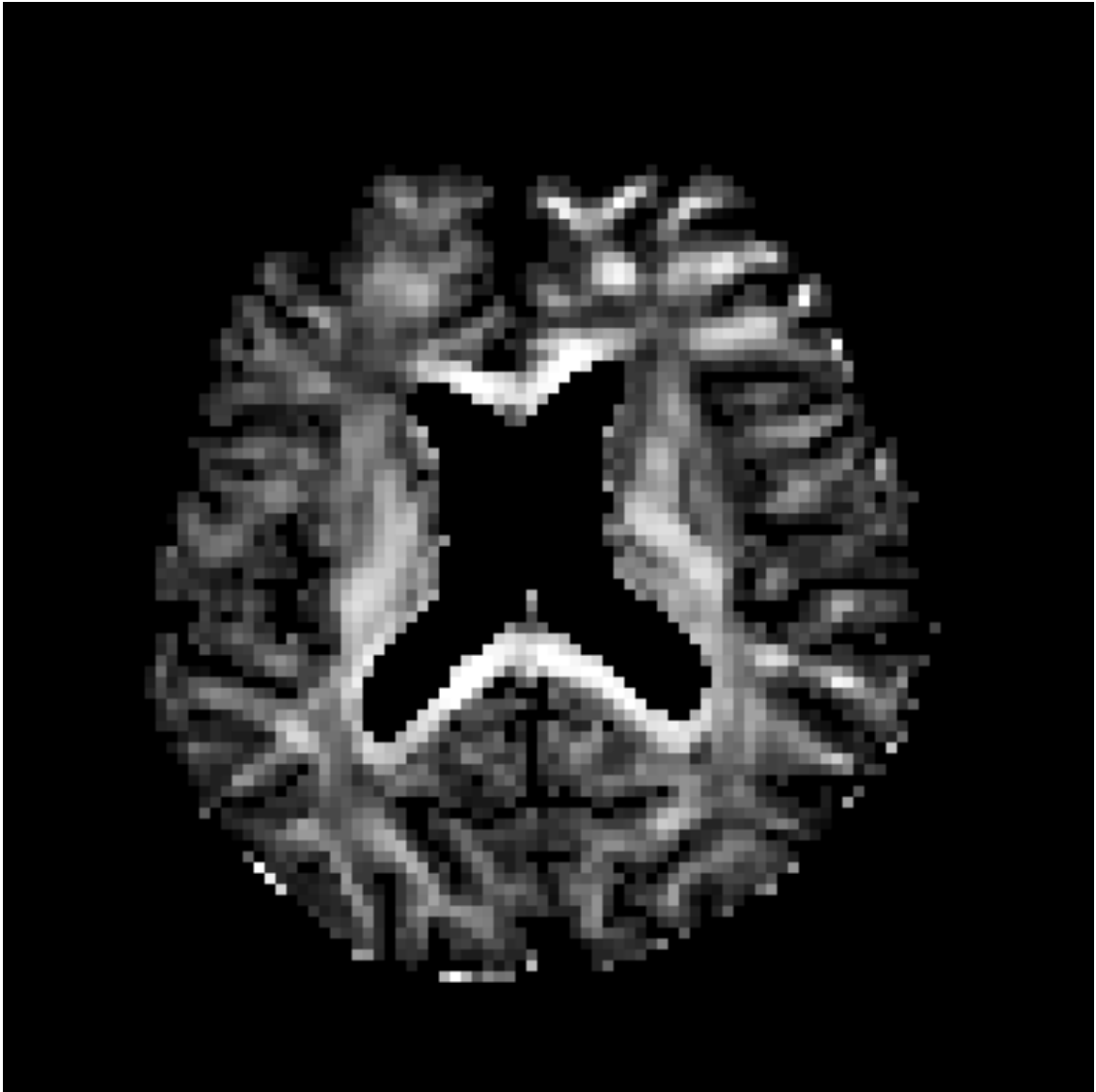}
    \end{subfigure}
    \begin{subfigure}{0.15\textwidth}
        \includegraphics[width=1\linewidth]{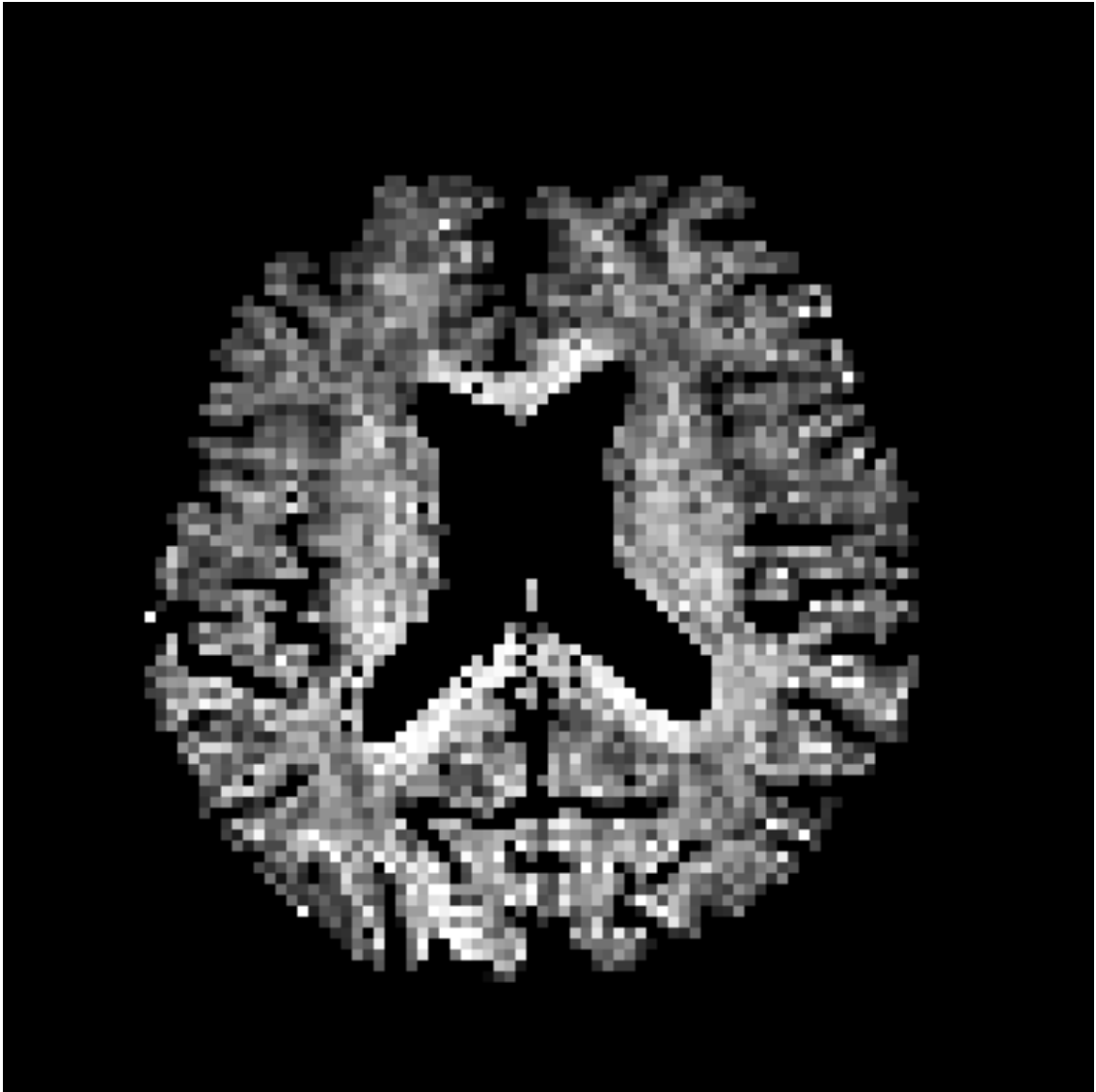}
    \end{subfigure}
    \begin{subfigure}{0.15\textwidth}
        \includegraphics[width=1\linewidth]{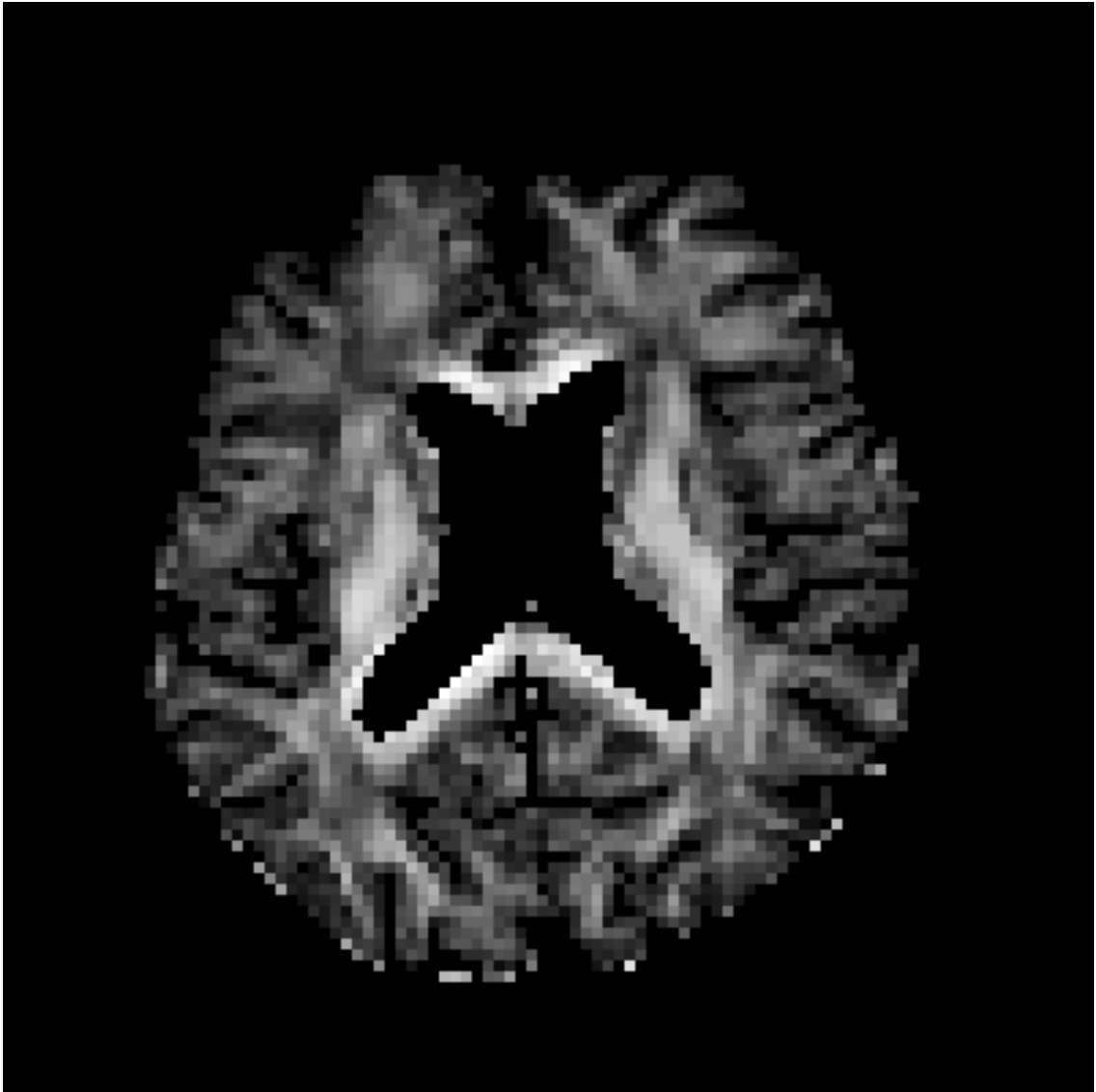}
    \end{subfigure}
    \begin{subfigure}{0.04\textwidth}
        \includegraphics[height=0.8in]{figures/fa_colorbar-eps-converted-to.pdf}
    \end{subfigure}
    \begin{subfigure}{0.02\textwidth}
        \raisebox{0in}{\rotatebox[origin=t]{-90}{\small FA}}
    \end{subfigure}

    \begin{subfigure}{0.02\textwidth}
        \raggedright
        \raisebox{0in}{\rotatebox[origin=t]{90}{No PF}}
    \end{subfigure}
    \begin{subfigure}{0.15\textwidth}
        \includegraphics[width=1\linewidth]{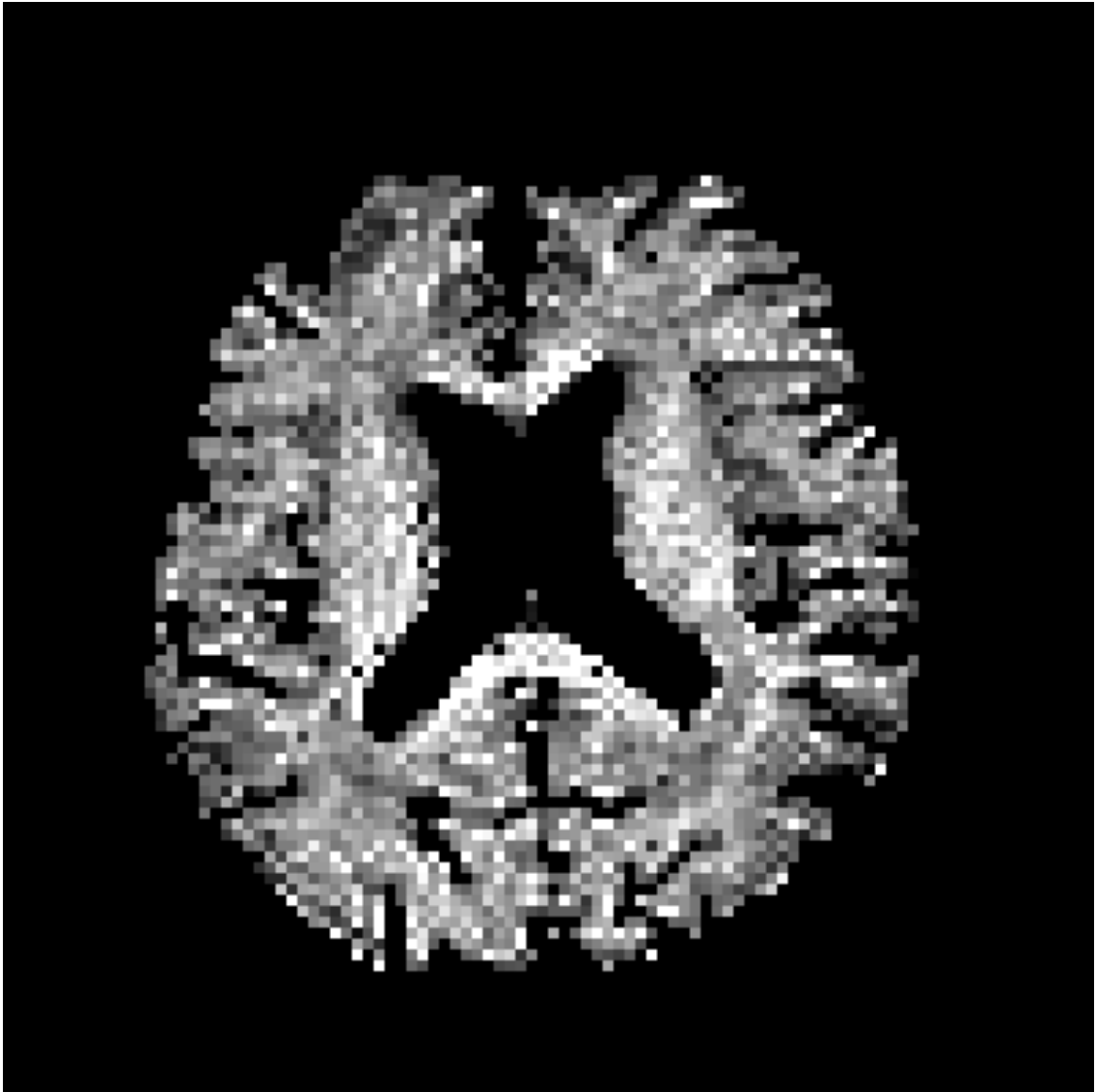}
    \end{subfigure}
    \begin{subfigure}{0.15\textwidth}
        \includegraphics[width=1\linewidth]{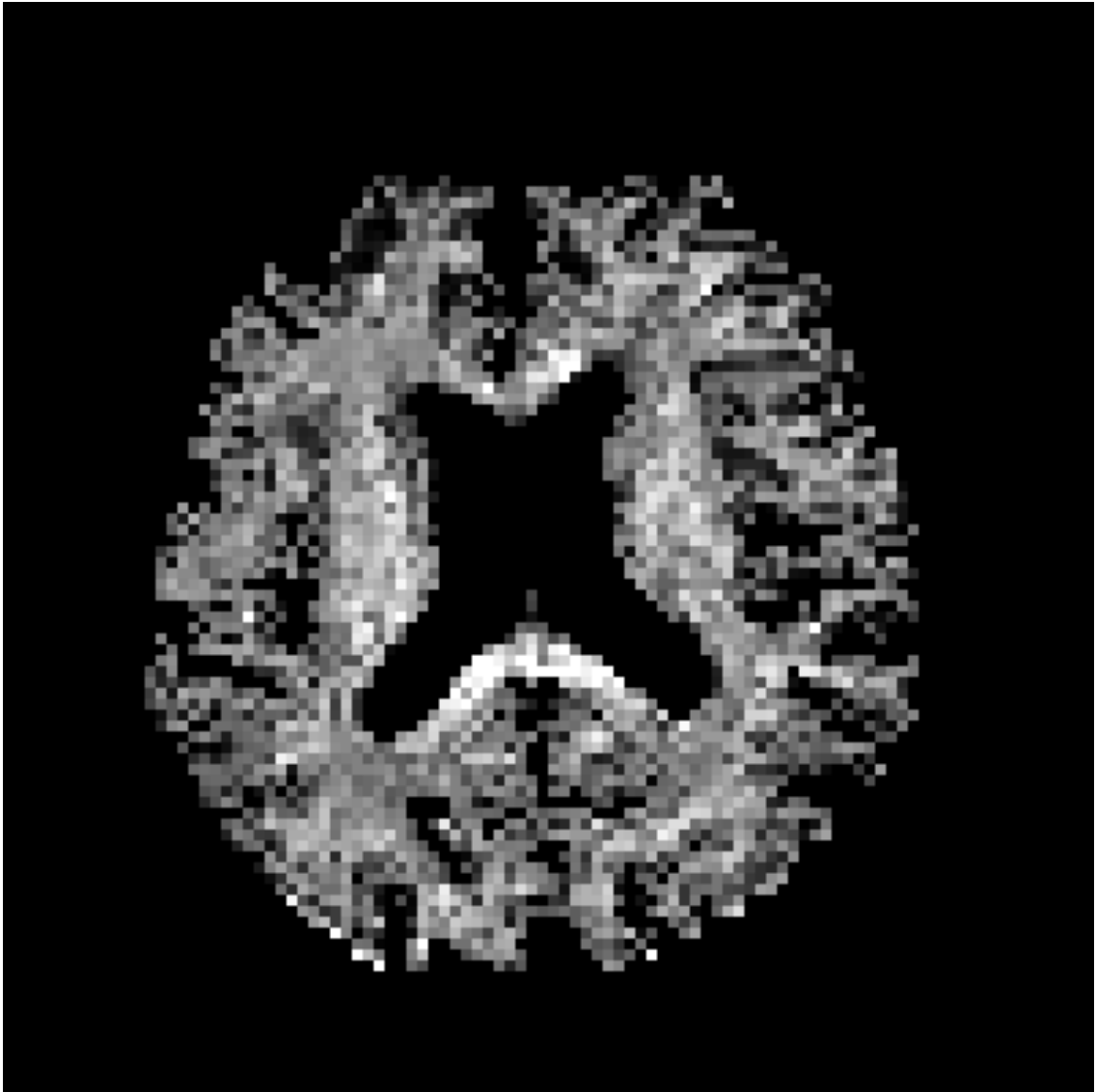}
    \end{subfigure}
    \begin{subfigure}{0.15\textwidth}
        \includegraphics[width=1\linewidth]{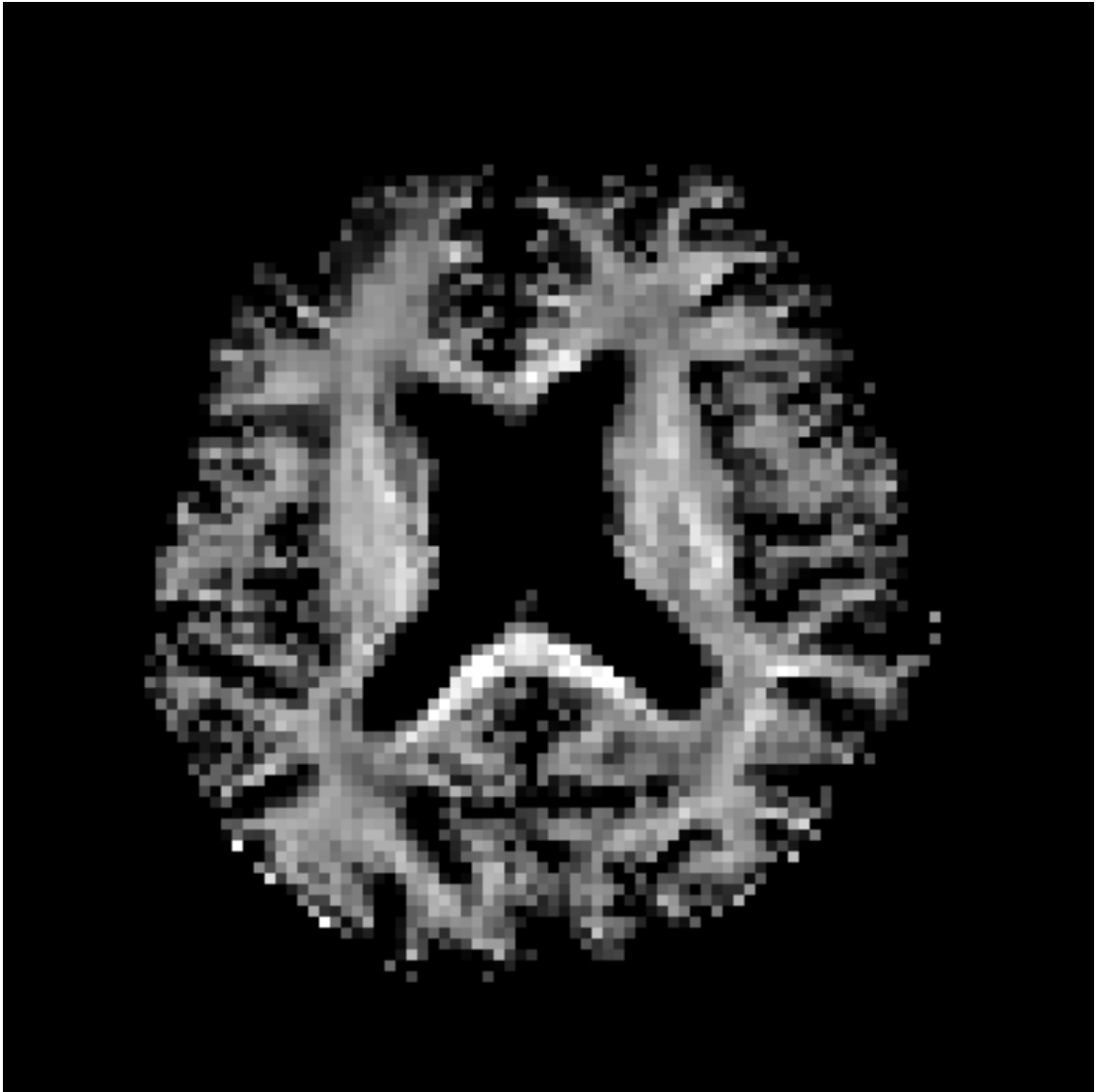}
    \end{subfigure}
    \begin{subfigure}{0.15\textwidth}
        \includegraphics[width=1\linewidth]{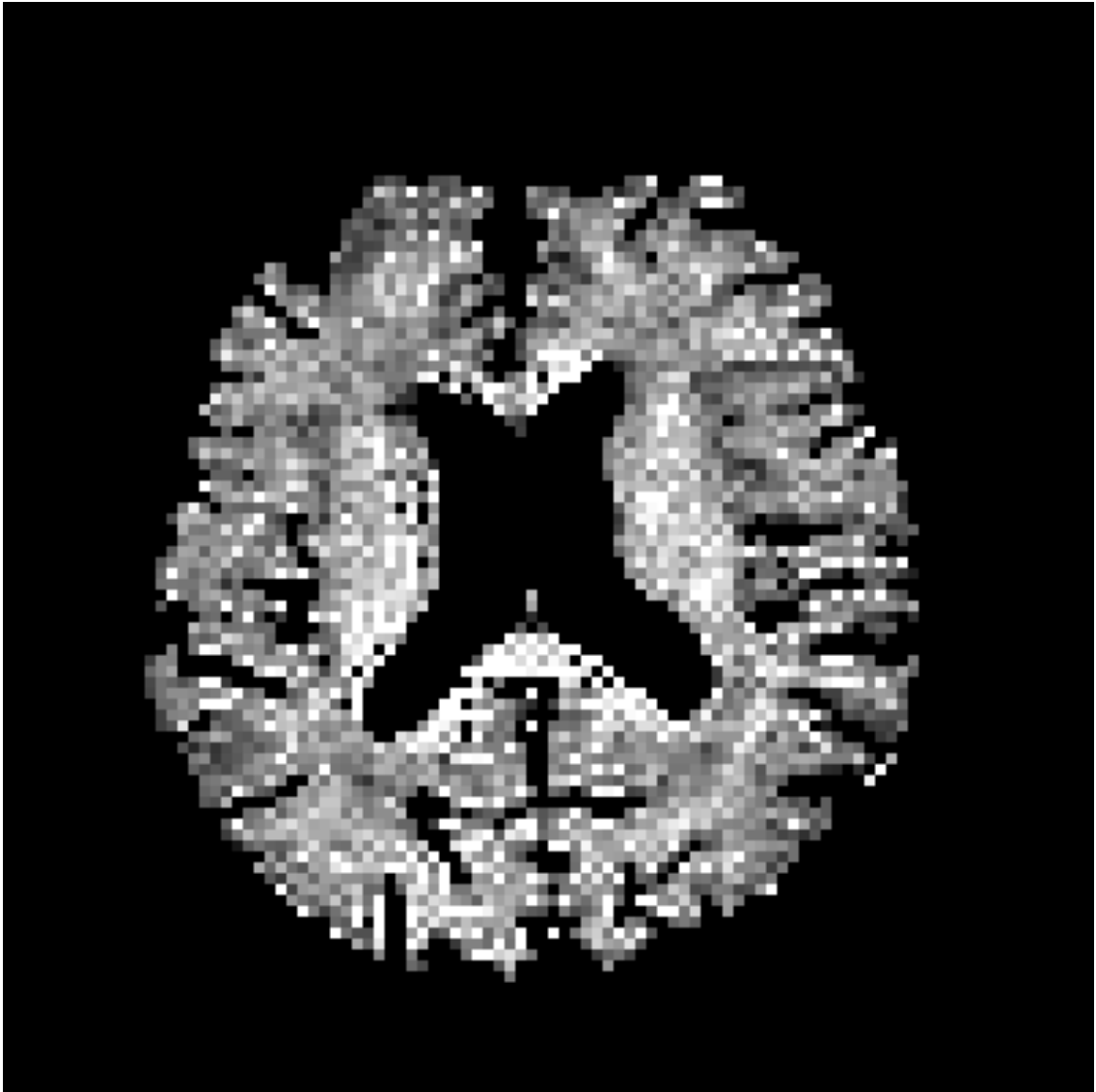}
    \end{subfigure}
    \begin{subfigure}{0.15\textwidth}
        \includegraphics[width=1\linewidth]{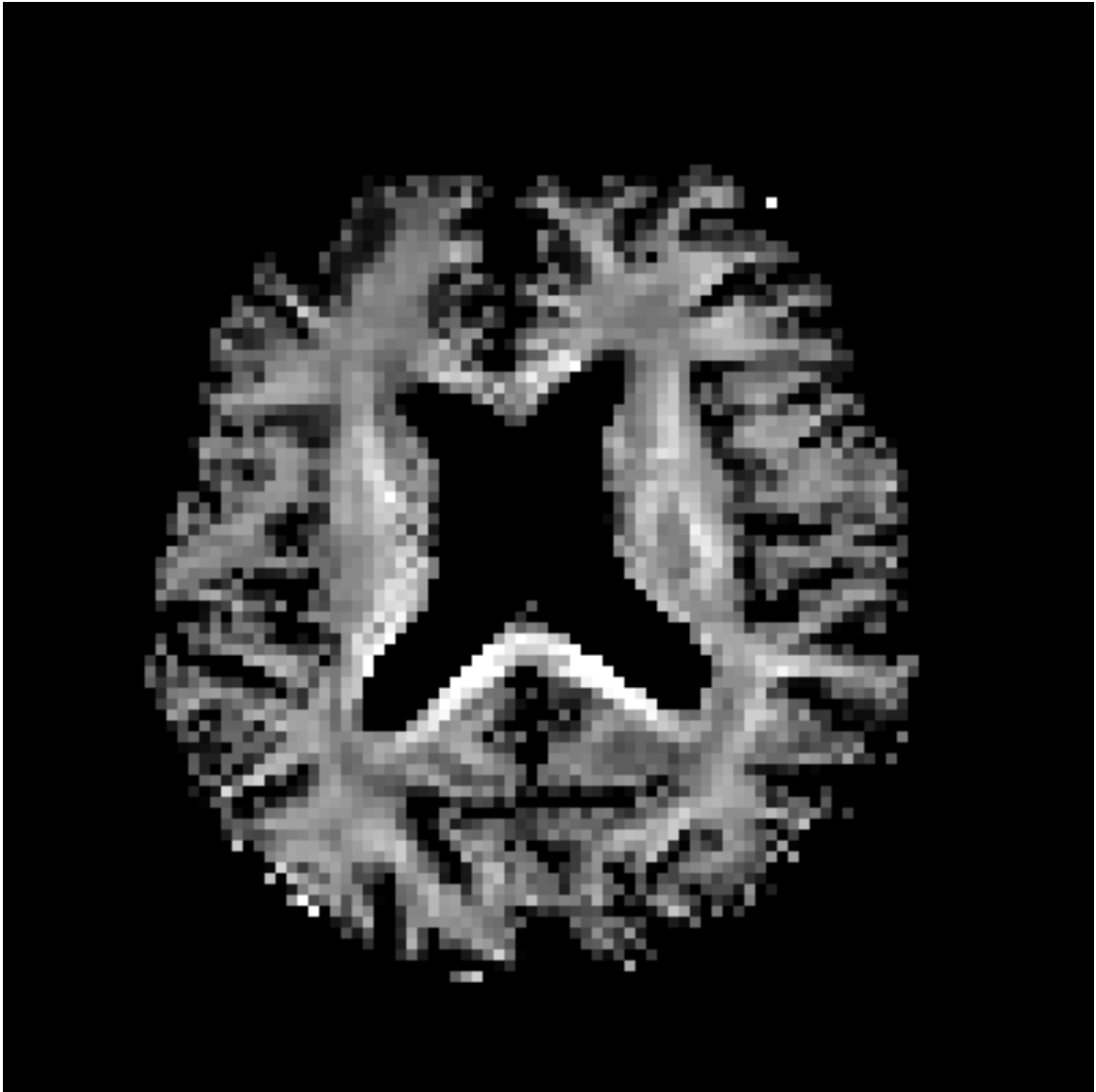}
    \end{subfigure}
    \begin{subfigure}{0.04\textwidth}
        \includegraphics[height=0.8in]{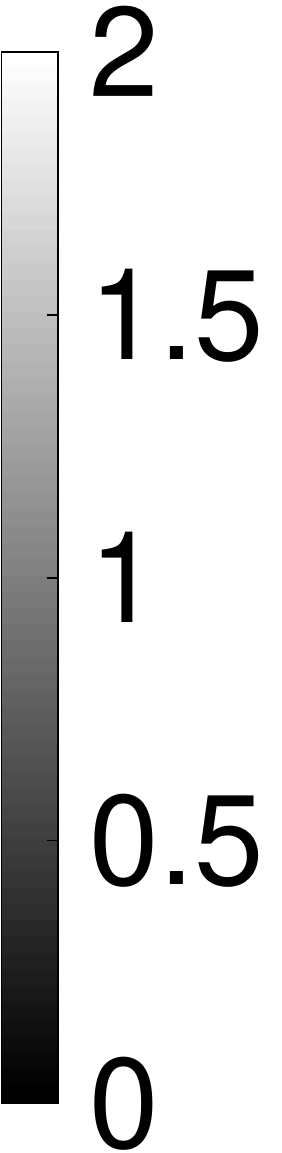}
    \end{subfigure}
    \begin{subfigure}{0.02\textwidth}
        \raisebox{0in}{\rotatebox[origin=t]{-90}{\small MK}}
    \end{subfigure}

    \begin{subfigure}{0.02\textwidth}
        \raggedright
        \raisebox{0in}{\rotatebox[origin=t]{90}{5/8 PF}}
    \end{subfigure}
    \begin{subfigure}{0.15\textwidth}
        \includegraphics[width=1\linewidth]{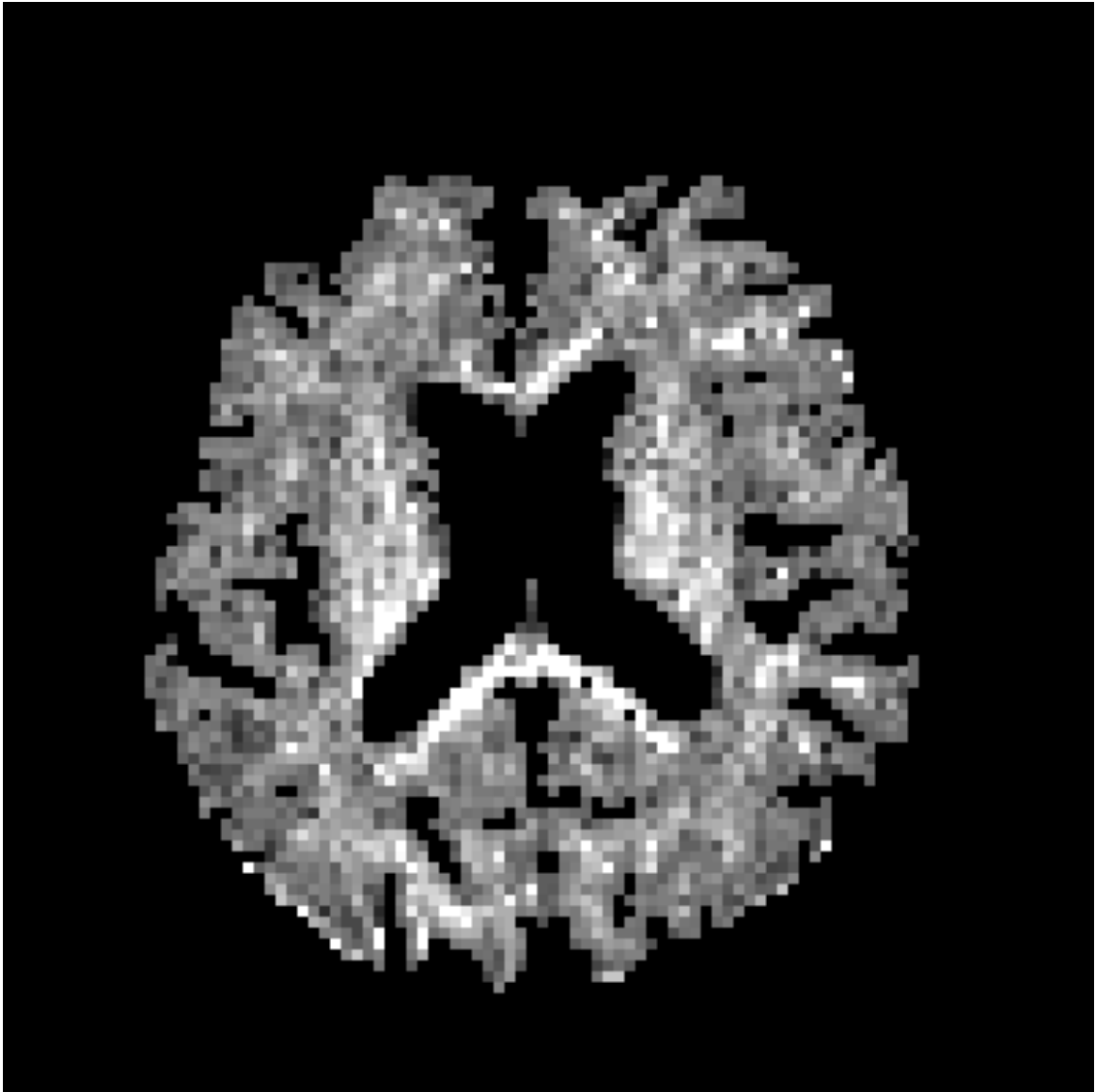}
    \end{subfigure}
    \begin{subfigure}{0.15\textwidth}
        \includegraphics[width=1\linewidth]{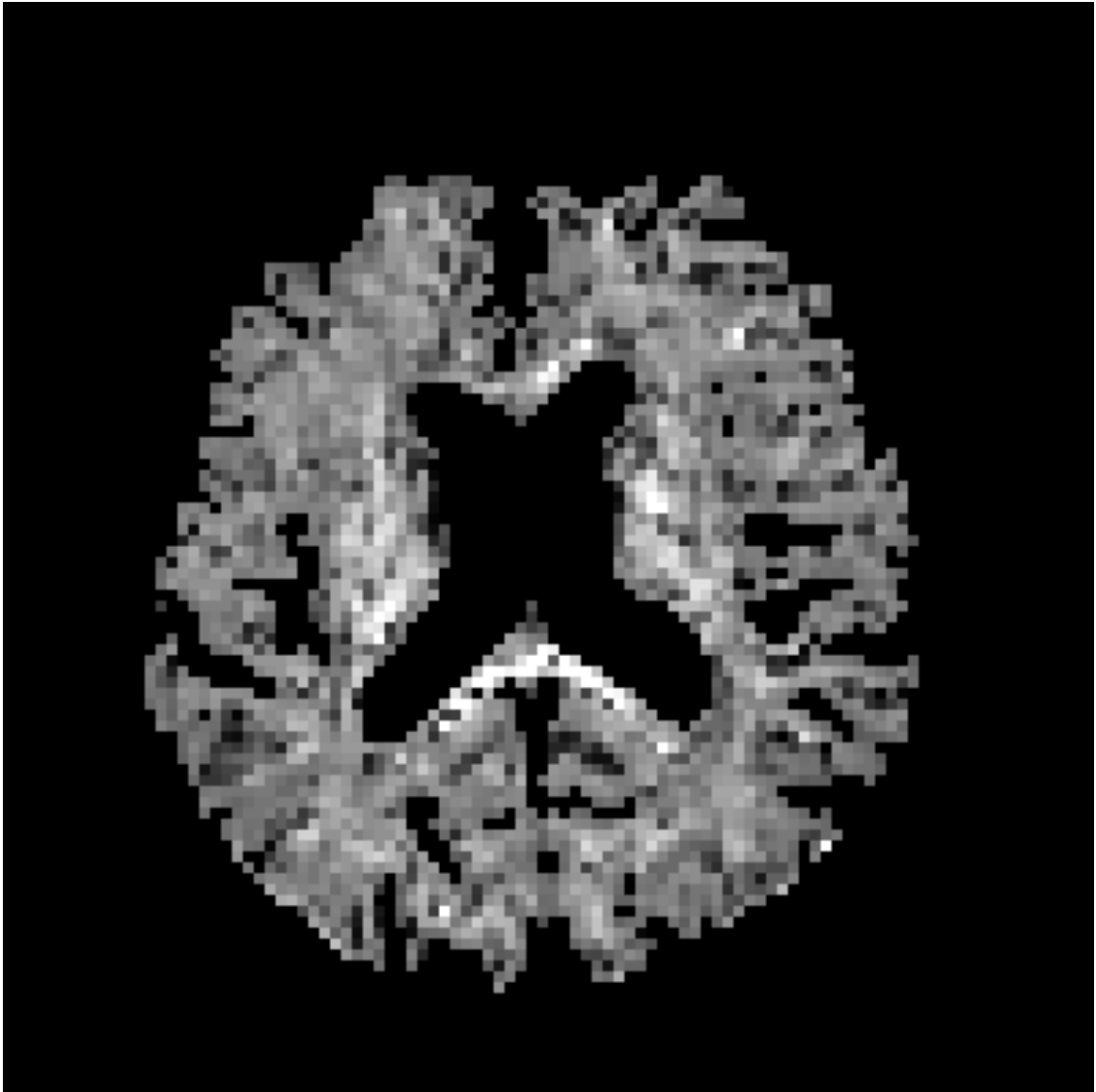}
    \end{subfigure}
    \begin{subfigure}{0.15\textwidth}
        \includegraphics[width=1\linewidth]{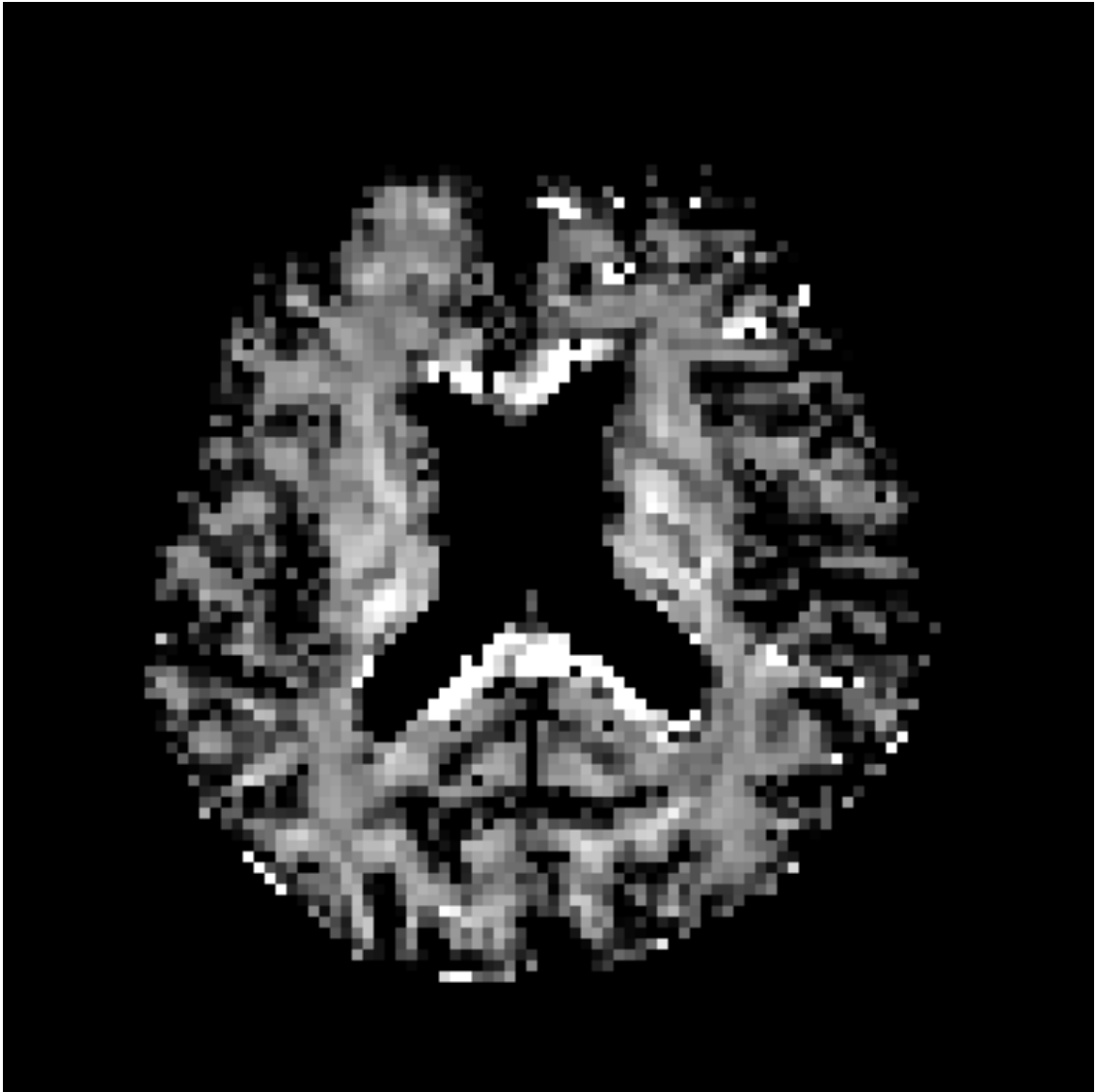}
    \end{subfigure}
    \begin{subfigure}{0.15\textwidth}
        \includegraphics[width=1\linewidth]{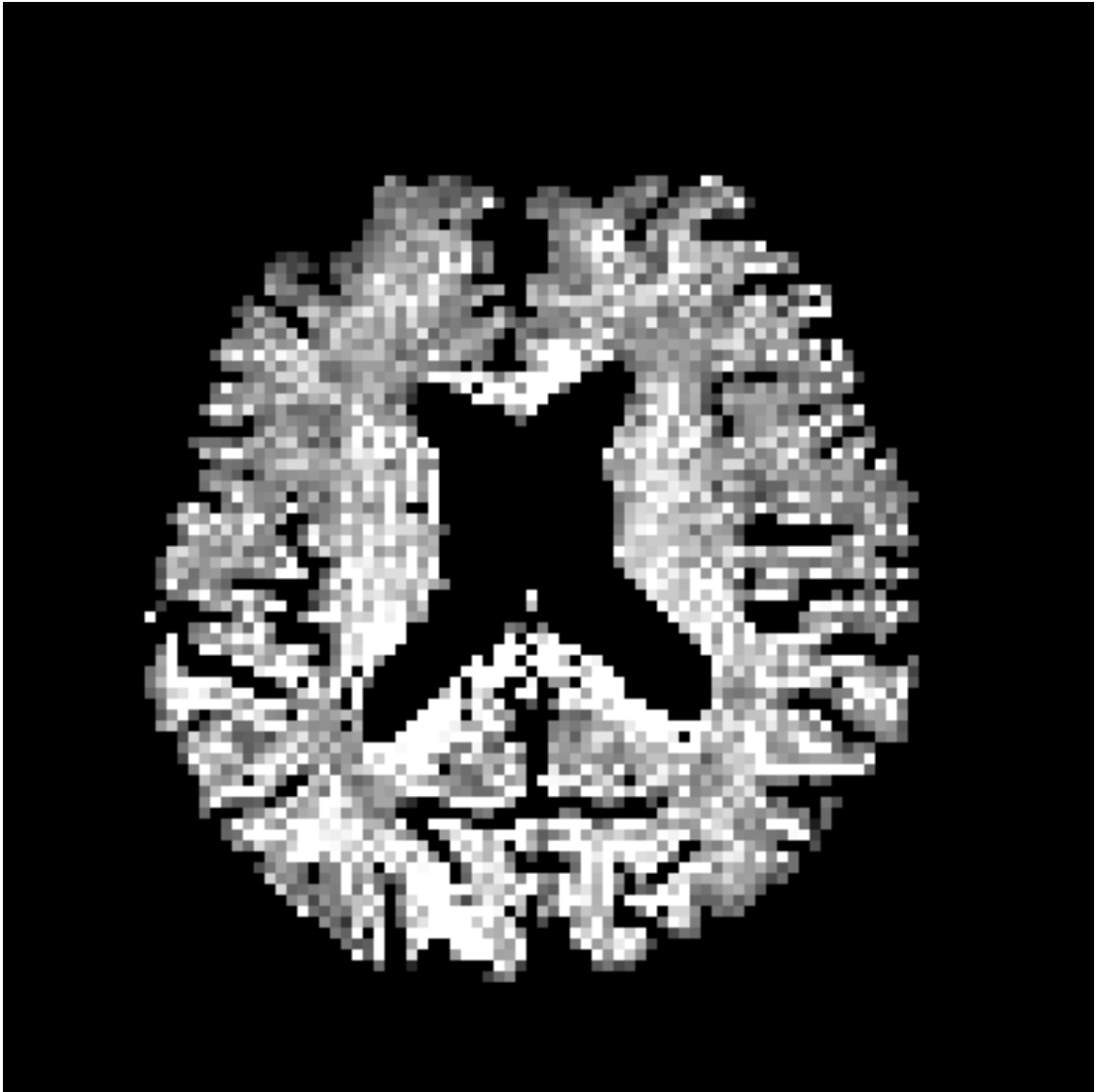}
    \end{subfigure}
    \begin{subfigure}{0.15\textwidth}
        \includegraphics[width=1\linewidth]{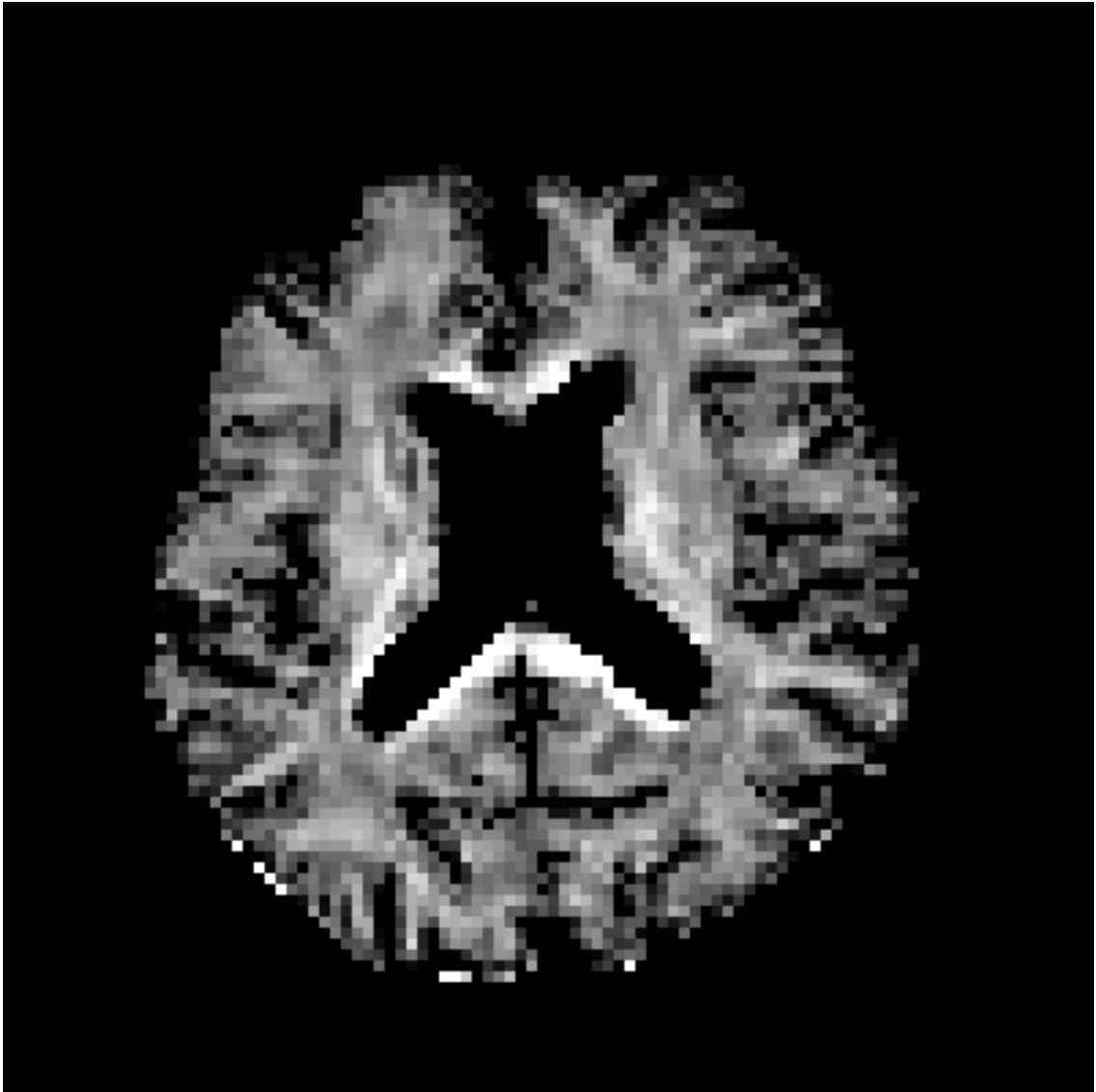}
    \end{subfigure}
    \begin{subfigure}{0.04\textwidth}
        \includegraphics[height=0.8in]{figures/mk_colorbar-eps-converted-to.pdf}
    \end{subfigure}
    \begin{subfigure}{0.02\textwidth}
        \raisebox{0in}{\rotatebox[origin=t]{-90}{\small MK}}
    \end{subfigure}
    \caption{Parameter maps from various methods (with CSF masks for FA and MK). At the top are shown the $b=0$ images from Raw, state-of-the-art (SoA), MCNN, standard partial Fourier, and CCNN methods. Rows 2 and 3 show results for mean diffusivity, rows 4 and 5 show results for fractional anisotropy, and rows 6 and 7 show results for mean kurtosis. The SoA method and both deep learning methods perform well without partial Fourier acceleration; however, at the 5/8ths partial Fourier factor, substantial artifacts are present for all methods other than the CCNN method.}
    \label{fig:meth_comp}
\end{figure*}

Figure \ref{fig:meth_comp} shows parameter maps for the different methods without partial Fourier (No PF) and with 5/8ths partial Fourier (5/8 PF). Mean diffusivities calculated from the raw DWI data (Raw) show notable noise and Gibbs ringing artifacts, while this is substantially removed with the state-of-the-art (SoA) method. However, the state-of-the-art method begins to lose its ability to compensate for the Gibbs ringing and resolution loss when partial Fourier is utilized in the acquisition. The effects of partial Fourier increases are primarily evident in the enlargement of the lateral ventricles and the presence of black lines in the vicinity of the lateral ventricles. The MCNN model is able to compensate somewhat for the ringing effects, but begins to introduce substantial artifacts at the 5/8ths partial Fourier factor, whereas the CCNN model continues to give high-quality mean parameter maps across all PF factors.
Similar trends are observed in the other diffusion parameter maps.

\begin{figure*}[htb]
    \centering
    \begin{subfigure}{0.02\textwidth}
        \raisebox{0in}{\rotatebox[origin=t]{90}{}}
    \end{subfigure}
    \begin{subfigure}{0.18\textwidth}
        \centering
        {No PF}
    \end{subfigure}
    \begin{subfigure}{0.18\textwidth}
        \centering
        {7/8 PF}
    \end{subfigure}
    \begin{subfigure}{0.18\textwidth}
        \centering
        {6/8 PF}
    \end{subfigure}
    \begin{subfigure}{0.18\textwidth}
        \centering
        {5/8 PF}
    \end{subfigure}
    \begin{subfigure}{0.04\textwidth}
        \hspace{1mm}
    \end{subfigure}
    \begin{subfigure}{0.02\textwidth}
        \raisebox{0in}{\rotatebox[origin=t]{-90}{\hspace{1mm}}}
    \end{subfigure}

    \begin{subfigure}{0.02\textwidth}
        \raggedright
        \raisebox{0in}{\rotatebox[origin=t]{90}{SoA}}
    \end{subfigure}
    \begin{subfigure}{0.18\textwidth}
        \includegraphics[width=1\linewidth]{figures/mppca_kellner_pf_8_8_md-eps-converted-to.pdf}
    \end{subfigure}
    \begin{subfigure}{0.18\textwidth}
        \includegraphics[width=1\linewidth]{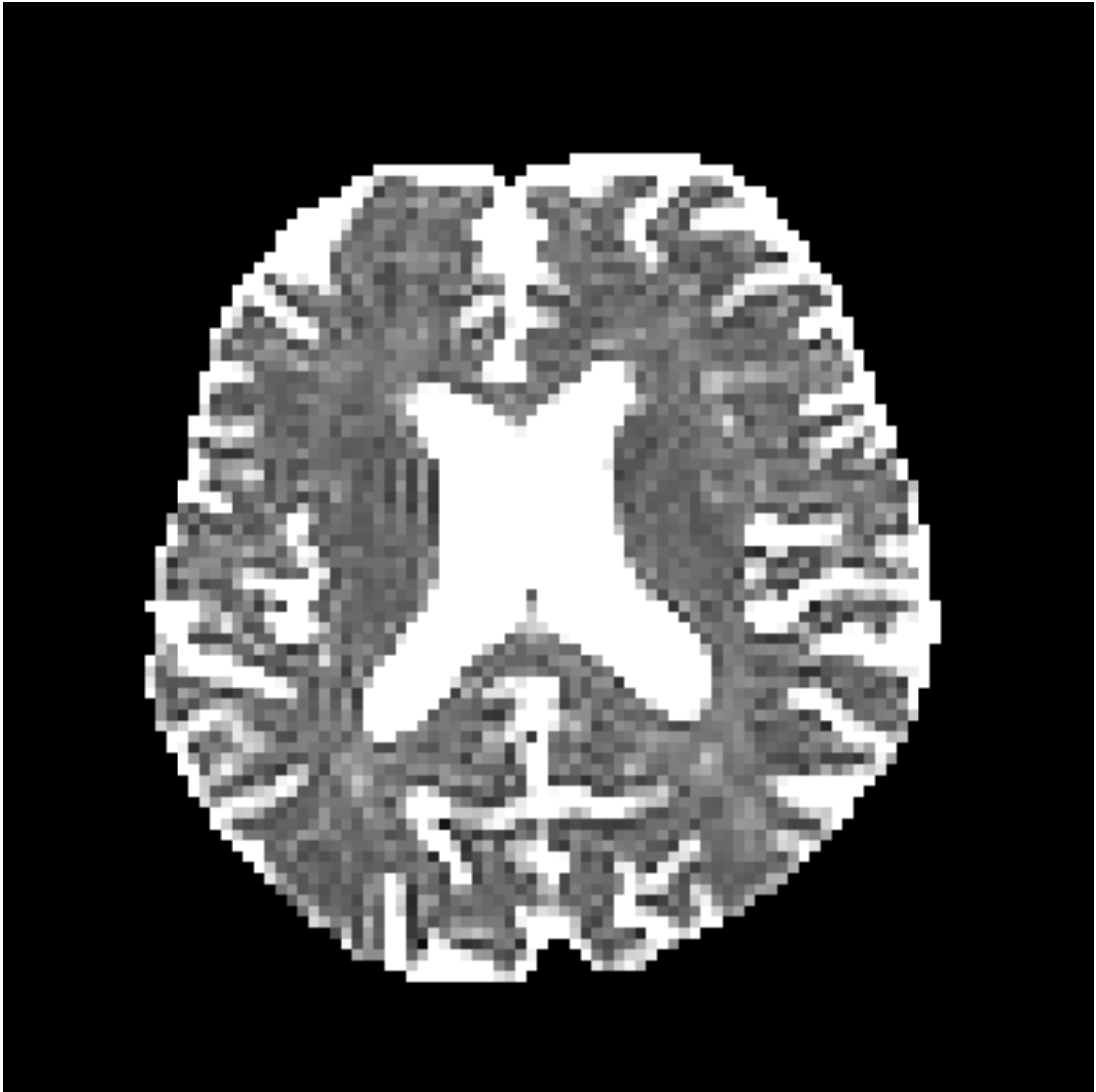}
    \end{subfigure}
    \begin{subfigure}{0.18\textwidth}
        \includegraphics[width=1\linewidth]{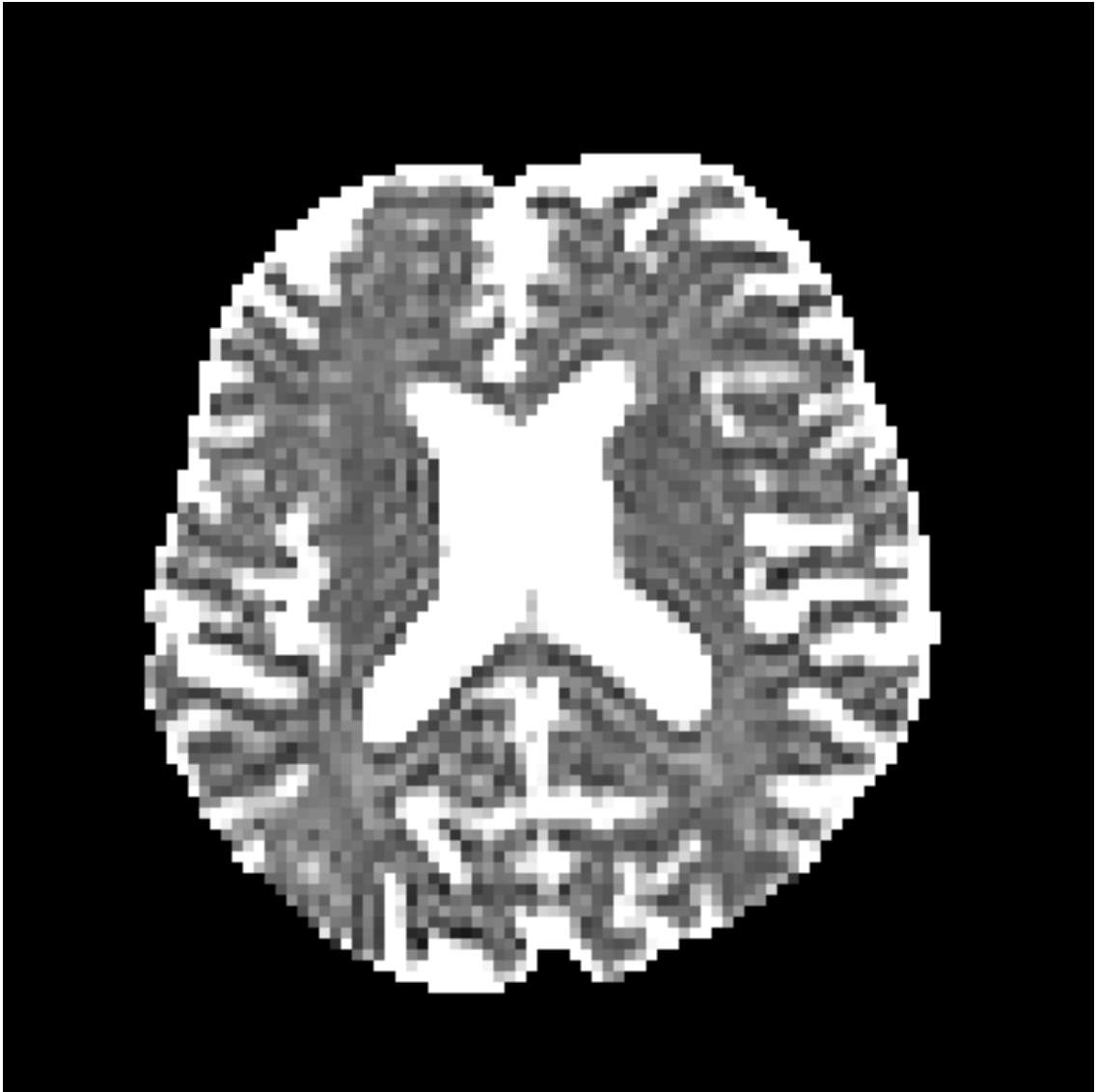}
    \end{subfigure}
    \begin{subfigure}{0.18\textwidth}
        \includegraphics[width=1\linewidth]{figures/mppca_kellner_pf_5_8_md-eps-converted-to.pdf}
    \end{subfigure}
    \begin{subfigure}{0.04\textwidth}
        \includegraphics[height=0.9in]{figures/md_colorbar-eps-converted-to.pdf}
    \end{subfigure}
    \begin{subfigure}{0.02\textwidth}
        \raisebox{0in}{\rotatebox[origin=t]{-90}{MD, $\mu$m$^2$/ms}}
    \end{subfigure}

    \begin{subfigure}{0.02\textwidth}
        \raggedright
        \raisebox{0in}{\rotatebox[origin=t]{90}{CCNN}}
    \end{subfigure}
    \begin{subfigure}{0.18\textwidth}
        \includegraphics[width=1\linewidth]{figures/dl_10layers_pf8_comp2mag_md-eps-converted-to.pdf}
    \end{subfigure}
    \begin{subfigure}{0.18\textwidth}
        \includegraphics[width=1\linewidth]{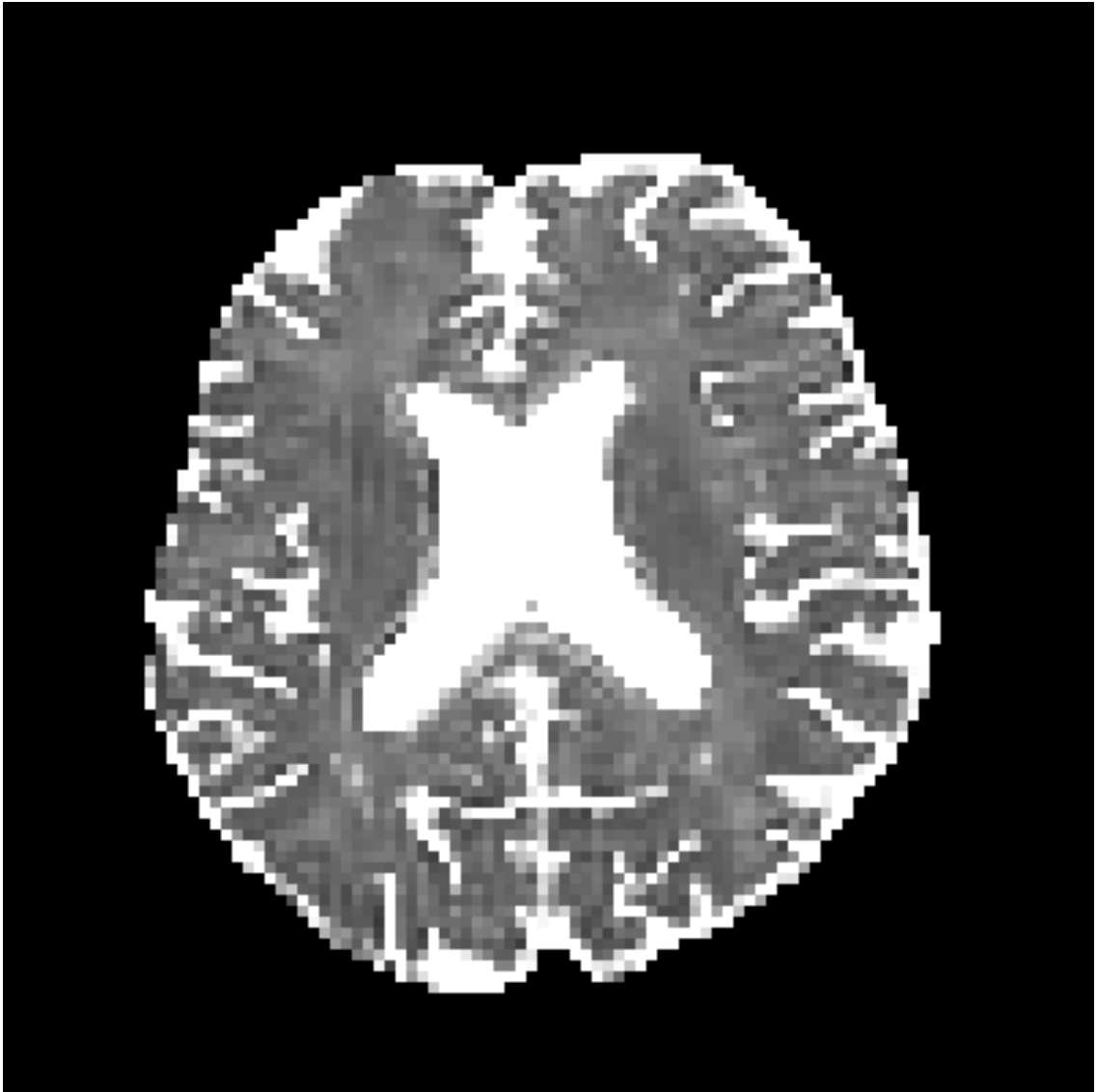}
    \end{subfigure}
    \begin{subfigure}{0.18\textwidth}
        \includegraphics[width=1\linewidth]{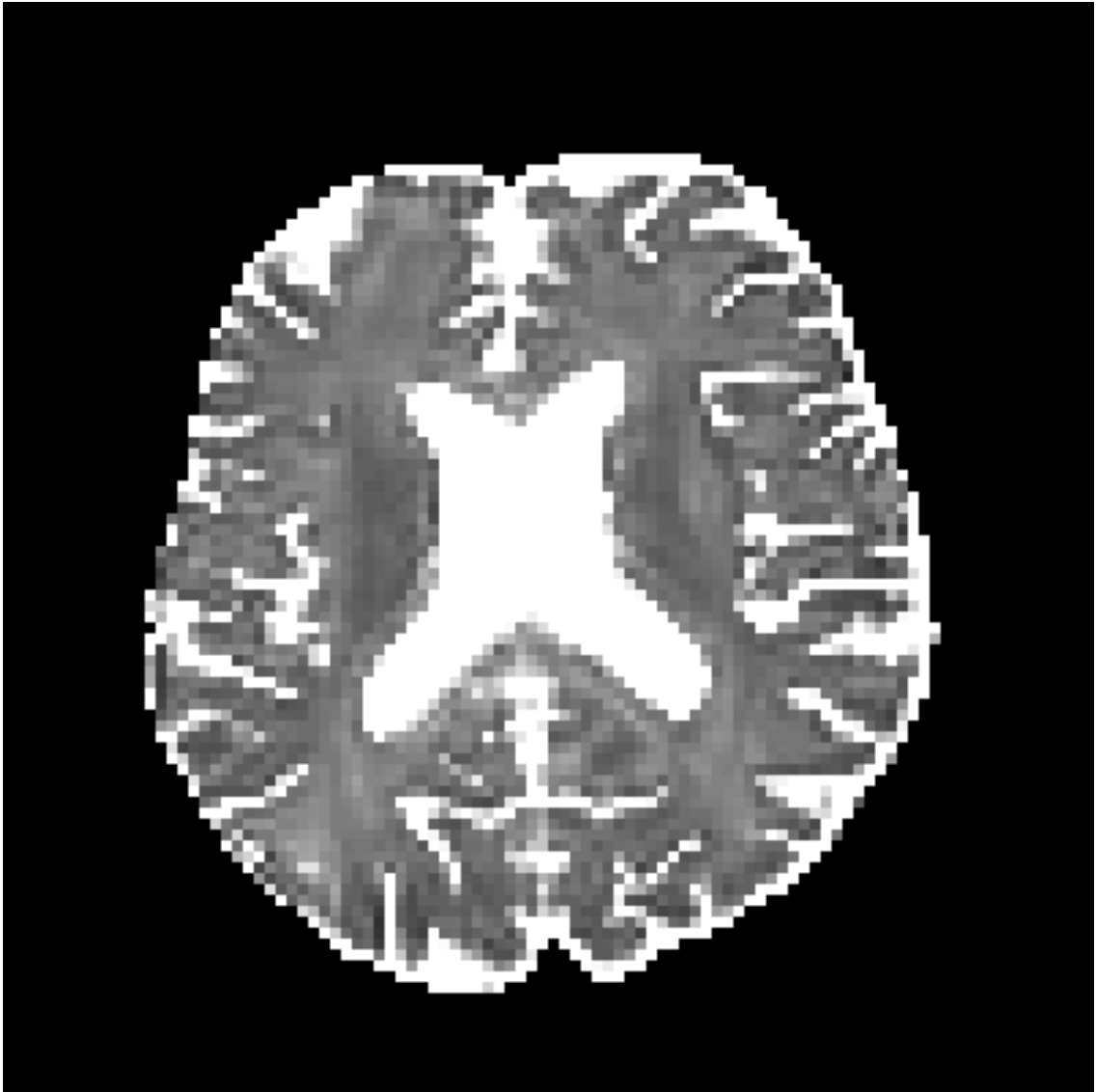}
    \end{subfigure}
    \begin{subfigure}{0.18\textwidth}
        \includegraphics[width=1\linewidth]{figures/dl_10layers_pf5_comp2mag_md-eps-converted-to.pdf}
    \end{subfigure}
    \begin{subfigure}{0.04\textwidth}
        \includegraphics[height=0.9in]{figures/md_colorbar-eps-converted-to.pdf}
    \end{subfigure}
    \begin{subfigure}{0.02\textwidth}
        \raisebox{0in}{\rotatebox[origin=t]{-90}{MD, $\mu$m$^2$/ms}}
    \end{subfigure}
    \caption{Comparison of mean diffusivity (MD) parameter maps across PF factors of 5/8ths, 6/8ths, 7/8ths, and without PF (No PF). The methods include state-of-the-art (SoA) and CCNN methods. Without partial Fourier, both methods are similar; however, as the PF factor increases, substantial artifacts are introduced in the SoA method, particularly around the lateral ventricles. The CCNN method is able to reduce the introduction of artifacts.}
    \label{fig:pf_fact}
\end{figure*}

Figure \ref{fig:pf_fact} compares mean diffusivity maps across various partial Fourier factors between the CCNN and state-of-the-art methods.
The methods perform similarly without partial Fourier acceleration, but as partial Fourier acceleration increases, the image is continually degraded in the SoA method, with "black voxels" appearing around the lateral ventricles. The CCNN method mitigates the appearance of these artifacts in the parameter maps.

\section{Discussion}
\label{sec:discussion}
Our results indicate that with carefully-crafted simulations, it is possible to train CNN models based solely on MR encoding operators that can be applied to in vivo MR imaging data. We characterized the performance of this method with canonical experiments, showing Gibbs removal capability with hard edges, as well as providing preliminary evidence that the network's denoising capability depends on signal-adaptive smoothing, and that this smoothing largely takes place at contrast-to-noise ratios below 1. In principle there are no restrictions on applying the method in new applications other than retraining based on the application image matrix size. Our chosen in vivo setting here was diffusion parameter mapping due to the relevance of the artifacts and the sensitivity of diffusion parameter mapping to Gibbs effects. Nonetheless, we would recommend conducting independent performance studies for each application prior to using the method.

We found that it is possible to recover the  high-resolution information missing in partial Fourier acquisitions while simultaneously reducing Gibbs artifacts and suppressing noise in a way that is robust enough for diffusion parameter estimation. Figures \ref{fig:cnrresults} and \ref{fig:spectresponse} suggest that the primary determinant of resolution with the proposed models is SNR rather than PF factor. The CCNN method performed better than the MCNN method in our experiments, both qualitatively and quantitatively. Although DWI images from both methods were superior to the raw, unprocessed images, in the presence of partial Fourier encoding the MCNN method introduced artifacts that were subtle in DWIs and readily apparent in the parameter maps in Figure \ref{fig:meth_comp}. This failure could have arisen due to systematic artifacts introduced by the MCNN - an effect that is known to be possible with the use of CNNs \cite{li2018detection}. The CCNN model, on the other hand, may be able to use the smoothness of the phase to estimate missing k-space areas - something which is not possible after the magnitude operation has been applied. It may be possible to adapt the MCNN method in the future to handle the ripple effects seen in Figures \ref{fig:dl_meth_dwi} and \ref{fig:meth_comp} - we leave this investigation to future work.

Practically, the success of the approach can be attributed to 1) having a sufficiently deep neural network able to adequately capture the difference between Gibbs-corrupted and ground-truth images and 2) to the heterogeneity of the synthetic training set and the richness of the encoding simulations providing ample training data. During training, the models saw over 10 million candidate phase maps simulated on over 1 million baseline images. Data sets of this size are difficult to acquire in medical imaging, and impossible to acquire for the purpose of simulating Gibbs artifacts in diffusion-weighted imaging. The use of millions of parameters in the design of CNNs, although standard practice in deep learning, includes overfitting risk. Training such a model requires careful consideration of the simulation process and its relationship with the test dataset. As we see in Figures \ref{fig:dl_meth_dwi} and \ref{fig:meth_comp}, subtle or benign artifacts in the DWIs are substantially magnified in the corresponding parameter maps. To consider the effects of overfitting, we designed our experiments to have not only separate training, validation, and test data sets as different groups of images, but to have the test phase of the experiments consider images of a completely different class. Prior to our diffusion imaging experiments, we did not use MR images. Implicitly, this suggests that we would observe the same nature of artifacts in healthy subjects as those with pathology. Furthermore, training in this way mitigates the method's susceptibility to selection bias in the construction of the dataset that would occur with clinical imaging.

Our results were achieved by processing each diffusion-weighted image separately. An advantage of this approach is that our CNN methods can be applied to each image independently for each MR imaging application. Our methods are also readily extendable to other non-diffusion applications that may benefit from Gibbs removal. However, a disadvantage is that the proposed methods do not fully leverage the repetitive correlations present in applications such as diffusion MRI. It is difficult to denoise images at high $b$-values without access to information from low $b$-value images. Such relations are considered in the random matrix denoising method \cite{veraart2016denoising}. This suggests that one avenue to explore in the future for diffusion would be to combine the random matrix theory-based denoising \cite{veraart2016denoising} as a pre-processing step, able to incorporate the correlations between different DWIs, with the subsequently improved performance of CCNN-based removal of Gibbs and other artifacts.

\section{Conclusion}
\label{sec:conclusion}
We have shown that Gibbs artifacts and noise can be substantially reduced by training a convolutional neural network with simulations of the image acquisitions. We demonstrated our method on canonical experiments and in vivo data, and showed its potential for artifact removal. The method can be applied independently on each slice of the imaging dataset, enabling its use in many clinical settings such as clinical diffusion MRI where few diffusion directions may be available. In the future, we will examine possible extensions, including consideration of 3D anatomical structures, correlations across repetitive acquisitions, explicit incorporation of physics into the model, and validation on other clinical applications.

\appendix
\section{Phase Simulation}
\label{app:phasesim}
Inspired by results from functional approximation theory \cite{powell1987radial,carr2001reconstruction} and machine learning \cite{broomhead1988radial}, we approximate the phase as being a summation of Gaussian radial basis functions:
\begin{equation}
    p_{\textit{sim}}(\boldsymbol{r}) = \text{exp} \left (i \sum_{s=1}^S \sum_{b=1}^{B_s} a_{b,s} e^{-\frac{1}{z_s} \mattnorm{\boldsymbol{r}-\boldsymbol{r}_{0,b,s}}^2_{2}} \right ),
    \label{eq:phasemodel}
\end{equation}
where $p_{\textit{sim}}(\boldsymbol{r})$ is the phase map, $\boldsymbol{r}$ is a 2-length vector for spatial position, $a_{b,s}$ is the amplitude of the $(b,s)$th basis function, $\boldsymbol{r}_{0,b,s}$ is the center of the $(b,s)$th basis function, and $z_s$ governs the width of the basis functions in the $s$th subset. In functional approximation theory, formulations like (\ref{eq:phasemodel}) are used to build a candidate for $p_{\textit{sim}}(\boldsymbol{r})$ given a set of points, $(a_{b,s}, \boldsymbol{r}_{0,b,s}), b \in [1, ..., B_s]$ ($z_s$ is usually chosen to be fixed). However, rather than estimate $p_{\textit{sim}}(\boldsymbol{r})$ from a measured cloud of points, our goal is to randomly generate an example from a synthetic cloud of points. For each training example we randomly generate $p_{\textit{sim}}(\boldsymbol{r})$ according to the following model:
\begin{align*}
    S &\sim \text{Poisson}\left ( \lambda_S \right ) \\
    B_s &\sim \text{Poisson}\left ( \lambda_B \right ) \\
    z_{s} &\sim |\mathcal{N}| \left ( \mu_z, \sigma_z \right ) \\
    a_{b,s} &\sim \mathcal{N} \left ( \mu_a, \sigma_a \right ) \\
    r_{0,b,s,j} &\sim \mathcal{U}\left ( r_{\textit{min},j}, r_{\textit{max},j} \right),
\end{align*}
where $\mathcal{N}$ indicates a Gaussian distribution, $|\mathcal{N}|$ is a folded Gaussian distribution, and $\mathcal{U}$ indicates a uniform distribution. The parameters $\lambda_S$, $\lambda_B$, $\mu_z$, $\sigma_z$, $\mu_a$, and $\sigma_a$ are set for training, and $r_{\textit{min},j}$ and $r_{\textit{max},j}$ are the minima and maxima for $\boldsymbol{r}$ in the $j$th imaging dimension. Contrary to practice in the approximation literature, we allow $z_s$ to vary to avoid the degenerate training case where networks create filters to detect Gaussian functions of a fixed width.

When $B_s$ and $S$ are not too large, $p_{\textit{sim}}(\boldsymbol{r})$ will typically be smooth as in real application settings; however, sharp changes can also occur during training when basis function centers are near each other. As a further form of data augmentation, we include a 1\% probability that no phase at all is simulated. Figure \ref{fig:phase_example} shows an example of simulated phase compared to real phase from in vivo data. For our experiments, we used $\lambda_S = 12$, $\lambda_B = 15$, $\mu_z = 64 \text{ pixels}^2$ $\sigma_z = 100 \text{ pixels}^2$, $\mu_a = 0 \degree$, and $\sigma_a = 5 \degree$. We set $r_{\textit{min},j}$ and $r_{\textit{max},j}$ to restrict the center of the Gaussian basis function to the image support in the $j$th dimension.

\section*{acknowledgements}
We would like to thank E. Owens, M. Drozdzal, M. Tygert, and the rest of the fastMRI team for many useful discussions.

\printendnotes

\bibliography{dl_pf}

\end{document}